\documentclass[12pt]{article}
\usepackage{amsfonts,amssymb,graphicx}
\usepackage{fullpage}
\usepackage[tight]{subfigure}
\usepackage{rotating}
\usepackage{sidecap}
\pdfoutput=1
\usepackage{ifpdf}
\ifpdf
  \usepackage{color}
  \definecolor{darkblue}{rgb}{0.3,0.3,0.6}
    \definecolor{darkgreen}{rgb}{0,0.6,0}
  \usepackage[pdftitle={Deformations},pdfauthor={Michael Blaszczyk, Gabriele Honecker, Isabel Koltermann},pdfsubject={String theory},pdfkeywords={string theory},colorlinks=true,linkcolor=black,urlcolor=darkblue,filecolor=darkblue,citecolor=darkblue,menucolor=darkblue,breaklinks=true,bookmarks=true,anchorcolor=black,unicode=true]{hyperref}
\else
  
\fi
\usepackage[numbers,compress]{natbib}
\usepackage{a4wide}
\usepackage{longtable}
\usepackage{graphicx}
\usepackage{subfigure}
\usepackage[centertags]{amsmath}
\usepackage{multirow}
\usepackage{slashed}
\usepackage{pdflscape}
\usepackage{caption} 
 
\newcommand{\bCentering}{\centering}
\newcommand{\bCaption}{\caption}

\newcommand{\unity}{{\footnotesize\mbox{1\!\!I}}}
\newcommand{\bC}{\mathbb{C}}

\def\muc{\multicolumn}

\def\bZ{\mathbb{Z}}
\def\Z{\mathbb{Z}}
\def\R{\mathbb{R}}
\def\unity{1\!\!{\rm I}}
\def\ov{\overline}
\def\N{\mathbf{N}}

\def\ov{\overline}
\def\1{{\bf 1}}
\def\2{{\bf 2}}
\def\3{{\bf 3}}
\def\4{{\bf 4}}
\def\6{{\bf 6}}
\def\8{{\bf 8}}
\def\OR{\Omega\mathcal{R}}

\def\targ#1#2{\genfrac{[}{]}{0pt}{}{#1}{#2}}
\def\tarh#1#2{\genfrac{(}{)}{0pt}{}{#1}{#2}}
\def\targ2#1#2{\genfrac{}{}{0pt}{}{#1}{#2}}

\definecolor{blus}{rgb}{0.1,0.1,0.8}
\definecolor{GreenYellow}{cmyk}{0.15,0,0.69,0}
\definecolor{Yellow}{cmyk}{0,0,1,0}
\definecolor{Goldenrod}{cmyk}{0,0.10,0.84,0}
\definecolor{Dandelion}{cmyk}{0,0.29,0.84,0}
\definecolor{Apricot}{cmyk}{0,0.32,0.52,0}
\definecolor{Peach}{cmyk}{0,0.50,0.70,0}
\definecolor{Melon}{cmyk}{0,0.46,0.50,0}
\definecolor{YellowOrange}{cmyk}{0,0.42,1,0}
\definecolor{Orange}{cmyk}{0,0.61,0.87,0}
\definecolor{BurntOrange}{cmyk}{0,0.51,1,0}
\definecolor{Bittersweet}{cmyk}{0,0.75,1,0.24}
\definecolor{RedOrange}{cmyk}{0,0.77,0.87,0}
\definecolor{Mahogany}{cmyk}{0,0.85,0.87,0.35}
\definecolor{Maroon}{cmyk}{0,0.87,0.68,0.32}
\definecolor{BrickRed}{cmyk}{0,0.89,0.94,0.28}
\definecolor{Red}{cmyk}{0,1,1,0}
\definecolor{OrangeRed}{cmyk}{0,1,0.50,0}
\definecolor{RubineRed}{cmyk}{0,1,0.13,0}
\definecolor{WildStrawberry}{cmyk}{0,0.96,0.39,0}
\definecolor{Salmon}{cmyk}{0,0.53,0.38,0}
\definecolor{CarnationPink}{cmyk}{0,0.63,0,0}
\definecolor{Magenta}{cmyk}{0,1,0,0}
\definecolor{VioletRed}{cmyk}{0,0.81,0,0}
\definecolor{Rhodamine}{cmyk}{0,0.82,0,0}
\definecolor{Mulberry}{cmyk}{0.34,0.90,0,0.02}
\definecolor{RedViolet}{cmyk}{0.07,0.90,0,0.34}
\definecolor{Fuchsia}{cmyk}{0.47,0.91,0,0.08}
\definecolor{Lavender}{cmyk}{0,0.48,0,0}
\definecolor{Thistle}{cmyk}{0.12,0.59,0,0}
\definecolor{Orchid}{cmyk}{0.32,0.64,0,0}
\definecolor{DarkOrchid}{cmyk}{0.40,0.80,0.20,0}
\definecolor{Purple}{cmyk}{0.45,0.86,0,0}
\definecolor{Plum}{cmyk}{0.50,1,0,0}
\definecolor{Violet}{cmyk}{0.79,0.88,0,0}
\definecolor{RoyalPurple}{cmyk}{0.75,0.90,0,0}
\definecolor{BlueViolet}{cmyk}{0.86,0.91,0,0.04}
\definecolor{Periwinkle}{cmyk}{0.57,0.55,0,0}
\definecolor{CadetBlue}{cmyk}{0.62,0.57,0.23,0}
\definecolor{CornflowerBlue}{cmyk}{0.65,0.13,0,0}
\definecolor{MidnightBlue}{cmyk}{0.98,0.13,0,0.43}
\definecolor{NavyBlue}{cmyk}{0.94,0.54,0,0}
\definecolor{RoyalBlue}{cmyk}{1,0.50,0,0}
\definecolor{Blue}{cmyk}{1,1,0,0}
\definecolor{Cerulean}{cmyk}{0.94,0.11,0,0}
\definecolor{Cyan}{cmyk}{1,0,0,0}
\definecolor{ProcessBlue}{cmyk}{0.96,0,0,0}
\definecolor{SkyBlue}{cmyk}{0.62,0,0.12,0}
\definecolor{Turquoise}{cmyk}{0.85,0,0.20,0}
\definecolor{TealBlue}{cmyk}{0.86,0,0.34,0.02}
\definecolor{Aquamarine}{cmyk}{0.82,0,0.30,0}
\definecolor{BlueGreen}{cmyk}{0.85,0,0.33,0}
\definecolor{Emerald}{cmyk}{1,0,0.50,0}
\definecolor{JungleGreen}{cmyk}{0.99,0,0.52,0}
\definecolor{SeaGreen}{cmyk}{0.69,0,0.50,0}
\definecolor{Green}{cmyk}{1,0,1,0}
\definecolor{ForestGreen}{cmyk}{0.91,0,0.88,0.12}
\definecolor{PineGreen}{cmyk}{0.92,0,0.59,0.25}
\definecolor{LimeGreen}{cmyk}{0.50,0,1,0}
\definecolor{YellowGreen}{cmyk}{0.44,0,0.74,0}
\definecolor{SpringGreen}{cmyk}{0.26,0,0.76,0}
\definecolor{OliveGreen}{cmyk}{0.64,0,0.95,0.40}
\definecolor{RawSienna}{cmyk}{0,0.72,1,0.45}
\definecolor{Sepia}{cmyk}{0,0.83,1,0.70}
\definecolor{Brown}{cmyk}{0,0.81,1,0.60}
\definecolor{Tan}{cmyk}{0.14,0.42,0.56,0}
\definecolor{Gray}{cmyk}{0,0,0,0.50}
\definecolor{Black}{cmyk}{0,0,0,1}
\definecolor{White}{cmyk}{0,0,0,0}

\definecolor{mygr}{rgb}{0,0.6,0}
\definecolor{mygrey}{rgb}{0,0.1,0.2}
\definecolor{myblue}{rgb}{0,0.5,0.9}
\definecolor{myblue2}{rgb}{0,0.5,0.5}
\definecolor{myorange}{rgb}{1,0.5,0}
\definecolor{mypurple}{rgb}{0.6,0,1}
\definecolor{mygolden}{rgb}{1,0.8,0.2}

\newcommand{\bCaptionfonts}{\small}
\makeatletter
\long\def\@makecaption#1#2{%
  \vskip\abovecaptionskip
  \sbox\@tempboxa{{\bCaptionfonts #1: #2}}%
  \ifdim \wd\@tempboxa >\hsize
    {\bCaptionfonts #1: #2\par}
  \else
    \hbox to\hsize{\hfil\box\@tempboxa\hfil}%
  \fi
  \vskip\belowcaptionskip}
\makeatother

\makeatletter
\let\ORIGINALlatex@openbib@code=\@openbib@code
\renewcommand{\@openbib@code}{\ORIGINALlatex@openbib@code\setlength{\itemsep}{1ex plus.5ex minus.5ex}\setlength{\parsep}{0pt}}
\makeatother

\def\mathtab#1#2#3{\begin{table}[th]\bCentering$#1$\bCaption{#3}\label{tab:#2}\end{table}}
\def\mathtabfix#1#2#3{\begin{table}[th]\bCentering\resizebox{\linewidth}{!}{$#1$}\bCaption{#3}\label{tab:#2}\end{table}}

\renewcommand{\arraystretch}{1.3}

\allowdisplaybreaks[4]

\begin{document}
\begin{center}
\begin{flushright}
{\small MITP/14-018\\ 
\today}

\end{flushright}

\vspace{25mm}
{\Large\bf Circling the Square:\\ Deforming fractional D-branes in Type II/$\OR$ orientifolds}
\vspace{12mm}

{\large Michael Blaszczyk${}^{\clubsuit}$, Gabriele Honecker${}^{\heartsuit}$ and Isabel Koltermann${}^{\spadesuit}$
}

\vspace{8mm}
{
\it PRISMA Cluster of Excellence \& Institut f\"ur Physik  (WA THEP), \\Johannes-Gutenberg-Universit\"at, D-55099 Mainz, Germany
\;$^{\clubsuit}${\tt blaszczyk@uni-mainz.de},~$^{\heartsuit}${\tt Gabriele.Honecker@uni-mainz.de},~$^{\spadesuit}${\tt kolterma@uni-mainz.de}}

\vspace{15mm}{\bf Abstract}\\[2ex]\parbox{140mm}{
We study complex structure deformations of special Lagrangian cycles associated to fractional D-branes at $\Z_2$ singularities in Type II/$\OR$ orientifold models. 
By means of solving hypersurface constraints, we show how to compute the volumes of the most simple D-brane configurations. These volumes are given as a function of the 
deformation parameters depending on the D-brane position relative to the smoothed out singularity. We observe which cycles keep the special Lagrangian property in various deformation scenarios and what orientifold involutions are allowed.

As expected, the volume and thus the tree level value of the gauge coupling hardly change for D-branes not wrapping the exceptional cycle on the deformed singularity,
whereas the volume of D-branes passing through the singularity depends on the deformation parameter by some power law.

}
\end{center}

\thispagestyle{empty}
\clearpage 

\tableofcontents
\setlength{\parskip}{1em plus1ex minus.5ex}
\section{Introduction}\label{S:intro}

Fractional D6-branes in Type IIA/$\OR$ orientifold models on orbifolds with $\Z_2$ subsymmetries play a key role in the quest for phenomenologically viable global string compactifications~\cite{Blumenhagen:2002wn,Blumenhagen:2002gw,Honecker:2004kb,Honecker:2004np,Blumenhagen:2005tn,Bailin:2006zf,Gmeiner:2007we,Gmeiner:2007zz,Bailin:2007va,Gmeiner:2008xq,Bailin:2008xx,Forste:2008ex,Forste:2010gw,Bailin:2011am,Honecker:2012qr,Honecker:2013kda,Bailin:2013sya}.\footnote{For more complete lists of references on intersecting D6-branes see~\cite{Blumenhagen:2006ci,Ibanez:2012zz}.}
While at the level of topological data, orbifold and Calabi-Yau models have been shown to agree~\cite{Blumenhagen:2002wn}, it remains an open question how physical quantities in the low-energy effective field theory are affected by complex structure deformations of the cycles wrapped by D6-branes. Most prominently, D6-branes may wrap the same bulk cycle inherited from the underlying torus, but have different contributions from $\Z_2$ fixed points. As a result, the tree-level gauge couplings are identical at the orbifold point, but one-loop corrections differ in general~\cite{Blumenhagen:2007ip,Gmeiner:2009fb,Honecker:2011sm,Honecker:2011hm}. We expect the difference already to show up at leading order once deformations away from the singular orbifold locus are turned on. More concretely, the gauge couplings need to be computed from
\begin{equation}
\frac{1}{g^2_{a,\text{tree}}} \propto \left[\text{Vol}(\Pi_a) + \text{Vol}(\Pi_{a'}) \right]
\,,
\end{equation}
where $\Pi_a$ denotes the cycle wrapped by D6-brane $a$ and $\Pi_{a'}$ its orientifold image.
For {\it special Lagrangian (sLag)} cycles the volume is given by the integral of the holomorphic three-form, but upon deformations some fractional cycles can loose the {\it sLag} property while staying {\it Lag}, such that
\begin{equation}
\text{Vol}(\Pi_a) = \int_{\Pi_a} \left| \Omega_3 \right| \neq \int_{\Pi_a} \Omega_3 = \overline{ \int_{\Pi_a'} \Omega_3 } \not\in \mathbb{R} \,.
\end{equation}
For which cycles this applies depends on the one hand on the choice of background six- or four-torus of four-dimensional Type IIA/$\OR$ or six-dimensional Type IIB/$\OR$ models, namely untilted (rectangular) or tilted tori~\cite{Blumenhagen:2000ea}.\footnote{See also~\cite{Bianchi:1991eu,Bianchi:1997rf,Witten:1997bs,Angelantonj:1999jh,Kakushadze:2000hm} for the T-dual formulation with $B$-field in Type IIB/$\Omega$ orientifolds.} On the other hand, any potential cancellation depends on the choice of exotic O6-plane in the case of $T^6/\Z_{2M} \times \Z_2$ backgrounds with discrete torsion. These are exactly the phenomenologically most promising Type IIA/$\OR$ models, since they admit rigid three-cycles implying the absence of unwanted matter in the adjoint representation~\cite{Blumenhagen:2005tn,Forste:2008ex,Forste:2010gw,Honecker:2012qr,Honecker:2013kda} as well as exponentially suppressed D-brane instanton generated couplings~\cite{Blumenhagen:2006xt,Ibanez:2006da,Cvetic:2007ku,Billo:2007py,Billo:2007sw,Blumenhagen:2009qh}.

To our best knowledge, complex structure deformations of {\it sLag} $n$-cycles~\cite{Joyce:2001xt,Joyce:2001nm} on $T^{2n}$ orbifolds have to date not been studied in the context of D-brane geometries and their  implications for  low-energy physics. An early discussion of the geometry of deformations of $T^6/\Z_2 \times \Z_2$ with discrete torsion~\cite{Vafa:1986wx} was performed in~\cite{Vafa:1994rv}.
 By contrast, blow-ups by means of K\"ahler moduli of codimension $(2n-4)$ singularities have been studied systematically, see e.g.~\cite{Lust:2006zh,Reffert:2006du,Reffert:2007im} for a large class of Type IIB/$\Omega {\cal I}$ orientifold compactifications with holomorphic involutions and~\cite{Cvetic:2007ju} for a $T^6/\Z_2 \times \Z_2$ model of Type IIA/$\OR$ without discrete torsion. In the context of heterotic orbifold\footnote{For a recent classification of semi-realistic models see e.g.\ \cite{Nilles:2014owa}. } models the blow-up procedure has been studied extensively in the recent years  \cite{Nibbelink:2007rd,Nibbelink:2009sp, Blaszczyk:2010db}.

This article is organised as follows: in section~\ref{S:sLags}, we briefly review the relevant ingredients of {\it sLag} cycles for D-brane model building on $T^4/(\Z_{2N} \times \OR)$ and $T^6/(\Z_{2M} \times \Z_2 \times \OR)$ orientifolds.
We set the notation for the subsequent discussion of deformations and identify the relevant complex structure moduli within the closed string spectrum. Several examples for fractional D-branes passing through $\Z_2$ singularities are given, and the expected difference in the value of the tree-level gauge coupling constant for fractional D-branes with identical bulk part after deformation is highlighted. 
Section~\ref{S:HypersurfaceFormalism} is devoted to the inspection of {\it sLag} deformations by the hypersurface formalism. Several examples of orbifold models with `horizontal' and `vertical' D-branes on square tori are discussed in detail, and additional technical complexities for deformations of a variety of more general D-brane configurations are showcased. 
Section~\ref{S:Conclusions} contains our conclusions, and in appendix~\ref{A:Z6} we collect phenomenologically interesting examples of global models with fractional D-branes, which expose the need of going beyond the special configurations discussed in the bulk of the present article.

\section[Special Lagrangian Cycles on $T^4/\Z_{2N}$ and $T^6/\Z_{2M} \times \Z_{2}$ Orbifolds]{Special Lagrangian Cycles on $\boldsymbol{T^4/\Z_{2N}}$ and \newline $\boldsymbol{T^6/\Z_{2M} \times \Z_{2}}$ Orbifolds}\label{S:sLags}

In this section, we briefly review the construction of {\it sLag} three-(two-)-cycles
on toroidal orbifolds $T^6/\Z_{2M} \times \Z_{2}$ with discrete torsion ($T^4/\Z_{2N}$), which are customarily used
for intersecting D6(D7)-brane model building. Our focus lies on the cycles and moduli spaces of those orbifolds with known
global particle physics models, and in section~\ref{S:HypersurfaceFormalism} deformations
of the corresponding {\it sLag} cycles away from the singular orbifold point will be discussed.

Intersecting D$p$-branes and O$p$-planes with $p=6$ (7) arise in Type IIA (IIB) orientifold compactifications 
on Calabi-Yau manifolds $CY_{3}$ $(CY_2=K3)$ if an antiholomorphic involution ${\cal R}: z_i \to \ov{z}_i$ for $i=1,2,3$
($i=1,2$) is introduced such that ${\cal R}(\Omega_{n})=\ov{\Omega}_{n}$ and ${\cal R}(J_{1,1}^{\text{K\"ahler}})=-J_{1,1}^{\text{K\"ahler}}$ for the holomorphic
volume form and the K\"ahler form on $CY_n$, respectively. The fixed loci of ${\cal R}$ form {\it sLag} submanifolds 
which support O$p$-planes and satisfy the {\it calibration} condition
\begin{equation}\label{Eq:sLag-cond}
J_{1,1}^{\text{K\"ahler}} \big|_{\text{\it sLag}}=0,
\qquad
\Im(\Omega_{n}) \big|_{\text{\it sLag}}=0,
\qquad
 \Re(\Omega_{n}) \big|_{\text{\it sLag}}>0 %
. 
\end{equation}
D$p$-branes preserve the same supersymmetry and are said to satisfy the same calibration as the O$p$-planes if they  satisfy
the {\it sLag} conditions~(\ref{Eq:sLag-cond}) for the same choice of holomorphic volume form $\Omega_{n}$, cf.~\cite{Joyce:2001xt,Joyce:2001nm,Blumenhagen:2002wn} for an extended discussion.
In the rest of this article, we say that a {\it sLag} cycle has calibration $\Re(\Omega_n)$ or is calibrated w.r.t\ $\Re(\Omega_n)$ if \eqref{Eq:sLag-cond} holds, and analogously a cycle has calibration $\Im(\Omega_n)$ if \eqref{Eq:sLag-cond} holds with $\Omega_n$ replaced by $i \Omega_n$. A cycle is called {\it Lag} if it satisfies only the first condition $J_{1,1}^{\text{K\"ahler}} \big|_{\text{\it Lag}}=0$ without specifying a calibration.

Denoting the {\it sLag} cycle wrapped by $N_a$ identical D$p$-branes $a$ by $\Pi_a$, its ${\cal R}$-image by $\Pi_{a'}$ and the {\it sLag} cycle of the O$p$-planes by $\Pi_{\text{O}p}$,
the global consistency or {\it RR tadpole cancellation} conditions can be cast in the following form,
\begin{equation}\label{Eq:RR-tcc}
\sum_a N_a \left( \Pi_a + \Pi_{a'} \right) + Q_{\text{O}p} \, \Pi_{\text{O}p}=0
\qquad
\text{with}
\qquad 
Q_{\text{O}p}=\left\{\begin{array}{cr}
-4 & p=6\\
-8 & p=7
\end{array}\right.
,
\end{equation}
and the {\it sLag} condition on each cycle ensures the simultaneous vanising of all NS-NS tadpoles.

For D$p$-brane model building purposes, it is necessary to be able to explicitly construct the {\it sLag}s and compute their intersection numbers and volumes.
This can in particular be done on orbifolds of the factorised six-torus $T^6=\bigotimes_{k=1}^3 T^2_{(k)}$ for Type IIA/$\OR$ orientifolds ($T^4=\bigotimes_{k=1}^2 T^2_{(k)}$ for Type IIB/$\OR$),
where a D$p$-brane wraps a factorisable three-(two-)-cycle,
\begin{equation}
\Pi^{\text{torus}} = \otimes_{k=1}^{3(2)} \left( n_k \pi_{2k-1} + m_k \pi_{2k}\right)
\qquad
\text{and}
\qquad
\Pi^{\text{bulk}} \equiv \sum_{\text{orbifold images}} \Pi^{\text{torus}}
,
\end{equation}
with the one-cycles $(\pi_{2k-1},\pi_{2k})$ spanning the $k^{\text{th}}$ two-torus $T^2_{(k)}$ (cf.\ figures~\ref{Fig:Z2-lattice} and~\ref{Fig:Z6lattice}) and the toroidal wrapping numbers $n_k,m_k \in \Z$ coprime. For $T^6$ ($T^4$), a basis of three-cycles (two-cycles) is provided by the products of one-cycles $\Pi_{ijk} := \pi_i \otimes \pi_j \otimes \pi_k$ ($\Pi_{ij} := \pi_i \otimes \pi_j$).  To obtain a basis of bulk cycles of the orbifold $T^6/\Z_N \times \Z_M$ ($T^4/\Z_N$), one has to sum over all orbifold images. Using the same notation, one gets the intersection numbers
\begin{equation}
 \label{Eq:BulkIntersection}
 \Pi_{ijk} \circ \Pi_{lmn} = N M \epsilon_{ijklmn} \,, \qquad
\left( \Pi_{ij} \circ \Pi_{kl} = N \epsilon_{ijkl} \right) \,.
\end{equation}

The {\it sLag} conditions for a given bulk cycle can be fully expressed in terms of the wrapping numbers $(n_k,m_k)$ and the complex structure parameters $\rho_k$ per two-torus $T^2_{(k)}$
by using the integrated and normalised version of~(\ref{Eq:sLag-cond}) (cf. e.g.~\cite{Forste:2010gw}), 
\begin{equation}\label{Eq:Def-Z}
\Re({\cal Z}_a) =  \frac{\text{Vol}(\Pi_a)}{\sqrt{\text{Vol}_{6(4)}}} >0
\quad 
\text{and}
\quad
\Im({\cal Z}_a)=0
\quad
\text{with}
\quad 
{\cal Z}_a \equiv \frac{1}{\sqrt{\text{Vol}_{6(4)}}} \int_{\Pi_a} \Omega_{n}
,
\end{equation}
as explicitly listed below  in section~\ref{Ss:T4Z2N} and~\ref{Ss:T6Z2Z2N} for each individual orbifold under consideration.
Besides from the {\it sLag} condition, for D6-branes the volume of the wrapped three-cycle also determines the value of the tree-level gauge coupling $g_{G_a}$
for the corresponding gauge group $G_a$,
\begin{equation}\label{Eq:gauge-vs-volumes}
\frac{4\pi}{g^2_{G_a}} = \frac{1}{2 \, k_ac_a \, g_{\text{string}}} \frac{\text{Vol}(\Pi_a) + \text{Vol}(\Pi_{a'})}{\ell_s^3}
\quad
\text{with}
\quad
k_a=\left\{\begin{array}{cr} 
1 & G_a = U(N_a) \\
2 & USp(2N_a) \text{ or } SO(2N_a)
\end{array}\right.
,
\end{equation}
and $c_a=1,2,4$ for bulk and the two kinds of fractional cycles defined below in equation~(\ref{Eq:Def-frac}), respectively.
While the formula with $\text{Vol}(\Pi_a) = \text{Vol}(\Pi_{a'})$ has been used extensively on the six-torus and at the orbifold point (see e.g.~\cite{Aldazabal:2000cn,Klebanov:2003my,Blumenhagen:2003jy,Blumenhagen:2006ci,Honecker:2011sm}), we will see in section~\ref{S:HypersurfaceFormalism} that this equality of orientifold image volumes is violated for fractional three-(two-)-cycles in the deformed phase.

In the presence of one or three $\Z_2$ symmetries, the cycle can be {\it fractional} and stuck at $\Z_2$ singularities,
\begin{equation}\label{Eq:Def-frac}
\Pi^{\text{frac}} = \frac{1}{2} \left( \Pi^{\text{bulk}} + \Pi^{\Z_2} \right)
\qquad
\text{or}
\qquad
\Pi^{\text{frac}} = \frac{1}{4} \bigl( \Pi^{\text{bulk}} + \sum_{k=1}^3 \Pi^{\Z_2^{(k)}} \bigr)
,
\end{equation}
respectively.
The form of $\Pi^{\Z_2}$ or $\Pi^{\Z_2^{(k)}}$ is determined by the following sets of choices:
\begin{enumerate}
\item
The even- \& oddness of wrapping numbers $(n_k,m_k) \in \{ \text{(even,odd)}, \, \text{(odd,even)},$ $\text{(odd,odd)}\}$ determines two possible sets $(\sigma_k=0,1)$, each of which contains two $\Z_2$ fixed points of $T^2_{(k)}$ that are traversed by the torus cycle without $(\sigma_k=0)$ or with a displacement  $(\sigma_k=1)$ from the origin (cf. below for details).
\item
For each $\Z_2^{(k)}$ twisted sector, the overall sign of $ \Pi^{\Z_2^{(k)}}$ is given by the $\Z_2^{(k)}$ eigenvalue \mbox{$(-1)^{\tau^{\Z_2^{(k)}}}=\pm 1$} with the constraint $\prod_{k=1}^3 (-1)^{\tau^{\Z_2^{(k)}}}=1$ for three $\Z_2$ symmetries
due to $\Z_2^{(2)} \simeq \Z_2^{(1)} \cdot \Z_2^{(3)}$, where $\Z_2^{(k)}$ denotes the $\Z_2$ symmetry acting on $T^4_{(k)} \equiv T^2_{(i)} \times T^2_{(j)}$ and leaving the two-torus $T^2_{(k)}$ invariant for some permutation $(ijk)$ of $(123)$.
\item
On each $T^2_{(k)}$, the relative sign $(-1)^{\tau_k}$ between the two $\Z_2$ fixed point contributions is parametrised by the discrete Wilson line $\tau_k \in \{0,1\}$.
\end{enumerate}
As an example, the exceptional part of a fractional two-cycle on the $T^4 / \Z_2$ orbifold reads
\begin{equation} \label{Eq:Def-PiZ2}
 \Pi^{\Z_2} = \left(-1\right)^{\tau^{\Z_2}} \cdot \left( e_{\alpha\beta} + \left(-1\right)^{\tau_1} e_{\hat{\alpha}\beta} +  \left(-1\right)^{\tau_2} e_{\alpha\hat{\beta}}  + \left(-1\right)^{\tau_1 + \tau_2} e_{\hat{\alpha}\hat{\beta}} \right)
\end{equation}
with $e_{\alpha\beta}$ an exceptional two-cycle at the $\Z_2$ fixed point $\alpha\beta$ on $T^2_{(1)} \times T^2_{(2)}$, where the pair $(\alpha,\hat{\alpha})$ of $T^2_{(1)}$ is set by
\begin{center}
 \begin{tabular}{|c||c|c|}
 \hline
 & $\underline{\sigma_1 = 0:}$ & $\underline{\sigma_1 = 1:}$ \\
$(n_1,m_1)$ & $(\alpha,\hat{\alpha})$ & $(\alpha,\hat{\alpha})$ \\
\hline
\hline
(odd,even) & (1,2) & (4,3) \\
\hline
(odd,odd) & (1,3) & (2,4) \\
\hline
(even,odd) & (1,4) & (2,3) \\
\hline
 \end{tabular} \,,
\end{center}
and $(\beta,\hat{\beta})$ of $T^2_{(2)}$ analogously (cf.\ figure~\ref{Fig:Z2-lattice} for the fixed point labels).
A {\it factorisable} fractional two-cycle on $T^4/\Z_{2N}$ with given wrapping numbers $(n_k,m_k)_{k=1,2}$ is thus fully specified by $2^5=32$ choices of $(\sigma_1,\sigma_2; \tau_1,\tau_2; \tau^{\Z_2})$,
and a  {\it factorisable} fractional three-cycle on $T^6/\Z_{2M} \times \Z_{2}$ with discrete torsion by $2^8=256$ choices of $(\sigma_k;\tau_k; \tau^{\Z_2^{(k)}})_{k=1,2,3}$. 
At the orbifold point, the {\it sLag} condition for a fractional cycle is fulfilled if the bulk part is {\it sLag} and the above combinatorics of adding exceptional cycles is respected.

The intersection numbers among bulk cycles are computed in equation \eqref{Eq:BulkIntersection}, and fixed points under $\Z_n$ along $T^2_{(i)} \times T^2_{(j)}$ support $(n-1)$ exceptional two-cycles with their intersection form given by minus the Cartan matrix of the Lie algebra $A_{n-1}$,
\begin{equation}
\label{Eq:ExceptionalCycleIntersection}
e_{\alpha_1\beta_1} \circ e_{\alpha_2\beta_2} = -  \text{C}(A_{n-1}) \, \delta_{\alpha_1\alpha_2} \delta_{\beta_1\beta_2} \stackrel{n=2}{=} -2 \, \delta_{\alpha_1\alpha_2} \delta_{\beta_1\beta_2}
\, .
\end{equation}
Details for some fractional D7-brane models on $T^4/\Z_{2N}$ orientifolds are given in section~\ref{Ss:T4Z2N} and appendix~\ref{Sss:T4Z6}.

The  exceptional contributions  $\Pi^{\Z_2^{(k)}}$ to three-cycles at $\Z_2^{(k)}$ fixed points  arise as tensor products of exceptional two-cycles along $T^4_{(k)} \equiv T^2_{(i)} \times T^2_{(j)}$ with a toroidal one-cycle
along $T^2_{(k)}$, summed over fixed point contributions and orbifold images,
\begin{equation}
\Pi^{\Z_2^{(k)}} \equiv  (-1)^{\tau^{\Z_2^{(k)}}} \sum_{\text{orbifold images}} \; \sum_{\text{set of }\alpha_p\beta_q} \left(\pm \, e_{\alpha_p\beta_q}^{\Z_2^{(k)}} \right) \otimes  \left( n_k \pi_{2k-1} + m_k \pi_{2k}\right)
,
\end{equation}
where $\pm$ denotes relative signs due to Wilson lines as in equation~(\ref{Eq:Def-PiZ2}). 
Details for some rigid D6-brane models are given below in section~\ref{Ss:T6Z2Z2N} and appendices~\ref{Sss:Z2Z6} and~\ref{Sss:Z2Z6p}  for the $T^6/\Z_{2M} \times \Z_{2}$ (with discrete torsion) examples of phenomenological interest.

The RR tadpole cancellation conditions~(\ref{Eq:RR-tcc}) contain ${\cal R}$-image cycles. Due to ${\cal R}(J_{1,1}^{\text{K\"ahler}})=-J_{1,1}^{\text{K\"ahler}}$, one can infer that exceptional two-cycles, which arise from the blow-up of different $\Z_n$ fixed points, are permuted depending on the background lattices, but the twist sector is preserved (also for $n \neq 2$~\cite{Blumenhagen:2002wn}),
\begin{equation}\label{Eq:R-on-e}
e_{\alpha\beta}^{\Z_n^{(k)}} \stackrel{{\cal R}}{\longrightarrow} - \, e_{\alpha' \beta'}^{\Z_n^{(k)}}
.
\end{equation}
The details again depend on the choice of orbifold and lattice orientation. The relevant data are summarised below in section~\ref{Ss:T4Z2N} and appendix~\ref{Sss:T4Z6} for the $T^4/\Z_{2}$ and $T^4/\Z_6$ examples, resepctively, and in section~\ref{Ss:T6Z2Z2N} and appendices~\ref{Sss:Z2Z6},~\ref{Sss:Z2Z6p} for the $T^6/\Z_{2} \times \Z_{2}$ and $T^6/\Z_6^{(\prime)} \times \Z_2$ examples with discrete torsion, respectively. In section~\ref{S:HypersurfaceFormalism}, we consider (complex structure) deformations of codimension two- and three- singularities away from the singular orbifold limit, respectively, for these models.

Before discussing various orbifolds explicitly, let us briefly summarise the stringy origin of the K\"ahler and complex structure moduli which encode the blow-ups and deformations away from the singular orbifold point. 
The worldsheet parity $\Omega = \pm$ of the massless bosonic field content of Type IIB and Type IIA  string theory in ten dimensions is for convenience given in table~\ref{Tab:WS-parities}.
\begin{SCtable}
\begin{tabular}{|c||c|c||c|c|}\hline
\muc{5}{|c|}{\text{\bf Worldsheet parity of  bosonic  states}}
\\\hline\hline
\text{sector} & \muc{2}{|c|}{  $\Omega =+$}  &   \muc{2}{|c|}{  $\Omega =- $}
\\\hline\hline
\text{\bf NS-NS} &  \muc{2}{|c|}{$\phi$,  $G_{MN}$ }  & \muc{2}{|c|}{$B_{MN}$}
\\\hline\hline
 &  \muc{2}{|c|}{\text{\bf IIB}} & \muc{2}{|c|}{\text{\bf IIA}}
\\\hline
& $\Omega =+$ & $\Omega =-$ &  $\Omega =+$ & $\Omega =- $
\\\hline
\text{\bf R-R} &  $C_2$ &  $C_0, C_4$  &  $C_3$ & $C_1$
\\\hline 
\end{tabular}
\caption{Worldsheet parity $\Omega = \pm$ of the ten-dimensional massless bosonic states in Type IIB and Type IIA string theory. The five-form field strength of $C_4$ in Type IIB string theory 
satisfies a self-duality relation.}
\label{Tab:WS-parities}
\end{SCtable}
The six- and four-dimensional massless bosonic closed string spectra arise by dimensional reduction and integration over $\OR$-even and $\OR$-odd cycles as follows:
the NS-NS sector of both Type IIB/$\OR$ on $K3$ and Type IIA/$\OR$ on $CY_3$ contains besides the dilaton $\phi$ and truncated metric $G_{\mu\nu}$ the volume moduli $\mathfrak{v}^i$ and their axionic partners $\mathfrak{b}^i$,
\begin{equation}
\mathfrak{v}^i =\int_{\Pi^{2-}_i} J_{1,1}^{\text{K\"ahler}} 
,
\qquad 
\mathfrak{b}^i =\int_{\Pi^{2-}_i} B_2^{\text{NS-NS}}
,
\end{equation}
which belong to hyper and chiral multiplets in six and four dimensions, respectively, as listed in table~\ref{Tab:closed-OR-spectrum}.
Here, $\Pi^{n-}_i$ denotes a ${\cal R}$-odd $n$-cycle, and $\Pi^{n+}_j$ a ${\cal R}$-even $n$-cycle.

The hyperk\"ahler property of the $K3$ surface implies that for a fixed Ricci-flat metric there is an ambiguity in the definition of the complex structure, see e.g.\ \cite{Aspinwall:1996mn}. More precisely, a complex structure is specified by the cohomology class of $\Omega_2$ as below in equation~\eqref{Eq:Omega2Topo}, whose real and imaginary part span a two-plane in $H^2(K3,\R)$. The K\"ahler form then spans a line in $H^2(K3,\R)$ which is orthogonal to that two-plane. The metric, however, only depends on the three-plane spanned by the two-forms $\Re(\Omega_2)$, $\Im(\Omega_2)$ and $J_{1,1}^\text{K\"ahler}$. The remaining freedom to rotate these three two-forms within the three-plane determines if a smoothed out singularity is blown up (using a K\"ahler modulus $\mathfrak{v}$) or deformed (using a complex structure modulus $\mathfrak{z}$).
%

\begin{table}[h!]
\renewcommand{\arraystretch}{1.3}
  \begin{center}
\begin{equation*}
\begin{array}{|c|c|c||c|c|c|}\hline
\multicolumn{6}{|c|}{\text{\bf Bosonic massless closed string d.o.f. in Type II/$\OR$ orientifolds on $K3$ and $CY_3$}}
\\\hline\hline
\muc{3}{|c||}{\text{\bf IIB/$\OR$ on $K3$}} & \muc{3}{|c|}{\text{\bf IIA/$\OR$ on $CY_3$}}
\\\hline
{\cal N}=1 \text{ multiplet in 6D} & \# & \text{bosons} & {\cal N}=1 \text{ multiplet in 4D} & \# & \text{bosons}
\\\hline\hline
\text{gravity} & 1 & (G_{\mu\nu},B_{\mu\nu}^{\text{s.d.}}) & \text{gravity} & 1 & (G_{\mu\nu})
\\\hline
\text{tensor} & 1+ h_{11}^+ & (B_{\mu\nu}^{\text{a-s.d.},j},\varphi^j) 
& \text{linear (dilaton-axion)} & 1 & (\phi,\xi^0)
\\ \cline{4-6}
& & \text{with } \varphi^0 \equiv \phi
& \text{vector} & h_{11}^+ & (A_{\mu}^j)
\\\hline
\text{hyper} & h_{11}^- & (\mathfrak{v}^i,\mathfrak{b}^i,\mathfrak{z}^i,\zeta^i)
& \text{chiral (K\"ahler moduli)} & h_{11}^- & (\mathfrak{v}^i,\mathfrak{b}^i)
\\\hline 
\muc{3}{|c||}{} & \text{chiral (complex structures)} & h_{21} & (\mathfrak{c}^k,\xi^k)
\\\hline
\end{array}
\end{equation*}
\end{center}
\caption{Bosonic matter content of Type II/$\OR$ compactifications on $K3$ and $CY_3$ with multiplicities $\#$ of each type of multiplet in terms of Hodge numbers.
}
\label{Tab:closed-OR-spectrum}
\end{table}
The RR sector of Type IIB/$\OR$ on $K3$ contains the truncation and splitting into self-dual and anti-self-dual part of the two-form $C_2 =B_{\mu\nu}^{\text{s.d.}} +  B_{\mu\nu}^{\text{a-s.d.},0}$ 
as well as one scalar $\varphi^j =  \int_{\Pi^{2+}_j} C_2$  per model-dependent tensor multiplet, while the model-dependent tensors $B_{\mu\nu}^{\text{a-s.d.},j}$ originate from integrals over anti-self-dual two-forms of the (self-dual) RR-form $C_4$. 
Here one has to use the fact that the hyperk\"ahler geometry on $K3$ contains three self-dual and 19 anti-self-dual two-forms, cf.\ e.g.\ the review~\cite{Aspinwall:1996mn}. 

The RR sector in Type IIA/$\OR$ orientifolds contains the vectors and axionic partners of the complex structure moduli (cf. e.g.~\cite{Grimm:2004ua,Forste:2010gw}),
\begin{equation}
A^j_{\mu} =  \int_{\Pi^{2+}_j} C_3
,
\qquad
\xi^k =  \int_{\Pi^{3+}_k} C_3
,
\end{equation}
with the complex structures $\mathfrak{c}^k$ associated to the $(h_{21}+1)$ $\OR$-even three-cycles on $CY_3$. In equation~\eqref{Eq:Omega3Topo} of
section~\ref{Ss:SquareT6Z22} these moduli, which at the orbifold point can be explicitly constructed, are used in the topological expansion of the holomorphic three-form.

\subsection[Two-cycles on $T^4/\Z_{2N}$]{Two-cycles on $\boldsymbol{T^4/\Z_{2N}}$}\label{Ss:T4Z2N}

Exceptional three-cycles in phenomenologically appealing orbifold compactifications to four dimensions arise as tensor products of exceptional divisors on $T^4/\Z_{2N}$  with toroidal one-cycles along an additional two-torus $T^2$.  Deformations away from the singular orbifold point  thus consist of deformations of $\Z_2$ singularities on $T^4$ times the relevant toroidal one-cycle on $T^2$. 
It is instructive to first study the technique of deformations in six-dimensional $T^4/\Z_{2N}$ models and discuss subtleties related to special background lattices and D-brane configurations there. 
We will in particular distinguish between untilted and tilted lattices and enhancements of gauge groups $U(N_a) \to USp(2N_a)$ in the $T^4/\Z_2$ context. In the associated appendix~\ref{Sss:T4Z6}, we briefly mention additional subtleties, e.g.\ the reduced number of independent deformations, related to orbifold image cycles under the $\Z_3$ subgroup in the $T^4/\Z_6$ case.

\subsubsection[$T^4/\Z_2$ with shift vector $\vec{v}=\frac{1}{2}(1,-1)$]{$\boldsymbol{T^4/\Z_2}$ with shift vector $\boldsymbol{\vec{v}=\frac{1}{2}(1,-1)}$}\label{Sss:T4Z2}

\begin{figure}[ht]
\begin{minipage}[b]{0.5\linewidth}
\hspace{-3mm}
\begin{center}\hspace{-3mm}
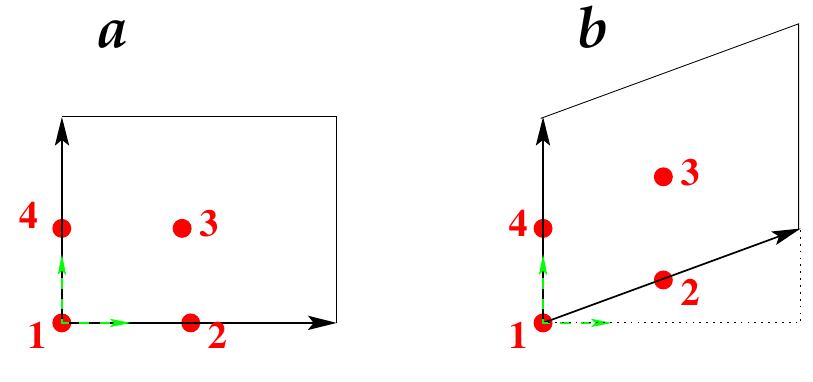
\end{center}
\caption{The $\Z_2$ invariant `untilted' {\textbf{\textit{a}}}-type and `tilted' 
{\textbf{\textit{b}}}-type  tori, which are parametrised by $b=0,\frac{1}{2}$, respectively,
with $\Z_2$ fixed points 1,2,3,4 and $\Re(z)$ axis along $\frac{\pi_{2i-1} -b \, \pi_{2i}}{1-b}$. 
The points 1,4 are invariant under ${\cal R}$, whereas 
\mbox{$(2,3) \stackrel{\cal R}{\longleftrightarrow} (2+2b, 3-2b)$}. 
The tilted torus for $R_2/R_1=2\sqrt{3},2/\sqrt{3}$ corresponds to the 
{\bf A}- and {\bf B}-type $\Z_6$ invariant lattice of  figure~\protect\ref{Fig:Z6lattice} 
with radii $r= R_2/\sqrt{3},R_2$, respectively. The $\Z_2$ fixed points are relabeled
as follows: $(1,2,3,4)_{\Z_2,{\textbf{\textit{b}}}} = (1,5,6,4)_{\Z_6,{\bf A}}$ and $(1,5,4,6)_{\Z_6,{\bf B}}$.
}
\label{Fig:Z2-lattice}
\end{minipage}
\hspace{0.5cm}
\begin{minipage}[b]{0.5\linewidth}
\begin{center}
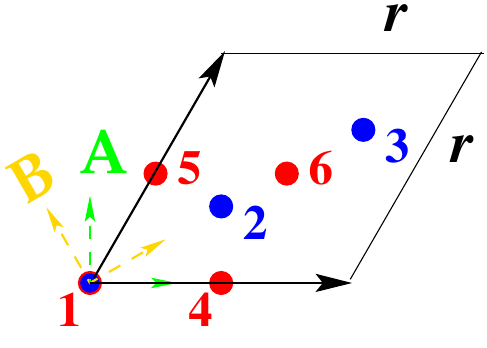
\end{center}
\caption{The $\Z_6$ invariant lattices.
For the {\bf A} orientation (green coordinate axes), $\pi_{2i-1}$ spans
the $\Re(z)$ axis, and on the {\bf B}-lattice (axes in yellow), the 
$\Re(z)$ axis extends along $\pi_{2i-1} + \pi_{2i}$. 
The $\Z_3$ invariant points $2\stackrel{\Z_2}{\longleftrightarrow} 3$
are exchanged under ${\cal R}$ on the {\bf A} orientation, but are invariant on the 
{\bf B}-lattice. The triplet of $\Z_2$ fixed points $4 \stackrel{e^{i \pi/3}}{\longrightarrow} 5 \stackrel{e^{i\pi/3}}{\longrightarrow} 6\stackrel{e^{i\pi/3}}{\longrightarrow} 4$
contains one point fixed under ${\cal R}$, while the other two are exchanged: $5 \stackrel{{\cal R} \text{ on } {\bf A}}{\longleftrightarrow} 6$, $4 \stackrel{{\cal R} \text{ on } {\bf B}}{\longleftrightarrow} 5$.
The origin 1 is fixed under the full ${\cal R}$ and $\Z_6$ symmetry.
}
\label{Fig:Z6lattice}
\end{minipage}
\end{figure}

The antiholomorphic involution ${\cal R}$ enforces each two-torus lattice to be oriented such that ${\cal R}$ acts crystallographically (cf.\ e.g.\ the reviews~\cite{Blumenhagen:2006ci,Ibanez:2012zz}), see figure~\ref{Fig:Z2-lattice} for our notation. 

Any {\it factorisable} bulk two-cycle on $T^4/\Z_2$ can be expanded as 
\begin{equation}
\Pi_a^{\text{bulk}} = X_1^a \, \Pi_{13}^{\text{bulk}} - X_2^a \, \Pi_{24}^{\text{bulk}} + Y_1^a \, \Pi_{14}^{\text{bulk}} + Y_2^a \, \Pi_{23}^{\text{bulk}}
\qquad 
\text{with}
\quad 
\left\{\begin{array}{cc}
X_1^a  \equiv n_1^a n_2^a & X_2^a  \equiv  -  m_1^a  m_2^a
\\
Y_1^a \equiv n_1^a m_2^a & Y_2^a  \equiv m_1^a n_2^a
\end{array}\right.
.
\end{equation}
In the following, we use the abbreviation $\tilde{m}_k^a \equiv m_k^a + b_k n_k^a$ with $b_k=0$ for untilted and $b_k=\frac 1 2$ for tilted tori, respectively, and correspondingly $\tilde{Y}_1^a= n_1^a \tilde{m}_2^a$ (etc.). Denoting the complex structure modulus per two-torus $T^2_{(k)}$ by $\rho_k \equiv \frac{R_2^{(k)}}{R_1^{(k)}}$, the bulk RR tadpole cancellation and bulk supersymmetry conditions for {\it fractional} D7-branes can be written as
\begin{equation}
\begin{aligned}
\underline{\text{RR tadpole:}}  \quad   \left\{\begin{array}{c} \sum_a N_a \tilde{X}_1^a  = 16 \\
 \sum_a N_a \frac{\tilde{X}_2^a}{\prod_{k=1}^2 (1-b_k)}=16
 \end{array}\right.
 ,
 \qquad
\underline{\text{SUSY:}} &  \quad   \left\{\begin{array}{c}  \tilde{X}_1^a + \rho_1 \rho_2 \, \tilde{X}_2^a >0 \\ 
\rho_1^{-1} \tilde{Y}_1^a + \rho_2^{-1} \tilde{Y}_2^a  =0
 \end{array}\right.
,
\end{aligned}
\end{equation}
where in the notation of equation~(\ref{Eq:Def-Z}) the quantity  ${\cal Z}_a = \prod_{k=1}^2 \frac{n_k^a R_1^{(k)} + i \, \tilde{m}_k^a R_2^{(k)}}{\sqrt{R_1^{(k)}R_2^{(k)}}}$ has been used.

For untilted tori,  equation~(\ref{Eq:R-on-e}) implies that the ${\cal R}$-image of any $\Pi^{\text{frac}}_a = \frac{1}{2} \left( \Pi^{\text{bulk}}_a + \Pi^{\Z_2}_a \right)$ is simply $\Pi^{\text{frac}}_{a'} = \frac{1}{2} \left( \Pi^{\text{bulk}}_{a'} - \Pi^{\Z_2}_a \right)$ and thus $\Pi^{\text{frac}}_a + \Pi^{\text{frac}}_{a'} =  \frac{1}{2} \left( \Pi_a^{\text{bulk}} + \Pi_{a'}^{\text{bulk}} \right)$ and no gauge enhancement $U(N_a) \to USp(2N_a)$ or $SO(2N_a)$ can occur. Since 
it is instructive to first study deformations of orbifold singularities in this simple case, we briefly review the generalised T-dual of the  {\bf Gimon-Polchinski} model~\cite{Gimon:1996rq} here: 
the RR tadpoles are cancelled by two stacks of D7-branes $a$ and $b$ parallel to the $\OR$-invariant and $\OR\Z_2$-invariant O7-planes, respectively, as specified in table~\ref{tab:GP-model}.
\mathtab{
\begin{array}{|c|c|c|c|c|c|c|c|}\hline
\muc{7}{|c|}{\text{\bf D7-branes of the generalised T-dual to the $T^4/\Z_2$ Gimon-Polchinski model}}
\\\hline\hline
x &  \tarh{n_1, m_1}{n_2 , m_2} & (\vec{\sigma}) & \Z_2 & (\vec{\tau}) & \begin{array}{c} \Pi_x^{\text{frac}} \\  \Pi_{x'}^{\text{frac}} \end{array} & \Pi_x^{\text{frac}} +  \Pi_{x'}^{\text{frac}}
\\\hline\hline
a &  \tarh{1,0}{1,0} &  \tarh{0}{0} & (-1)^{\tau^{\Z_2}_a}  &\tarh{\tau^a_1}{\tau^a_2}
& \begin{array}{c}  \frac{\Pi_{13}^{\text{bulk}} +  (-1)^{\tau^{\Z_2}_a}  \left[ e_{11} + (-1)^{\tau_1^a} e_{21} + (-1)^{\tau_2^a} e_{12}  + (-1)^{\tau_1^a + \tau_2^a} e_{22} \right]}{2} \\
 \frac{\Pi_{13}^{\text{bulk}} -  (-1)^{\tau^{\Z_2}_a}  \left[ e_{11} + (-1)^{\tau_1^a} e_{21} + (-1)^{\tau_2^a} e_{12}  + (-1)^{\tau_1^a + \tau_2^a} e_{22} \right]}{2}
\end{array}
& \Pi_{13}^{\text{bulk}}
\\\hline
b & \tarh{0,1}{0,-1} &  \tarh{0}{0} & (-1)^{\tau^{\Z_2}_b}   &\tarh{\tau^b_1}{ \tau^b_2}
& \begin{array}{c}  \frac{-\Pi_{24}^{\text{bulk}} +  (-1)^{\tau^{\Z_2}_b}  \left[ e_{11} + (-1)^{\tau_1^b} e_{41} + (-1)^{\tau_2^b} e_{14}  + (-1)^{\tau_1^b + \tau_2^b} e_{44} \right]}{2} \\
\frac{-\Pi_{24}^{\text{bulk}} -  (-1)^{\tau^{\Z_2}_b}  \left[ e_{11} + (-1)^{\tau_1^b} e_{41} + (-1)^{\tau_2^b} e_{14}  + (-1)^{\tau_1^b + \tau_2^b} e_{44} \right]}{2}
\end{array}
& -\Pi_{24}^{\text{bulk}}
\\\hline
\end{array}
}{GP-model}{D7-brane configuration of a global model on $T^4/\Z_2$ with untilted tori. The displacements $(\vec{\sigma}^{\, x})$ are for the sake of notational briefness set to $(\vec{0})$, while arbitrary choices 
of $\Z_2$ eigenvalues $(-1)^{\tau^{\Z_2}_x}$ and discrete Wilson lines $(\vec{\tau}^{\, x})$ are allowed. Switching on different values of $(\vec{\sigma}^{\, x})$ results in permuting the fixed point sets $(\alpha,\hat{\alpha}) \in \{(1,2),(4,3)\}$ for stack $a$ 
and $\{(1,4),(2,3)\}$ for stack $b$ per two-torus, as explained around equation~(\ref{Eq:Def-PiZ2}).
}
The model is supersymmetric for any choice of complex structure of the two-tori; this includes in particular the option of square tori ($ \rho_1 = \rho_2 = 1$), on which we will perform explicit calculations of deformations away from the singular orbifold point in section \ref{Ss:SquareT4Z2}. The resulting gauge group is $U(16)_a \times U(16)_b$ with (up to hermitian conjugation ${\bf 16}_b \to \ov{\bf 16}_b$)  unique massless matter field content
for any choice $(x \in \{a,b\}$)  of $\Z_2$ eigenvalues $(-1)^{\tau^{\Z_2}_x}$, discrete displacements $(\sigma^x_1,\sigma^x_2)$ and discrete Wilson lines $(\tau^x_1,\tau^x_2)$,
\begin{equation}
2 \times ({\bf 120}_a,\1) + 2 \times (\1,{\bf 120}_b) + ({\bf 16}_a,{\bf 16}_b)
.
\end{equation}
As first geometrically interpreted in~\cite{Blumenhagen:2002wn}, by giving suitable {\it vev}s to the antisymmetric representations $({\bf 120})$, the corresponding D7-branes and their orientifold 
images pairwise recombine into  pure bulk D7-branes, which can then be moved away from the ${\cal R}$ and $\Z_2$-invariant loci.  In this way,  the gauge group 
is broken to $USp(16)_a \times USp(16)_b$, or by  splitting each of these bulk D7-brane stacks into two parallelly displaced ones, the gauge group $\prod_{i=1}^2 USp(8)_{a_i} \times \prod_{j=1}^2 USp(8)_{b_j}$ 
of the {\bf Bianchi-Sagnotti} model~\cite{Bianchi:1990yu} is obtained.

The models with tilted tori clearly differ from those with untilted tori as can be seen from the number of tensor multiplets in the closed string sector, $n_T=1,5,7 $  for the {\textbf{\textit{aa}}}, {\textbf{\textit{ab}}}, {\textbf{\textit{bb}}}
lattice~\cite{Blumenhagen:2000ea,Blumenhagen:2002wn}, respectively. The closed string spectrum furthermore contains $n_H^{\text{closed}, U}=4$ hyper multiplets from the untwisted sector comprising the 
size- and shape-modulus of each two-torus as well as $n_H^{\text{closed},\Z_2}=16,12,10$ hyper multiplets on the {\textbf{\textit{aa}}}, {\textbf{\textit{ab}}}, {\textbf{\textit{bb}}} lattice, respectively,
from the $\Z_2$ twisted sector containing the blow-up and deformation moduli. The number $n_H^{\text{closed},\Z_2}$ depends on the choice of  tilted background 
lattices by the pairwise identification of $\Z_2$ fixed points under the antiholomorphic involution ${\cal R}$ as detailed in figure~\ref{Fig:Z2-lattice}.

The massless closed string spectrum of Type IIB/$\OR$ on $T^4/\Z_2$ can be explicitly computed following the technique described e.g. in chapter 15.2 of~\cite{Blumenhagen:2013fgp}.
With the abbreviated notation $\pm \frac{1}{2} \equiv \pm$, the bosonic matter content in the $\Z_2$ twisted sector is obtained from the following list,
\begin{equation}\label{Eq:T4Z2-IIB+OR-twisted-spectrum}
\mbox{\resizebox{0.95\textwidth}{!}{%
$
\begin{array}{l|l}
|0 0 + +\rangle|0 0 + +\rangle_{\text{NSNS}}^{(\alpha\beta)} + |0 0 - -\rangle|0 0 - -\rangle_{\text{NSNS}}^{(\alpha'\beta')}  
& |+ +~0 0\rangle |- -~0 0\rangle_{\text{RR}}^{(\alpha\beta)} - |- -~0 0\rangle |+ +~0 0\rangle_{\text{RR}}^{(\alpha'\beta')}  
\\
|0 0 + +\rangle |0 0 - -\rangle_{\text{NSNS}}^{(\alpha\beta)} + |0 0 + +\rangle|0 0 - -\rangle_{\text{NSNS}}^{(\alpha'\beta')}   
& |+ +~0 0\rangle |+ +~0 0\rangle_{\text{RR}}^{(\alpha\beta)} - |+ +~0 0\rangle |+ +~0 0\rangle_{\text{RR}}^{(\alpha'\beta')}  
\\
|0 0 - -\rangle|0 0 + +\rangle_{\text{NSNS}}^{(\alpha\beta)} + |0 0 - -\rangle|0 0 + +\rangle_{\text{NSNS}}^{(\alpha'\beta')}  
& |- -~0 0\rangle |- -~0 0\rangle_{\text{RR}}^{(\alpha\beta)} - |- -~0 0\rangle|- -~0 0\rangle_{\text{RR}}^{(\alpha'\beta')}  
\end{array}
$}}
\end{equation}
where $(\alpha\beta)$ denotes the $\Z_2$ fixed point along $T^2_{(1)} \times T^2_{(2)}$ and $(\alpha'\beta')$ its orientifold image. 
For $(\alpha\beta)=(\alpha'\beta')$, the NS-NS sector provides three scalars with the fourth scalar in the hyper multiplet stemming from the R-R sector
in agreement with $e_{\alpha\beta}$ being $\OR$-odd, cf.\ table~\ref{Tab:closed-OR-spectrum}.
If  $(\alpha\beta) \neq (\alpha'\beta')$, the NS-NS sector contains four scalars and the R-R sector one more scalar plus an anti-self-dual tensor in agreement with $(e_{\alpha\beta} \pm e_{\alpha'\beta'})$
being $\OR$-odd $(+)$ and $\OR$-even $(-)$. This confirms the counting $(1+h^+_{11})$ and $h^-_{11}$ of multiplicities of tensor and hyper multiplets, respectively, in table~\ref{Tab:closed-OR-spectrum}.
Note, however, that from the present construction, we cannot infer how the NS-NS and R-R scalars located at the exceptional two-cycles $e_{\alpha\beta} \pm e_{\alpha'\beta'}$ are distributed over the hyper $(\mathfrak{v}^i,\mathfrak{b}^i,\mathfrak{z}^i,\zeta^i)$ and tensor $(\varphi^j)$  multiplet.

 As an additional special feature of tilted tori, for D7-branes parallel to one of the O7-planes $\OR$ or $\OR\Z_2$, orientifold invariant {\it fractional} two-cycles $\Pi_x^{\text{frac}} = \Pi_{x'}^{\text{frac}}$ 
exist for  $b_1 \sigma_1^x\tau_1^x \neq b_2 \sigma_2^x \tau_2^x$, which requires that at least one two-torus is tilted and a discrete displacement and Wilson line occur simultaneously. 
This can e.g.\ be seen by truncating the orientifold invariance condition for $T^6/\Z_2 \times \Z_2$ with discrete torsion in~\cite{Forste:2010gw} to $T^4/\Z_2$. 

\noindent As an exemplary {\bf model with tilted tori} we study the configuration in table~\ref{tab:Z2-tilted} with four stacks of orientifold invariant D7-branes, two of which are parallel to the $\OR$-invariant O7-plane and the other two to the $\OR\Z_2$-invariant O7-plane. The twisted RR tadpoles are cancelled by choosing pairwise opposite $\Z_2$ eigenvalues. Supersymmetry holds for any choice of toroidal complex structure $\rho_i$. Thus the results for the square tori (at $\rho_i =2$ in the notation of figure~\ref{Fig:Z2-lattice}) from section \ref{Ss:SquareT4Z2} apply.
\mathtab{
\begin{array}{|c|c|c|c|c|c|c|c|}\hline
\muc{6}{|c|}{\text{\bf  Orientifold invariant D7-branes on tilted tori}}
\\\hline\hline
x & \tarh{ n_1, m_1}{n_2 , m_2} & (\vec{\sigma}) & \Z_2 & (\vec{\tau}) & \Pi_x^{\text{frac}} = \Pi_{x'}^{\text{frac}} 
\\\hline\hline
a_{i,i=1,2} &  \tarh{2,-1}{2,-1} & \tarh{1}{1} & \pm 1  & \tarh{0}{1} & \frac{ 4 \Pi_{13}^{\text{bulk}} -2 \Pi_{14}^{\text{bulk}} - 2 \Pi_{23}^{\text{bulk}} + \Pi_{24}^{\text{bulk}} \pm  \left[ e_{22} + e_{32} - e_{23} - e_{33} \right]}{2}
\\\hline
b_{i,i=1,2} &  \tarh{0,1}{0,-1} & \tarh{1}{1} & \pm 1  & \tarh{1}{0} & \frac{-\Pi_{24}^{\text{bulk}} \pm  \left[  e_{22} - e_{32} + e_{23} - e_{33} \right] }{2}
\\\hline
\end{array}
}{Z2-tilted}{D7-brane configuration of a global $T^4/\Z_2$ model with tilted tori ($b_1=b_2=\frac{1}{2}$). Any combination of bulk cycles parallel to the $\OR$- or $\OR\Z_2$-invariant plane with $\sigma^x_1 \tau^x_1 \neq \sigma^x_2 \tau^x_2$ 
leads to orientifold invariant D7-branes. For notational concreteness $(\vec{\sigma})=(\vec{1})$ is presented. Other choices of displacements lead to permutations of the fixed point sets $(\alpha,\hat{\alpha}) \in \{(1,4),(2,3)\}$ per two-torus.
}
The gauge groups are due to the orientifold invariance enhanced to $USp(4)_{a_1} \times USp(4)_{a_2} \times USp(4)_{b_1} \times USp(4)_{b_2}$, and  the open string matter spectrum consists of
\begin{equation}
(\4,\4;\1,\1) + (\1,\1;\4,\4) + (\4,\1;\4,\1)+(\4,\1;\1,\4) + (\1,\4;\4,\1) + (\1,\4;\1,\4)
,
\end{equation}
which together with the 14 hyper and 7 tensor multiplets from the closed string sector satisfies the condition $0=  273 - 29 \, n_T + n_V - n_H$ of a vanishing gravitational anomaly as required for any globally consistent model  in six dimensions.

In this example, the volumes of $a_1=a_1'$ and $a_2=a_2'$ ($b_1=b_1'$ and $b_2=b_2'$) are identical at the singular orbifold point, but differ upon deformation of (some of) the ten independent $\Z_2$ fixed points as shown in section~\ref{S:HypersurfaceFormalism}.
Since in four-dimensional models, the volumes are related to the gauge couplings by equation~(\ref{Eq:gauge-vs-volumes}), the present observation constitutes the first indication that tree-level gauge couplings that were identical at the orbifold point for two generic fractional D-branes are expected to differ in magnitude upon deformations away from the singular locus.

For $T^4/\Z_{2N}$ orbifold models with $2N >2$, the number of independent $\Z_2$ fixed point deformations is further reduced by $\Z_{N}$ identifications as detailed in appendix~\ref{Sss:T4Z6} for $T^4/\Z_6$. This reduction can again be seen as a toy example for the phenomenologically appealing but more complicated $T^6/\Z_{2M} \times \Z_2$ geometries and their complex structure deformations.

\subsection[Three-cycles on $T^6/\Z_{2M} \times \Z_{2}$ with discrete torsion]{Three-cycles on $\boldsymbol{T^6/\Z_{2M} \times \Z_{2}}$ with discrete torsion}\label{Ss:T6Z2Z2N}

Factorisable $T^6/\Z_{2M} \times \Z_{2}$ orbifolds with discrete torsion and rigid D6-branes on {\it sLag} three-cycles play an important role in the search for phenomenologically appealing and fully computable 
Type II string theory vacua. The $T^6/\Z_2 \times \Z_2$ orbifold with $h_{21}^{\Z_2} =48$ possesses the maximal possible number of three-cycles, which for the $T^6/\Z_6 \times \Z_2$ and $T^6/\Z_6' \times \Z_2$
orbifolds  are reduced to $h_{21}^{\Z_2} =14$ and 15, respectively, due to the additional related $\Z_3$ subsymmetry as detailed in appendices~\ref{Sss:Z2Z6} and~\ref{Sss:Z2Z6p}.

A new feature of the $T^6/\Z_{2M} \times \Z_{2}$ orbifolds with discrete torsion compared to $T^4/\Z_{2N}$  stems from the fact that worldsheet consistency requires one of the four O6-plane orbits to be `exotic', 
i.e.\ $(\eta_{\OR},\eta_{\OR\Z_2^{(1)}},\eta_{\OR\Z_2^{(2)}},\eta_{\OR\Z_2^{(3)}}) = (\underline{-1,1,1,1})$ with the underline denoting all possible permutations. As explained in detail e.g.\ in~\cite{Blumenhagen:2005tn,Forste:2010gw}, the exceptional divisors in each $\Z_2^{(k)}$ twisted sector might pick up an additional minus sign $\eta_{(k)} \equiv \eta_{\OR} \cdot \eta_{\OR\Z_2^{(k)}}$ under the orientifold projection depending on the choice of exotic O6-plane. 
As an important consequence, orientifold invariant three-cycles also exist without invoking tilted tori as exemplified in section~\ref{Sss:Z2Z2}, and gauge couplings $g_{G_a}^{-2} \sim \text{Vol}(\Pi_a + \Pi_{a'})$ can depend 
on only one $\Z_2^{(k)}$ twisted sector as demonstrated in section~\ref{Sss:Z2Z2} and an example of phenomenological interest in appendix~\ref{Sss:Z2Z6p}. 
We use here the full classification of orientifold invariant three-cycles for any choice of tilted or untilted factorisable tori that was given in~\cite{Forste:2010gw}.

\subsubsection[$T^6/\Z_2 \times \Z_2$ with shift vectors $\vec{v}=\frac{1}{2}(1,-1,0)$ and $\vec{w}=\frac{1}{2}(0,1,-1)$]{$\boldsymbol{T^6/\Z_2 \times \Z_2}$ with shift vectors $\boldsymbol{\vec{v}=\frac{1}{2}(1,-1,0)}$ and $\boldsymbol{\vec{w}=\frac{1}{2}(0,1,-1)}$}\label{Sss:Z2Z2}

The $T^6/\Z_2 \times \Z_2$ orbifold with discrete torsion has the following Hodge numbers~\cite{Vafa:1994rv},
\begin{equation}
\label{Eq:T6Z22HodgeNumbers}
h_{11} =3 =  3_{\text{bulk}},
\qquad\quad
h_{21} = 51 = 3_{\text{bulk}} + (3 \times 16)_{\Z_2} ,
\end{equation}
and exceptional three-cycles are constructed as 
\begin{equation}
\varepsilon_{\alpha\beta}^{(k)} \equiv 2 \, e_{\alpha\beta}^{(k)} \otimes \pi_{2k-1} 
,
\qquad\quad
\tilde{\varepsilon}_{\alpha\beta}^{(k)} \equiv  2 \, e_{\alpha\beta}^{(k)} \otimes \pi_{2k} ,
\end{equation}
located at the $\Z_2^{(k)}$ fixed point $\alpha\beta$ on $T^2_{(i)} \times T^2_{(j)} \equiv T^4_{(k)}$.
The orientifold image cycles depend on the choice of untilted or tilted tori as well as exotic O6-plane as summarised in table~\ref{Tab:Z2Z2-ex-OR-images}.
\begin{table}[h!]
\renewcommand{\arraystretch}{1.3}
  \begin{center}
\begin{equation*}
\begin{array}{|c|c|}\hline
\multicolumn{2}{|c|}{\OR \; \text{\bf  on exceptional 3-cycles on $T^6/\Z_2 \times \Z_2$}}
\\\hline\hline
 \OR ( \varepsilon^{(k)}_{\alpha\beta}) &  \OR (\tilde{\varepsilon}^{(k)}_{\alpha\beta})
 \\\hline\hline
  \eta_{(k)}\,  \left(- \varepsilon^{(k)}_{\alpha'\beta'} + 2 \,  b_k \,   \tilde{\varepsilon}^{(k)}_{\alpha'\beta'}    \right) &  \eta_{(k)} \, \tilde{\varepsilon}^{(k)}_{\alpha'\beta'} 
 \\\hline 
\end{array}
\end{equation*}
\end{center}
\caption{Orientifold images of exceptional three-cycles on the $T^6/\Z_2 \times \Z_2$ orbifold with discrete torsion on a factorisable six-torus.
The ${\cal R}$-images $\alpha'\beta'$ of the fixed points $\alpha\beta$ depend on the tilt parameters $(b_i,b_j)$ along $T^2_{(i)} \times T^2_{(j)}$
as detailed in figure~\protect\ref{Fig:Z2-lattice}. $\eta_{(k)} = \eta_{\OR} \cdot \eta_{\OR\Z_2^{(k)}} =\pm 1$ depends on the choice of exotic O6-plane.}
\label{Tab:Z2Z2-ex-OR-images}
\end{table}
As in the $T^4/\Z_2$ case of section~\ref{Sss:T4Z2}, for untilted tori all 48 deformations of codimension two singularities are independent and associated to $\varepsilon_{\alpha\beta}^{(k)} $
for $\eta_{(k)}=-1$ or $\tilde{\varepsilon}_{\alpha\beta}^{(k)} $ for $\eta_{(k)}=1$. However, differently from $T^4/\Z_{2N}$, the set of $b_3=2 (h_{21}+1)$ three-cycles can for any choice of tilted tori
be split into a symplectic basis of $(h_{21}+1)$ ${\cal R}$-even and  $(h_{21}+1)$ ${\cal R}$-odd three-cycles. Tilted tori thus do not reduce the number of independent complex structure deformations
in four-dimensional models.

As an illustrative example to study the effect of complex structure deformations on the three-cycle volumes and corresponding gauge couplings, 
we focus here on untilted tori and a D6-brane $a$ parallel to the $\OR\Z_2^{(1)}$-invariant planes,
\begin{equation}\label{Eq:Z2Z2-rigid-example}
\begin{aligned}
\Pi^{\text{rigid}}_a = \frac{-\Pi^{\text{bulk}}_{146}}{4}  & + (-1)^{\tau^{\Z_2^{(1)}}_a} \frac{\varepsilon^{(1)}_{11} + (-1)^{\tau_2^a} \varepsilon^{(1)}_{41} + (-1)^{\tau_3^a}\varepsilon^{(1)}_{14} + (-1)^{\tau_2^a+\tau_3^a} \varepsilon^{(1)}_{44} }{4}\\
& + (-1)^{\tau^{\Z_2^{(2)}}_a}  \frac{\tilde{\varepsilon}^{(2)}_{11}  + (-1)^{\tau_1^a} \tilde{\varepsilon}^{(2)}_{21} + (-1)^{\tau_3^a}\tilde{\varepsilon}^{(2)}_{14} + (-1)^{\tau_1^a+\tau_3^a} \tilde{\varepsilon}^{(2)}_{24}  }{4}
\\
& - (-1)^{\tau^{\Z_2^{(3)}}_a}  \frac{\tilde{\varepsilon}^{(3)}_{11} + (-1)^{\tau_1^a} \tilde{\varepsilon}^{(3)}_{21} + (-1)^{\tau_2^a}\tilde{\varepsilon}^{(3)}_{14} + (-1)^{\tau_1^a+\tau_2^a} \tilde{\varepsilon}^{(3)}_{24}   }{4}
.
\end{aligned}
\end{equation}
The orientifold image cycle $a'$ is given by 
\begin{equation}
\begin{aligned}
\Pi^{\text{rigid}}_{a'} = \frac{-\Pi^{\text{bulk}}_{146}}{4}  & -\eta_{(1)} (-1)^{\tau^{\Z_2^{(1)}}_a} \frac{\varepsilon^{(1)}_{11} + (-1)^{\tau_2^a} \varepsilon^{(1)}_{41} + (-1)^{\tau_3^a}\varepsilon^{(1)}_{14} + (-1)^{\tau_2^a+\tau_3^a} \varepsilon^{(1)}_{44} }{4}\\
& + \eta_{(2)} (-1)^{\tau^{\Z_2^{(2)}}_a}  \frac{\tilde{\varepsilon}^{(2)}_{11}  + (-1)^{\tau_1^a} \tilde{\varepsilon}^{(2)}_{21} + (-1)^{\tau_3^a}\tilde{\varepsilon}^{(2)}_{14} + (-1)^{\tau_1^a+\tau_3^a} \tilde{\varepsilon}^{(2)}_{24}  }{4}
\\
& - \eta_{(3)}(-1)^{\tau^{\Z_2^{(3)}}_a}  \frac{\tilde{\varepsilon}^{(3)}_{11} + (-1)^{\tau_1^a} \tilde{\varepsilon}^{(3)}_{21} + (-1)^{\tau_2^a}\tilde{\varepsilon}^{(3)}_{14} + (-1)^{\tau_1^a+\tau_2^a} \tilde{\varepsilon}^{(3)}_{24}   }{4}
.
\end{aligned}
\end{equation}
The discussion below is independent of the choice of displacement parameters $(\vec{\sigma})$, and for simplicity we have taken $(\vec{0})$. Switching on $\sigma_1^a=1$ merely results in replacing $\alpha \in \lbrace 1,2 \rbrace$ by $\lbrace 4,3 \rbrace$ while $\sigma_2^a$ or $\sigma_3^a=1$ implies $\alpha \in \lbrace 2,3 \rbrace$ instead of $\lbrace 1,4 \rbrace$.

For the choice $\eta_{\OR\Z_2^{(1)}}=-1$ of exotic O6-plane, $\Pi^{\text{rigid}}_a$ is orientifold invariant and has (up to permutation of two-tori) e.g.\ been used in the context of D2-brane $O(1)$ instantons in~\cite{Cvetic:2007ku}.  Any other choice of exotic O6-plane implies $\Pi^{\text{rigid}}_a \neq \Pi^{\text{rigid}}_{a'}$,
\begin{equation}\label{Eq:Z2Z2-frac-Pis-one-Z2-only}
\begin{aligned}
\Pi^{\text{rigid}}_a + \Pi^{\text{rigid}}_{a'} = \frac{-\Pi^{\text{bulk}}_{146}}{2}  +
\left\{\begin{array}{cc}
(-1)^{\tau^{\Z_2^{(1)}}_a} \frac{\varepsilon^{(1)}_{11} + (-1)^{\tau_2^a} \varepsilon^{(1)}_{41} + (-1)^{\tau_3^a}\varepsilon^{(1)}_{14} + (-1)^{\tau_2^a+\tau_3^a} \varepsilon^{(1)}_{44} }{2}& \eta_{\OR}=-1 \\
- (-1)^{\tau^{\Z_2^{(3)}}_a}  \frac{\tilde{\varepsilon}^{(3)}_{11} + (-1)^{\tau_1^a} \tilde{\varepsilon}^{(3)}_{21} + (-1)^{\tau_2^a}\tilde{\varepsilon}^{(3)}_{14} + (-1)^{\tau_1^a+\tau_2^a} \tilde{\varepsilon}^{(3)}_{24}}{2}&  \eta_{\OR\Z_2^{(2)}}=-1 \\
(-1)^{\tau^{\Z_2^{(2)}}_a}  \frac{\tilde{\varepsilon}^{(2)}_{11}  + (-1)^{\tau_1^a} \tilde{\varepsilon}^{(2)}_{21} + (-1)^{\tau_3^a}\tilde{\varepsilon}^{(2)}_{14} + (-1)^{\tau_1^a+\tau_3^a} \tilde{\varepsilon}^{(2)}_{24}  }{2}&   \eta_{\OR\Z_2^{(3)}}=-1
\end{array}\right.
,
\end{aligned}
\end{equation}
and such a stack for $\eta_{\OR}=-1$ has been used e.g.\ in~\cite{Blumenhagen:2005tn}  to support the $U(4)$ gauge group of a global Pati-Salam model, see also~\cite{Forste:2008ex} for a similar model.

In the notation of equation~(\ref{Eq:Def-Z}), the calibration condition at the orbifold point is encoded in the complex quantity ${\cal Z}_a = \prod_{k=1}^3 \frac{n_k^a R_1^{(k)} + i \, \tilde{m}_k^a R_2^{(k)}}{\sqrt{R_1^{(k)}R_2^{(k)}}}$,
and the massless states in the $\Z_2^{(1)}$ twisted sector of Type IIA/$\OR$ on $T^6/\Z_{2M} \times \Z_2$ with discrete torsion are explicitly given by 
\begin{equation}
\mbox{\resizebox{0.95\textwidth}{!}{%
$
\begin{array}{l|l}  |00++\rangle|00++\rangle_{\rm NSNS}^{(\alpha\beta)} + \eta_{(1)} |00--\rangle|00--\rangle_{\rm NSNS}^{(\alpha'\beta')}  & 
 |++00\rangle|-+00\rangle_{\rm RR}^{(\alpha\beta)} - \eta_{(1)} |--00\rangle|+-00\rangle_{\rm RR}^{(\alpha'\beta')}  \end{array}\!\!\! 
 $}}
 \end{equation}
 in analogy to the $T^4/\Z_{2}$ twisted states in equation~(\ref{Eq:T4Z2-IIB+OR-twisted-spectrum}).  
 The bosonic field content of one NS-NS scalar $\mathfrak{c}^k$ plus one R-R axionic scalar $\xi^k$ per $\Z_2$ fixed point $(\alpha\beta)$ is again in agreement with the general spectrum in table~\ref{Tab:closed-OR-spectrum}.
Matter states in the other $\Z_2^{(k)}$ twisted sectors are obtained by permutation of two-torus indices and the corresponding entries in $SO(8)$ weight states. 
Since we do not consider twisted sector states of other order in this article,  we refer the interested reader to the full list of states in each twist sector of $T^6/\Z_{2M} \times \Z_2 \times \OR$ in~\cite{Forste:2010gw}.
The reduced number of independent complex structure deformations due to some $\Z_{2M}$ symmetry with $2M>2$ is detailed in appendices~\ref{Sss:Z2Z6} and~\ref{Sss:Z2Z6p}, with the latter containing a global model
exhibiting key features of the deformation dependences discussed above.

\section{Deformations of Orbifold Singularities and Hypersurface Formalism}\label{S:HypersurfaceFormalism}

This section discusses the hypersurface formalism which we use to describe {\it sLag} cycles on the deformed orbifolds. We start by introducing {\it sLag} cycles on elliptic curves, i.e.\ two-tori, in section~\ref{Sss:EllipticCurves}, and at deformed isolated (local) singularities in section~\ref{Ss:sLagLocalDeform}. These results will be carried over to the global deformation geometries in section~\ref{Ss:SquareT4Z2} for $T^4/\Z_2$  and in section~\ref{Ss:SquareT6Z22} for $T^6/\Z_2\times\Z_2$ with discrete torsion. There, we look at different deformation scenarios and analyze the structure of the {\it sLag} cycles that arises. In section~\ref{Sss:SquareT4Z2Integrals}, we numerically compute integrals of the holomorphic two-form over such cycles, depending on the deformation parameters. All computations in this section are performed for (untilted) square tori and the two basic choices of {\it sLag} cycles per two-torus.
In section~\ref{Ss:MoreGeneral}, we comment on the increase in computational complexity for generalizations required to analyze phenomenologically appealing global rigid D6-brane models.

\subsection{The geometric setup}\label{Ss:Geometry}

\subsubsection{Lagrangian lines on elliptic curves}\label{Sss:EllipticCurves}

Since the starting point for defining toroidal orbifolds is the torus, we start with the description of a two-torus as a hypersurface, namely an elliptic curve $E$ in the complex weighted projective space $\mathbb{P}^2_{112}$. This space is spanned by homogeneous coordinates $x,v,y$ which are subject to scalings $(x,v,y) \sim (\lambda x, \lambda v, \lambda^2 y)$. The elliptic curve is defined as the zero locus of the polynomial
\begin{equation}
 \label{Eq:DefEllipticCurve}
 \begin{aligned}
 E &= \{ f = 0 \} \,, \qquad f = -y^2 + F(x,v) \,, \\
  F(x,v) &= 4\, x^3v - g_2 \, xv^3 - g_3 \, v^4 = 4 \, v \, ( x-\epsilon_2 v) \cdot ( x- \epsilon_3 v ) \cdot (x - \epsilon_4 v) \,.
 \end{aligned}
\end{equation}
It appears useful to write the polynomial $F(x,v)$ in factorised form which shows the zeros of $x/v$ at $\epsilon_i, i=2,3,4$\footnote{This unusual enumeration is chosen because this way the $i^\text{th}$ fixed point (according to the convention of figure \ref{Fig:Z2-lattice}) lies at $x/v = \epsilon_i$.} and $\epsilon_1 = \infty$, with $\sum_i \epsilon_i = 0$, $g_2 = 4 \sum_{i<j} \epsilon_i \epsilon_j$ and $g_3 = 4 \epsilon_2 \epsilon_3 \epsilon_4$. 

To get a map from the ordinary definition of a torus to the elliptic curve, we will use the Weierstrass $\wp$-function. We define the torus as $T^2 = \bC / \Lambda$ with a lattice $\Lambda = \omega_1 \bZ \oplus \omega_2 \bZ$ with complex structure $\tau := \omega_2 / \omega_1 \not\in \R$. Then the Weierstrass $\wp$-function is given by
\begin{equation}
 \label{Eq:WeierstrassP}
 \wp(z) = \frac1{z^2} + \sum_{\Lambda \ni v \neq 0} \left( \frac1{( z- v )^2} - \frac1{v^2} \right) \,, 
 \end{equation}
and fulfils the differential equation
\begin{equation}
 \label{Eq:WeierstrassDiffEq}
  {\wp'(z)}^2 = 4 \, \wp(z)^3 - g_2 \, \wp(z) - g_3 \,.
\end{equation}
Thus, if we identify $\wp(z) = x/v $ and $\wp'(z) = y/v^2$, we get a bijective map from $T^2$ to $E$. The Klein invariant is then given by $j(\tau) = g_2^3 / ( g_2^3 - 27 \, g_3^2)$. Although $g_2$ and $g_3$ are fully determined on the lattice $\Lambda$, $j(\tau)$ uniquely represents a choice of complex structure and is thus invariant under the modular group $SL(2,\bZ)$.
Note that the $\Z_2$ reflection $z \mapsto -z$, which is a symmetry of every torus, translates to $(x,v,y) \mapsto (x,v,-y)$.

\paragraph{Conditions for Lagrangian lines}

We will describe Lagrangian lines as fixed sets under an antiholomorphic involution $\sigma$ of the space in question. In this way, they can - besides cycles wrapped by D-branes - potentially represent orientifold planes in a given string compactification. In case of a two-torus, one can always rotate the lattice $\Lambda$ such that this involution takes the form\footnote{We denotel a general antiholomorphic involution by $\sigma$, whereas $\sigma_{\cal R}$ refers to the orientifold involution.} 
$\sigma: z \mapsto \overline{z}$. This involution is a symmetry of $T^2$ if the lattice is invariant, i.e.\ $\sigma\Lambda\sigma = \Lambda$, or if $\tau$ and $\overline\tau$ are related by $SL(2,\bZ)$. 
Now, since $j(\tau)$ is meromorphic with real coefficients, this is the case precisely if $j(\tau) = j(\overline\tau) = \overline{j(\tau)}$ or $j(\tau) \in \R$. Then we can rotate the lattice such that $g_2$ and $g_3$ become real. We can distinguish between two cases:
\begin{itemize}
 \item Tori with $ 1 \le j(\tau) $ are the untilted ({\textbf{\textit{a}}}-type) tori of figure~\ref{Fig:Z2-lattice} since the complex structure can be represented by $\tau \in i \R$. The zeros $\epsilon_i$ lie on the real line, and we order them as $\epsilon_4 < \epsilon_3 < \epsilon_2$ to be consistent with the convention of figure \ref{Fig:Z2-lattice}. At the boundary, we find the square torus ($j=1$ or $g_3=0$) and the degenerate torus ($j = \infty$). 
 \item Tori with $ 1 \ge j(\tau)$ are the tilted ({\textbf{\textit{b}}}-type) tori of figure~\ref{Fig:Z2-lattice} where $\tau \in i \R + 1/2$. Here, two zeros of $F(x,v)$ are complex conjugate to each other, $\epsilon_4 = \overline \epsilon_3 =: \epsilon$ implying that the third one is $\epsilon_2 = - 2 \Re (\epsilon)$. This agrees with the convention e.g.\ in figure \ref{Fig:Z6lattice}. 
The boundary cases are the same as for the untilted tori, where the square torus is rotated by $\frac{\pi}{4}$. In addition, we find the hexagonal torus at $j(\tau) = 0$ ($g_2 = 0$) which has an extra $\Z_3$ symmetry given by $x \mapsto e^{2\pi i /3} x$.
\end{itemize}

On each two-torus, we can find Lagrangian lines for each set of winding numbers $(n,m)$ which are relatively prime and through any point on the torus. In the language of the homogeneous coordinates $x,v,y$, it is in general hard to write down an equation for them, so we focus on those Langrangian lines which are represented by antilinear (i.e.\ antiholomorphic and linear) maps $\sigma$ of the form
\begin{equation}
\label{Eq:OrientifoldEllipticCurve}
 \begin{pmatrix}
  x \\ v \\
 \end{pmatrix}
\longmapsto A
 \begin{pmatrix}
  \bar x \\ \bar v \\
 \end{pmatrix}\,, \qquad y \longmapsto e^{i \beta} \bar y \,,
\end{equation}
where $A$ is a complex $2 \times 2$ matrix. In order to be a valid involution,  the map must satisfy the two conditions
\begin{itemize}
 \item $ \overline{A} A = \unity$ ,
 \item $\sigma(F(x,v)) = e^{- 2 i \beta } F(\overline x, \overline v)$ .
\end{itemize}

\begin{landscape}
\begin{table}[th]
\bCentering
\begin{tabular}{|c|c|c|c|c|c|}\hline
\muc{6}{|c|}{\bf Antilinearly realised Lagrangian lines per square two-torus}
\\\hline\hline
Condition & A & $\beta$ & Equation in $\wp, \wp'$ & Label & \\
\hline\hline
\multirow{4}{*}{ 
\raisebox{-70pt}{$\begin{array}{c} \text{untilted tori} \\ j \ge 1 \\ \end{array}$ } } & \multirow{2}{*}{ \raisebox{-20pt}{$
\begin{pmatrix}
 1 & 0 \\
0 & 1
\end{pmatrix}
$}} & 0  & 
$\begin{array}{c} {\bf I_a}: \epsilon_2 \le \wp \le \infty\,, \\ {\bf I_b}: \; \epsilon_4 \le \wp \le \epsilon_3 \,, \\  \end{array}$  $\Im (\wp') =0$ & 
${\bf I_a}$ ,${\bf I_b}$  & \raisebox{-17pt}{\includegraphics[scale=0.5]{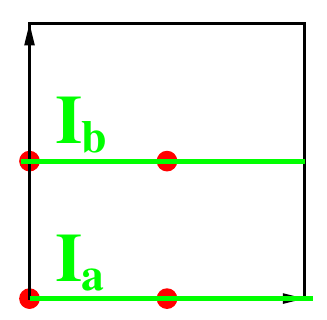}}  \\
\cline{3-6}
 & &  $\pi$ &  
$\begin{array}{c} {\bf II_a}: -\infty \le \wp \le \epsilon_4\,, \\ {\bf II_b}: \;\;\;\; \epsilon_3 \le \wp \le \epsilon_2 \,, \\  \end{array}$  $\Re (\wp') =0$ & 
${\bf II_a}$ ,${\bf II_b}$ & \raisebox{-17pt}{ \includegraphics[scale=0.5]{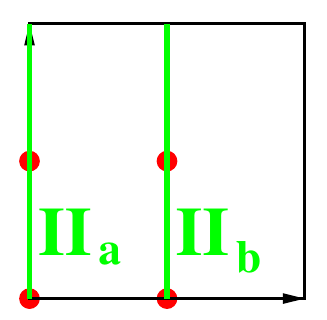}}\\
\cline{2-6}
& $\frac{1}{\sqrt{2 \epsilon_4^2 + \epsilon_3 \epsilon_2 }} 
 \begin{pmatrix}
 \epsilon_4 & \epsilon_4^2 + \epsilon_3 \epsilon_2 \\
1 & - \epsilon_4 
\end{pmatrix}$
& $\pi$  & $\left| \wp - \epsilon_4 \right|^2 = 2 \epsilon_4^2 + \epsilon_3 \epsilon_2$ \,, $\Re\left( (\wp - \epsilon_4)/ \wp' \right) =0$ & ${\bf III}$&  \raisebox{-17pt}{ \includegraphics[scale=0.5]{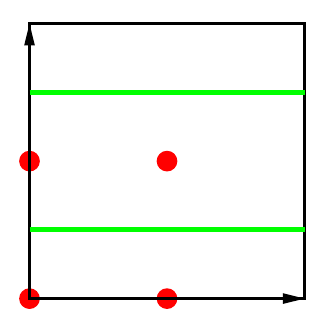}}\\
\cline{2-6}
& $\frac{1}{\sqrt{2 \epsilon_2^2 + \epsilon_4 \epsilon_3 }} 
 \begin{pmatrix}
 \epsilon_2 & \epsilon_2^2 + \epsilon_4 \epsilon_3 \\
1 & - \epsilon_2 
\end{pmatrix}$
& $0$  & $\left| \wp - \epsilon_2 \right|^2 = 2 \epsilon_2^2 + \epsilon_4 \epsilon_3$ \,, $\Im\left( (\wp - \epsilon_2) / \wp'\right) =0$ & ${\bf IV}$&  \raisebox{-17pt}{ \includegraphics[scale=0.5]{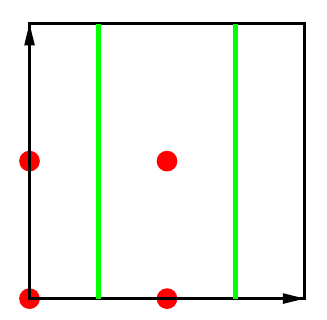}}\\
 \hline\hline
\multirow{4}{*}{
\raisebox{-70pt}{$ \begin{array}{c} \text{square tori} \\ j = 1 \\ \end{array}$  }} & \multirow{2}{*}{\raisebox{-20pt}{
$ \begin{pmatrix}
 1 & 0 \\
0 & - 1
\end{pmatrix}$}
} & $\frac\pi2$  & $\Re (\wp) = 0$ \,, $\Im ( e^{-i\pi/4} \wp') =0$ & {\bf V}&  \raisebox{-17pt}{\includegraphics[scale=0.5]{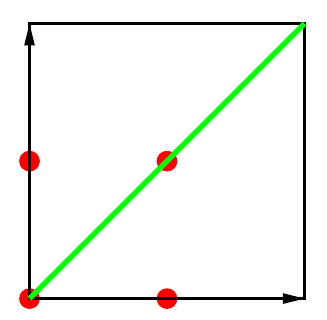}}\\
\cline{3-6}
 & & $\frac{3\pi}2$  & $\Re (\wp) = 0$ \,, $\Im ( e^{i\pi/4} \wp') =0$ & {\bf VI}&  \raisebox{-17pt}{\includegraphics[scale=0.5]{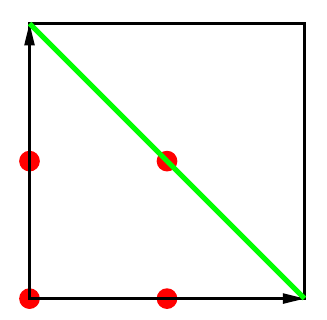}}\\
\cline{2-6}
 & \multirow{2}{*}{\raisebox{-20pt}{$ 
 \begin{pmatrix}
 0 & 1 \\
1 &  0
\end{pmatrix} $}
}& $\frac\pi2$  & $| \wp |^2 = 1$ \,, $\Im ( e^{-i\pi/4} \wp'/ \wp) =0$ & {\bf VII}&  \raisebox{-17pt}{\includegraphics[scale=0.5]{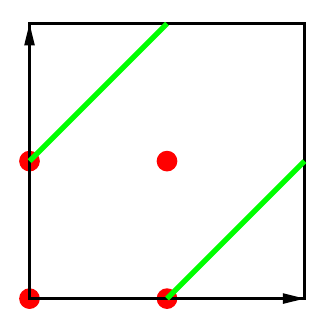}}\\
\cline{3-6}
 & & $\frac{3\pi}2$  &  $| \wp |^2 = 1$ \,, $\Im ( e^{i\pi/4} \wp'/ \wp) =0$ & {\bf VIII}&  \raisebox{-17pt}{\includegraphics[scale=0.5]{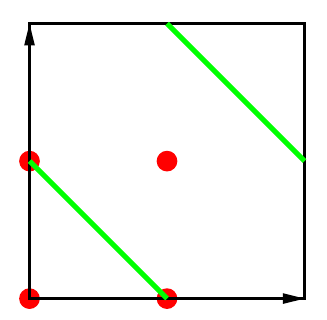}}\\
 \hline
\end{tabular}
\bCaption{Classification of antilinearly realised Lagrangian lines on untilted and square tori. Note that the pairs of cycles ${\bf III_{a/b}}$ as well as ${\bf IV_{a/b}}$ are connected by the $\Z_2$ symmetry on the orbifold, see section \ref{Sss:SquareT4Z2Cycles}. The subscripts ${\bf a}$ and ${\bf b}$ are not related to the two types \textbf{\textit{a}} and \textbf{\textit{b}} of two-tori, see figure \ref{Fig:Z2-lattice}.
}
\label{tab:T2LagrangianLinesUntilted}
\end{table}
\end{landscape}

\begin{table}[th]
\bCentering
\begin{tabular}{|c|c|c|c|c|}\hline
\muc{5}{|c|}{\bf Antilinearly realised Lagrangian lines per hexagonal two-torus}
\\\hline\hline
A & $\beta$ & Equation in $\wp, \wp'$ & Label & \\
\hline\hline
\multirow{2}{*}{
\raisebox{-20pt}{$ \begin{pmatrix}
 1 & 0 \\
0 & 1
\end{pmatrix}$}
} & $0$  & $\Im (\wp) = 0$ \,, $\Im (\wp') =0$ & {\bf I} & \raisebox{-17pt}{\includegraphics[scale=0.5]{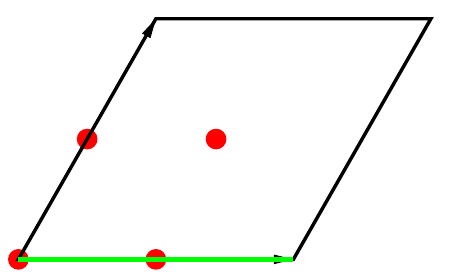}}\\
\cline{2-5}
 & $\pi$  & $\Im (\wp) = 0$ \,, $\Re (\wp') =0$ & {\bf II} & \raisebox{-17pt}{\includegraphics[scale=0.5]{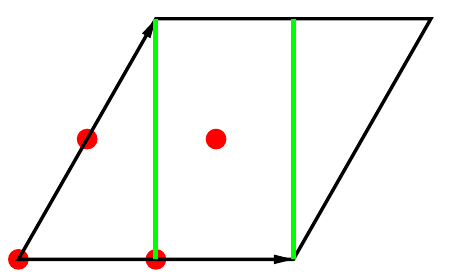}}\\
\cline{1-5}
\multirow{2}{*}{
\raisebox{-20pt}{$\frac1{\sqrt{9 \epsilon_\Re^2 + \epsilon_\Im^2}}
\begin{pmatrix}
 -2 \epsilon_\Re & 5 \epsilon_\Re^2 + \epsilon_\Im^2 \\
1 &  2\epsilon_\Re
\end{pmatrix}$}
}& $0$  &
$
\begin{array}{c}
\left| \wp + 2 \epsilon_\Re \right|^2 = 9 \epsilon_\Re^2 + \epsilon_\Im^2 \\ \Im( (\wp + 2 \epsilon_\Re) / \wp') =0
   \end{array} $ &
{\bf III} & \raisebox{-17pt}{\includegraphics[scale=0.5]{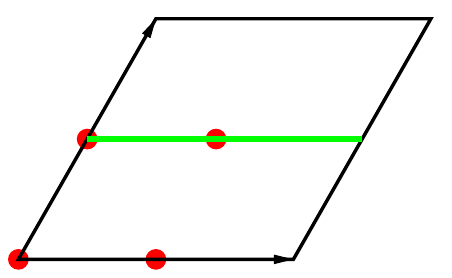}}\\
\cline{2-5}
 & $\pi$  & 
 $ \begin{array}{c}
\left| \wp + 2 \epsilon_\Re \right|^2 = 9 \epsilon_\Re^2 + \epsilon_\Im^2 \\ \Re( (\wp + 2 \epsilon_\Re) / \wp') =0
 \end{array}$ & 
 {\bf IV} & \raisebox{-17pt}{\includegraphics[scale=0.5]{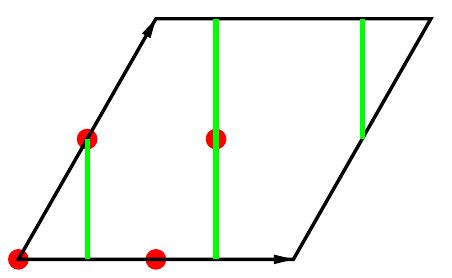}}\\
 \hline
\end{tabular}
\bCaption{Classification of antilinearly realised Lagrangian lines on tilted tori. For hexagonal tori ($g_2=0$)  as special cases one gets additional lines by the symmetry transformation $ x \mapsto e^{2\pi i/3} x$. For shortness we write $\epsilon = \epsilon_\Re + i \epsilon_\Im$. }
\label{tab:T2LagrangianLinesTilted}
\end{table}


Note that for each choice of $A$, we get two values of $\beta$ which differ by addition of $\pi$. Furthermore, matrices $A$ which differ by an overall complex phase correspond to the same Lagrangian line. For testing the {\it sLag} property, we need the holomorphic one-form,
\begin{equation}
\label{Eq:HolomorphicOneForm}
\Omega_1 := d z = \frac{d \wp}{\wp'} = \frac{v \cdot dx - x \cdot dv}{y}  \stackrel{v \equiv 1}{\longrightarrow} \frac{dx}y \,.
\end{equation}
This tells us that for each matrix $A$, the two choices for $\beta$ differ in their calibration. The last expression is given since we will often work in the coordinate chart $v \equiv 1$. It turns out that for each choice of $g_2$ and $g_3$, a finite number of Lagrangian lines arises in this way. The results are summarised in tables \ref{tab:T2LagrangianLinesUntilted} and \ref{tab:T2LagrangianLinesTilted}.

\subsubsection{Lagrangians on local deformations}\label{Ss:sLagLocalDeform}

As a next step to finding {\it sLags} on deformed global geometries, we first look for them in local deformations. We start with the deformation of a $\bC^2 / \Z_2$ singularity and then move on to $\bC^3 / \Z_2 \times \Z_2$.

The construction of a $\boldsymbol{\bC^2 / \Z_2}$ singularity as a hypersurface proceeds as follows. If $z_1,z_2$ are the coordinates with $\Z_2$ acting as $(z_1,z_2) \mapsto (-z_1,-z_2)$, we can form a basis of invariant monomials by $x_i = z_i^2\, (i=1,2)$ and $y=z_1 z_2$. These coordinates obey one relation, so we find
\begin{equation}
 \label{Eq:C2Z2Hypersurface}
 \bC^2 / \Z_2 = \{ f = - y^2 + x_1 x_2 =0 \} \subset \bC^3 \,.
\end{equation}
The singularity is now indicated by solutions to $f = df = 0$ which happens at the origin. Thus, the deformation is obtained by adding terms of lower order than in \eqref{Eq:C2Z2Hypersurface}. Linear terms in $y$ and $x_i$ can be absorbed by proper redefinitions so that we only add a constant term with complex parameter $\varepsilon$,
\begin{equation}
\label{Eq:DeformC2Z2}
 \text{Def}(\bC^2 / \Z_2) = \{ f = -y^2 + x_1 x_2 - \varepsilon = 0 \} \subset \bC^3 \,.
\end{equation}
For $\varepsilon \neq 0$, there are no solutions to $f = df = 0$; so \eqref{Eq:DeformC2Z2} describes a smooth space. In case of the local resolution, we can choose $\varepsilon \in \mathbb{R}$ by rotations of the coordinates. In the case $\varepsilon > 0$ we look at the solutions of the antiholomorphic equation $x_1 =  \overline{x}_2$ which represents two real equations, i.e.\ its solutions are Lagrangian surfaces. Inserting the relation into \eqref{Eq:DeformC2Z2} implies that $y^2 \in i\R$. 
Choosing the branch $y \in \R$, we end up with a cycle $E^+$ that fulfils the equation of a two-sphere whose size is controlled by $\varepsilon$, namely $\Im(y)^2 + |x_1|^2 = \varepsilon$. We define the holomorphic two-form such that it reproduces the familiar result in the singular limit,
\begin{equation}\label{Eq:HolomorphicTwoForm}
\Omega_2 = \, \frac{dx_1 \wedge dx_2}{4y}\ \stackrel{\varepsilon \to 0}{\longrightarrow}\  \, dz_1 \wedge dz_2 \,.
\end{equation}
Note the similarity of the form of $\Omega_2$ with the holomorphic form in equation~\eqref{Eq:HolomorphicOneForm} on the elliptic curve. The imaginary part $\Im(\Omega_2)$ vanishes on the cycle, so the cycle is in fact {\it sLag}. Integrating $\Re(\Omega_2)$ over the exceptional cycle at the origin, we find its volume to be equal to $2 \pi \sqrt{\varepsilon}$, which differs from the naive guess $4 \pi \varepsilon$.
If, however, $\varepsilon<0$, the resulting exceptional two-cycle is given by $E^- = \left\{ x_1 = -\overline{x}_2 \,,\ y \in i\R \right\}$, one finds that $\Omega_2$ restricted to it becomes imaginary. Thus $E^-$ has a different calibration from $E^+$, and in particular it is odd under the orientifold, to match the generalised Gimon--Polchinski model as displayed in table \ref{tab:GP-model}. To sum up, depending of whether the deformation is done in positive or negative direction in $\varepsilon$, the exceptional cycle is either even or odd under $\Omega\mathcal{R}$.

The step to the singularity of $\boldsymbol{\bC^3/ \Z_2 \times \Z_2}$ is straightforward at the level of the hypersurface description. But when one considers deformations and {\it sLag} cycles, new issues arise. 
As in the two-dimensional case, we start with coordinates $z_i\,, i=1,2,3$ of $\bC^3$ and form a basis of $\Z_2 \times \Z_2$ invariant monomials by $x_i = z_i^2 \;  (i=1,2,3)$ and $y=z_1z_2z_3$. Since they obey one relation, we recover the local singularity as a hypersurface,
\begin{equation}
 \label{Eq:C3Z22Hypersurface}
 \bC^3 / \Z_2 \times \Z_2 = \{ f = - y^2 + x_1 x_2 x_3 =0 \} \subset \bC^4 \,.
\end{equation}
Here the singular locus $f=df=0$ consists of three singular lines $\{x_i = x_j = x_k = 0\, | i \neq j \neq k \neq i \}$ which intersect at the origin. In order to deform these singularities, we again add terms of lower order, where terms linear in $y$ and terms of the form $x_ix_j$ can be absorbed by redefinitions. Thus, we are left with terms linear in $x_i$ and a constant term, which gives a total of four terms. However, as argued in \cite{Vafa:1994rv}, string theory only provides three deformation parameters, and a generic deformation will always leave a codimension three singularity which is a conifold. This will put one restriction on the four deformation parameters, and we find,
\begin{equation}
\label{Eq:DeformC3Z22}
 \text{Def}(\bC^3 / \Z_2 \times \Z_2) = \{ f = -y^2 + x_1 x_2 x_3  - \varepsilon_1 x_1 - \varepsilon_2 x_2 - \varepsilon_3 x_3 + 2 \sqrt{\varepsilon_1 \varepsilon_2 \varepsilon_3}  = 0 \} \subset \bC^4 \,.
\end{equation}
One easily sees that the conifold singularity lies at $x_i = \sqrt{\varepsilon_1 \varepsilon_2 \varepsilon_3 } / \varepsilon_i$. Note that the involution $x_i \mapsto \overline x_i$ not only requires $\varepsilon_i \in \R$, but also $\varepsilon_1 \varepsilon_2 \varepsilon_3 >0$ since otherwise the constant term becomes imaginary.

We first discuss the simpler case where just one of the fixed planes is deformed, e.g.\ $\varepsilon_1 \neq 0 \,, \varepsilon_2=\varepsilon_3=0$. One can then factor out $x_1$ in \eqref{Eq:DeformC3Z22}  and recover the deformed $\bC^2 / \Z_2$ singularity when $x_1 \neq 0$. For $x_1 = 0$, we find a singular line at $x_2 x_3 = \varepsilon_1$ which shows that the singular lines $x_2=0$ and $x_3=0$ have merged together. 
The Lagrangian three-cycles, that arise from the deformation, are a product of the Lagrangian two-cycle in the $x_2$-$x_3$ plane and a one-cycle in the $x_1$ plane of the form $x_1 = e^{i \phi} \, \overline x_1$. Due to the non--compactness of $\bC^3/\Z_2 \times \Z_2$, we find a whole family of Lagrangian cycles parametrised by the phase $\phi$ such that for given calibration, one of them is special Lagrangian. 

The case of a generic deformation $\varepsilon_i \neq 0\, \forall i$ is a bit more involved. By rotations we choose all $\varepsilon_i$ to be real and positive. We want to find an equation for a Lagrangian cycle similar to the one described above for just one deformation. In the $\Z_2^{(1)}$ twisted sector, the proper antiholomorphic equations are $\varepsilon_2 x_2 = \varepsilon_3 \overline x_3$ and $x_1 = \overline x_1$. Then \eqref{Eq:DeformC3Z22} implies $y^2 \in \R$, and we choose the branch $y \in i\R$. With the additional assumption $x_1 \neq 0$ we can rewrite \eqref{Eq:DeformC3Z22} as
\begin{equation}
\label{Eq:LagrangianDefC3Z22}
\left| \frac{\varepsilon_2}{\varepsilon_3} x_2 - \frac{\sqrt{\varepsilon_2 \varepsilon_3}}{x_1} \right|^2 + \left( \frac{\Im(y)}{\sqrt{x_1}} \right)^2 = \left( \sqrt{\varepsilon_1} - \frac{\sqrt{\varepsilon_2 \varepsilon_3}}{x_1} \right)^2 \,.
\end{equation}
This describes a fibration of a two-sphere over the line $x_1 \ge \sqrt{\varepsilon_2 \varepsilon_3 / \varepsilon_1}$ with radius $\sqrt{\varepsilon_1} - \sqrt{\varepsilon_2 \varepsilon_3}/x_1$. At the boundary $x_1 = \sqrt{\varepsilon_2 \varepsilon_3 / \varepsilon_1}$, the radius goes to zero, and the sphere degenerates to a point which closes the cycle. This point coincides with the conifold singularity found above. Furthermore, we define the holomorphic three-form such that it reproduces the familiar result in the singular limit,
\begin{equation}
\Omega_3 = i \, \frac{dx_1 \wedge dx_2 \wedge dx_3}{8y}\ \stackrel{\varepsilon_i \to 0}{\longrightarrow}\ i \, dz_1 \wedge dz_2 \wedge dz_3\,.
\end{equation}
The three-cycle is {\it sLag} w.r.t.\ $\Omega_3$ and is thus the exceptional cycle resulting from the deformation. Moreover, we can define other three-cycles analogously by permuting the homogeneous coordinates $x_i$. Thus, in the generic deformed space, we find three exceptional {\it sLag} cycles associated to the $\Z_2^{(i)}$ ($i=1,2,3$) twisted sectors, which intersect in the conifold singularity. However, the topological intersection number of these cycles is 
- in agreement with the orbifold language of section~\ref{S:sLags} -  zero as can be seen by deforming the conifold. Analogously to the two-dimensional case, the sign of the deformation parameters $\varepsilon_i$ will determine the calibration of the exceptional three-cycle and thus its behaviour under the orientifold.

\subsubsection{Global deformations as hypersurfaces}\label{Ss:GlobalDeform}

Finally, we can put the ingredients together to construct a hypersurface description of the global orbifolds that allow for deformations. We again start with $T^4 / \Z_2$: We take two elliptic curves as in section \ref{Sss:EllipticCurves} with homogeneous coordinates $y_i$, $x_i$, $v_i$, $i=1,2$ describing a $T^4$. The $\Z_2$ acts as $(y_1,y_2) \mapsto (-y_1,-y_2)$ while leaving the other coordinates invariant. 
We thus define $y = y_1 \cdot y_2$ as an invariant coordinate. In other words, we can consider this $T^4$ as a fibration of four points over $\mathbb{P}^1 \times \mathbb{P}^1$ where the $\mathbb{P}^1$'s are spanned by $(x_i,v_i)$ and the points are the solutions of the equation $y_i^2 = F_i(x_i,v_i)$. 
Then the $\Z_2$ acts only in the fibre such that the quotient fibre is described by just one equation $f = -y^2 + F_1(x_1,v_1) \cdot F_2(x_2,v_2) =0$. Instead of a complete intersection of hypersurfaces in a product of simple weighted projective spaces $\mathbb{P}^2_{112}$, we now have one hypersurface in a more general toric space \cite{Hori:2003ic,Fulton:1993} with weights $q_i$ given by the following table,
\begin{center}
 \begin{tabular}{|c|c|c|c|c|}
 \hline
   & $x_1,v_1$ & $x_2,v_2$ & $y$ & $f$ \\
   \hline
   $q_1$ & $1$ & $0$ & $2$ & $4$ \\
   \hline
   $q_2$ & $0$ & $1$ & $2$ & $4$  \\
   \hline
 \end{tabular} \,.
\end{center}
The Calabi--Yau condition is fulfilled since the weight of the hypersurface polynomial $f$ equals the sum of the weights of all coordinates. The singular locus ($f=df=0$) is given by $F_1 = F_2 = 0$ and consists of $4 \times 4 = 16$ isolated points since each $F_i$ is of degree four. To find the generic deformation, we need to count all monomials of the same degree as $f$. 
The coefficient of the $y^2$ term is scaled to one. Terms linear in $y$ are absorbed by shifting $y$ accordingly to complete the square. The remaining terms are monomials of bidegree $(4,4)$ in $x_1,v_1$ and $x_2,v_2$. For each pair $x_i,v_i$ we construct a basis of degree four monomials, which form a five-dimensional vector space. 
One basis element is $F_i(x_i,v_i)$. We call the other basis elements $\delta F_i^{(\alpha)}(x_i,v_i)$, where the index $\alpha=1,\ldots,4$ stands for the four fixed points per $T^2$. The  $\delta F_i^{(\alpha)}$  are chosen such that they have the same zeros as $F_i$ except for the $\alpha^\text{th}$ one. For example we can choose,
\begin{equation}
 \label{Eq:DeformMonomialBasis}
 \begin{array}{rl}
  F_i(x_i,v_i) &= 4 v_i \cdot \left( x_i - \epsilon_2 v_i \right) \cdot \left( x_i - \epsilon_3 v_i \right) \cdot \left( x_i - \epsilon_4 v_i \right) \,, \\
  \delta F_i^{(1)}(x_i,v_i) &= -4 \left( x_i - \epsilon_2 v_i \right) \cdot \left( x_i - \epsilon_3 v_i \right)^2 \cdot \left( x_i - \epsilon_4 v_i \right) \,, \\
  \delta F_i^{(2)}(x_i,v_i) &= 4 v_i \cdot \left( x_i - \epsilon_3 v_i \right) \cdot \left( x_i - \epsilon_4 v_i \right)^2 \,, \\
  \delta F_i^{(3)}(x_i,v_i) &= 4 v_i^2 \cdot \left( x_i - \epsilon_2 v_i \right) \cdot \left( x_i - \epsilon_4 v_i \right) \,, \\
  \delta F_i^{(4)}(x_i,v_i) &= -4 v_i \cdot \left( x_i - \epsilon_2 v_i \right)^2 \cdot \left( x_i - \epsilon_3 v_i \right) \,.
 \end{array}
\end{equation}
Now we can count the independent deformations. The coefficient of the $F_1 F_2$ term can be set to one by rescaling. Terms of the form $F_i \cdot \delta F_j^{(\alpha)}$ have four parameters, three of which can be absorbed by $PGL(2,\bC)$ transformations on $(x_j,v_j)$. The remaining parameter represents the complex structure of the $j^\text{th}$ two-torus. We are left with 16 terms of the form $\delta F_1^{(\alpha_1)} \cdot \delta F_2^{(\alpha_2)}$, and we will see that each such term deforms exactly one singularity. To sum up, we give the explicit form of the hypersurface equation with the 16 deformation parameters explicitly,
\begin{equation}
 \label{Eq:DefT4Z2Hypersurface}
 \text{Def}(T^4 / \Z_2) = \left\{ y^2 = F_1 \cdot F_2 + \sum_{a_1,a_2 = 1}^4 \varepsilon_{a_1 a_2} \delta F_1^{(\alpha_1)} \cdot \delta F_2^{(\alpha_2)} \right\} \,.
\end{equation}
The two more complex structure parameters of the tori are hidden in the expressions for $F_i(x_i,v_i)$. We define the two-form on the global deformation space as
 \begin{equation}
  \label{Eq:HolomorphicTwoFormGlobal}
  \Omega_2 = \frac{\left( v_1 \cdot dx_1 - x_1 \cdot dv_1 \right) \wedge \left( v_2 \cdot dx_2 - x_2 \cdot dv_2 \right) }{y} \,.
 \end{equation}
More precisely, this definition describes a family of holomorphic two-forms, where the parameters can be seen in the precise expression for $y$ from \eqref{Eq:DefT4Z2Hypersurface}. The definition~(\ref{Eq:HolomorphicTwoFormGlobal}) reproduces the formula for the local deformation \eqref{Eq:HolomorphicTwoForm} in the limit $v_i \to \infty$ as well as the orbifold result in the singular limit $\varepsilon_{a_1,a_2} \to 0$, where we use the product of one-forms on two elliptic curves \eqref{Eq:HolomorphicOneForm} with the orbifold identification $y = y_1y_2$. Most of the time, we will work in coordinate patches of the form $v_i = 1$ or $x_i = 1$ where \eqref{Eq:HolomorphicTwoFormGlobal} simplifies accordingly.

The construction of a deformable $T^6 / \Z_2 \times \Z_2$ orbifold with discete torsion works analogously. One starts with three elliptic curves, parametrised by homogeneous coordinates $x_i$,$v_i$,$y_i$ ($i=1,2,3$) and replaces the $y_i$ by the $\Z_2 \times \Z_2$ invariant $y := y_1 y_2 y_3$ which is subject to one constraint,
\begin{equation}
\label{Eq:T6Z22Hypersurface}
 f = -y^2 + F_1(x_1,v_1) \cdot  F_2(x_2,v_2) \cdot  F_3(x_3,v_3) = 0\,.
\end{equation}
 This equality describes a hypersurface in the toric space given by the weights $q_i$ as follows,
\begin{center}
 \begin{tabular}{|c|c|c|c|c|c|}
 \hline
   & $x_1,v_1$ & $x_2,v_2$ & $x_3,v_3$ & $y$ & $f$ \\
   \hline
   $q_1$ & $1$ & $0$ & $0$ & $2$ & $4$ \\
   \hline
   $q_2$ & $0$ & $1$ & $0$ & $2$ & $4$ \\
   \hline
   $q_3$ & $0$ & $0$ & $1$ & $2$ & $4$ \\
   \hline
 \end{tabular} \,.
\end{center}
The singular locus is $F_i = F_j = 0$ with $x_k,v_k$ free for all $(ijk)$ permutations of $(123)$; it represents the $3 \times 4 \times 4 = 48$ fixed lines of $\Z_2 \times \Z_2$, i.e.\ singularities of codimension two. These lines intersect in 64 codimension-three singularities at $F_1 = F_2 = F_3 = 0$. To go to generic deformations, we again classify the monomials which are allowed to modify $f$. As in the case with four compact dimensions, the important terms do not contain $y$. We distinguish between the following structures:
\begin{itemize}
 \item $F_1 F_2 F_3$ is the term whose coefficient is set to one.
 \item $F_i F_j \delta F_k^{(\alpha)}$, with $(ijk)$ a cyclic permutation of $(123)$, give rise to the three complex structure moduli of the three two-tori after absorbing $PGL(2,\bC)$ transformations in $(x_k,v_k)$. From the orbifold perspective in section~\ref{S:sLags}, their coefficients are the untwisted moduli.
 \item $F_i \, \delta F_j^{(\alpha_j)} \delta F_k^{(\alpha_k)}$ are $3 \times 16$ terms which correspond $1:1$ to deformations of the $(a_j,a_k)$ singularity in the $j-k$ plane. In the orbifold construction of section~\ref{S:sLags}, the coefficients of these terms appear as twisted complex structure moduli.
  \item $\delta F_1^{(\alpha_1)} \delta F_2^{(\alpha_2)}\delta F_3^{(\alpha_3)}$ would give 64 additional deformation parameters. However, for given values of all other parameters, these are determined by the requirement that the whole space contains 64 conifold singularities, as for the local deformation. This follows again from the fact that string theory only provides moduli that deform the codimension two singularities, but not those of codimension three. The precise expressions for these parameters are in general not as simple as in \eqref{Eq:DeformC3Z22}.
\end{itemize}
To sum up, the equation for the space with generically deformed fixed planes is $f=0$ with
\begin{equation}
 \label{Eq:DeformT6Z22}
 f= -y^2 + F_1 F_2 F_3 - \!\!\!\!\!\!\!\! 
 \sum_{(ijk) = \sigma(123) \atop \sigma \text{ cyclic}} \sum_{\alpha,\beta=1}^4 \varepsilon_{\alpha\beta}^i F_i \delta F_j^{(\alpha)} \delta F_k^{(\beta)} + \!\!\!\!
 \sum_{\alpha,\beta,\gamma=1}^4 \varepsilon_{\alpha\beta\gamma}\left( \varepsilon_{\alpha\beta}^i \right) \delta F_1^{(\alpha)}\delta F_2^{(\beta)}\delta F_3^{(\gamma)} \,,
\end{equation}
with $\varepsilon_{\alpha\beta\gamma} \left( \varepsilon_{\alpha\beta}^i \right)$ denoting the functions of  $\varepsilon_{\alpha\beta}^i$, which ensure the existence of the $64$ conifold singularities.

In section \ref{S:sLags}, we introduced three types of Lagrangian cycles on toroidal orbifolds, the bulk cycles $\Pi^\text{bulk}$, the exceptional cycles $\Pi^{\Z_2}$ and fractional cycles $\Pi^\text{frac}$. The fractional cycles are the most interesting ones because of their many discrete parameters. The questions that arise are, which of these cycles can be found in the hypersurface formalism for the resolution, and what are the restrictions on the parameters of the hypersurface equation. To answer these questions we take the ans\"atze for the {\it (s)Lag} cycles from the elliptic curves in section \ref{Sss:EllipticCurves} and the local deformations in section \ref{Ss:sLagLocalDeform} and transfer them to the global deformation. 

For concreteness, we fix the complex structure of the two-tori to the square (i.e.\ $\rho = 1$ for untilted tori in the notation of section~\ref{S:sLags}) which is done by setting $g_3=0$ in \eqref{Eq:DefEllipticCurve}. Then by a rescaling, we choose $g_2=4$ such that the zeros of $F_i(x_i,v_i)$ are at $x_i / v_i = -1, 0, +1, \infty$. The advantages of the square torus are that it can be seen as either tilted  or untilted (i.e.\ $\rho = 2,1$ in the notation of figure~\ref{Fig:Z2-lattice}), depending on how the lattice is embedded into $\bC^2$. It therefore offers a bigger variety of Lagrangian lines defined by an antilinear involution than an arbitrary rectangular torus, see cases {\bf V} to {\bf VIII} in table \ref{tab:T2LagrangianLinesUntilted}.

\subsection[sLags on the deformed $T^4 / \Z_2$ on square tori]{sLags on the deformed $\boldsymbol{ T^4 / \Z_2}$ on square tori}\label{Ss:SquareT4Z2}

We start with {\it sLag} cycles on the deformation of $T^4 / \Z_2$, where we focus on cycles which appear in the T-dual of the Gimon--Polchinski model as described in section \ref{Sss:T4Z2}. Cycles which descend from {\it sLag} lines on the elliptic curves are already present at the orbifold point. Thus, we will first analyse them on the singular space and then check what happens when deformations are turned on. In order to specify a {\it sLag} cycle, we choose one antilinear involution for each pair of homogeneous torus coordinates $x_i,v_i$ as in \eqref{Eq:OrientifoldEllipticCurve}. This is always possible at the orbifold point if it is possible on the underlying two-tori. Therefore such {\it sLags} can appear as O7-planes. In this case, every deformation of the singularities must be compatible with the involution $\sigma_{\cal R}$. Those {\it sLags} on which D7-branes sit need a priori not to be fixed loci of antilinear involutions since the {\it sLag} condition is formulated using the K\"ahler form $J_{1,1}^{\text{K\"ahler}}$ and the holomorphic two-form $\Omega_2$, see \eqref{Eq:sLag-cond}. This also holds for the exceptional cycles which need the deformation to have finite volume. 
Their local counterparts were introduced as fixed cycles of antilinear involutions in section \ref{Ss:sLagLocalDeform}.

\subsubsection{Cycle structure}\label{Sss:SquareT4Z2Cycles}
The first class of cycles corresponds to a product of {\it Lag} lines on the two two-tori. We denote them here as ${\bf N_1} \otimes {\bf N_2}$ where ${\bf N_i} \in \{ {\bf I_a} , {\bf I_b}, {\bf II_a}, \ldots, {\bf VIII} \}$, cf.\ table \ref{tab:T2LagrangianLinesUntilted}. From the position of the lines in table \ref{tab:T2LagrangianLinesUntilted}, we can already conclude whether the corresponding two-cycle on the orbifold or its deformation will be a bulk cycle or a fractional cycle.
  
Cycles which do not pass through the $\Z_2$ fixed points are representatives of bulk cycles. These are cycles which contain {\bf III} or {\bf IV} as one or two factors\footnote{Although bulk cycles on the orbifold have moduli that allow them to change their position, the cycles {\bf III} and {\bf IV} are just the representatives at position $\Re(z)=1/4 , 3/4$ or $\Im(z)=1/4, 3/4$, respectively, see table \ref{tab:T2LagrangianLinesUntilted}. As a consequence these cycles are always bulk cycles.}. We can deduce their topology from their behaviour in the two complex $x_i$ planes\footnote{Unless stated otherwise we work in the patch $v_i \equiv 1$.}. 
\begin{itemize}
 \item The cycles\footnote{Mixed cycles like ${\bf III} \otimes {\bf IV}$ are {\it sLag}, but calibrated with $\Im(\Omega_2)$. Nevertheless, the same discussion applies to them.}  ${\bf III}\otimes{\bf III}$ and ${\bf IV}\otimes{\bf IV}$ do not pass through the zero locus of $y$, and we can fix the branch cuts of the hypersurface equation \eqref{Eq:DefT4Z2Hypersurface} such that the cycles do not intersect the branch cuts. In this way, we see that each cycle consists in fact of two disconnected components, which are distinguished by the sign of $y$. Since each factor is a circle in the $x_i$ plane, the whole cycle has the topology of a $T^2$.
 \item If one of the factors is not {\bf III} or {\bf IV}, the projection of the cycle on the corresponding $x_i$ plane is a real curve whose endpoints are zeros of $y$. Thus, the two components $y = \pm \sqrt{F_1 \cdot F_2}$ are glued together at these points and form an $S^1$. Altogether these cycles are one copy of $T^2$ since the ${\bf a}$ and ${\bf b}$ components in the other factor (${\bf I_i}$ or ${\bf II_i}$) have been identified.
\end{itemize}

\begin{figure}[ht]
\begin{minipage}[h]{0.5\linewidth}
\hspace{-3mm}
\begin{center}\hspace{-3mm}
\includegraphics[scale=0.4]{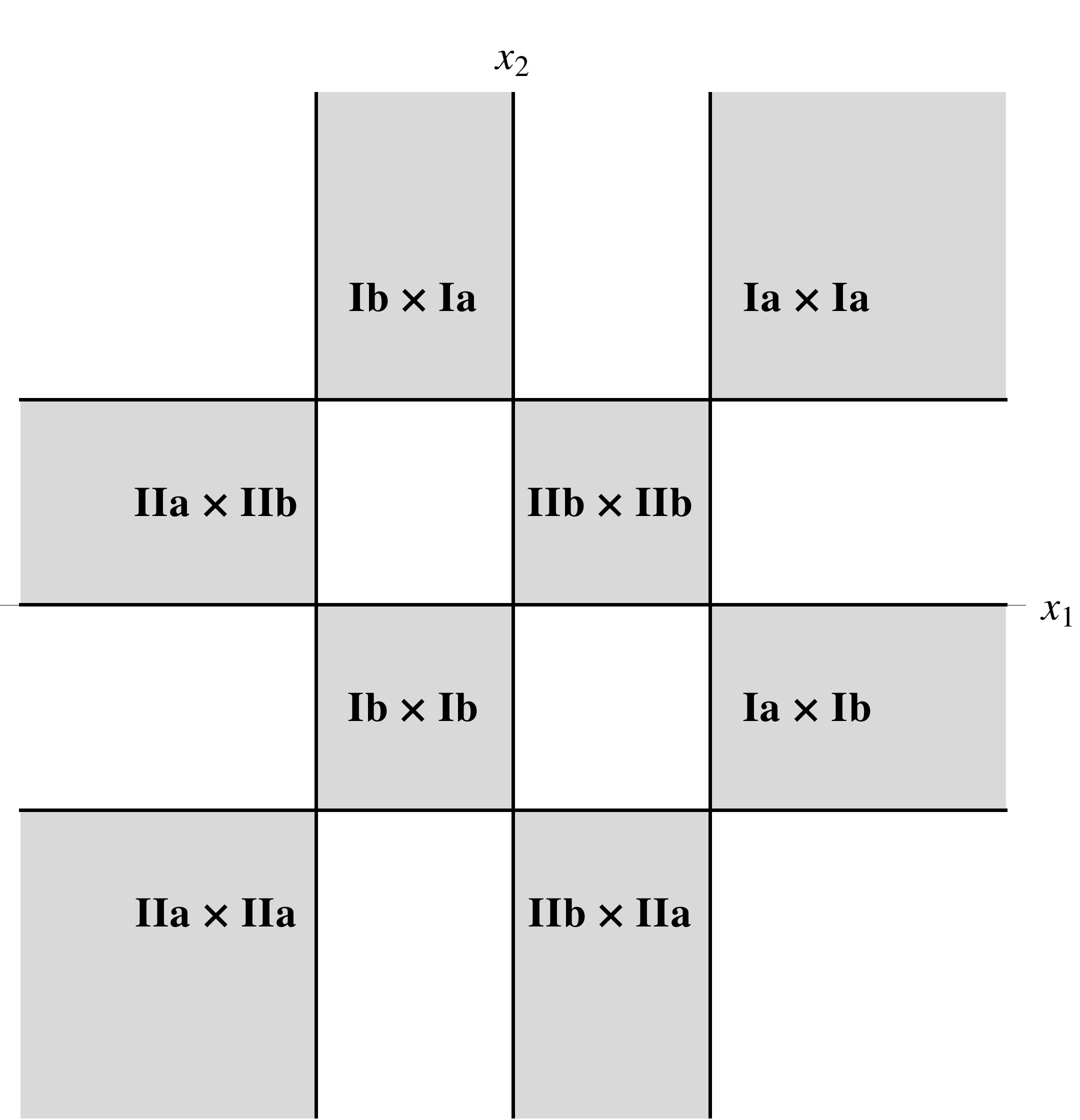} 
\end{center}
\caption{Projection of horizontal and vertical orbifold {\it sLag} cycles on the $x_1$-$x_2$ plane with calibration $\Re(\Omega_2)$. The lines are the zeros of $y$ which factorise on the orbifold. The white rectangles represent cycles which are {\it sLag} with respect to the calibration $\Im(\Omega_2)$.}
\label{Fig:LagsInX1X2Plane} 
\end{minipage}
\hspace{0.5cm}
\begin{minipage}[h]{0.5\linewidth}
\vspace{5mm}
\begin{center}
\includegraphics[scale=0.4]{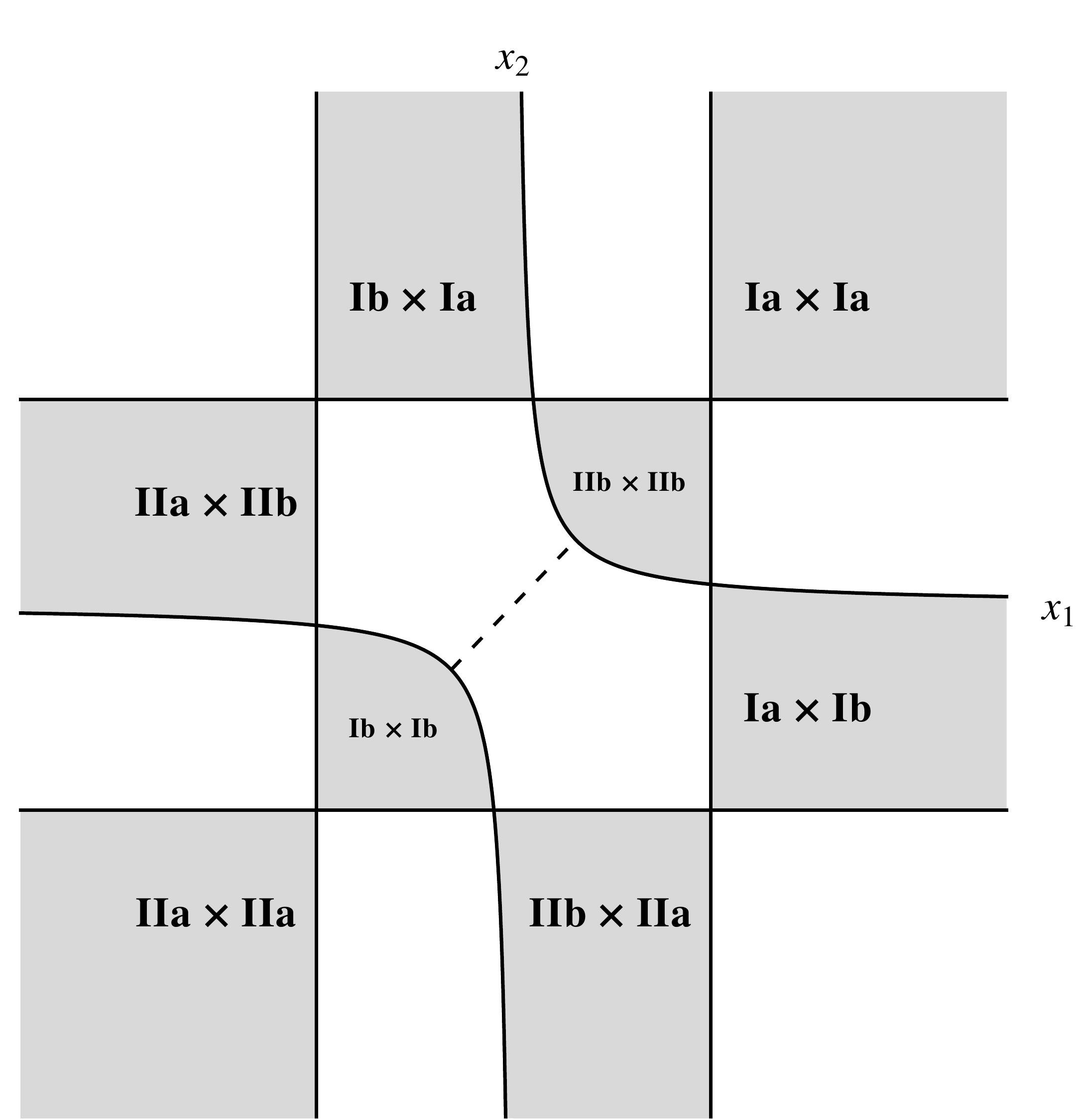} 
\end{center}
\caption{Same as figure \ref{Fig:LagsInX1X2Plane} but with fixed point $(3,3)$ of $T^2 \times T^2$ at the origin in the $x_1$-$x_2$ plane deformed. The dashed diagonal line through the origin represents the exceptional cycle that acquired a non-zero volume.}
\label{Fig:LagsInX1X2PlaneDeformed1} 
\end{minipage}
\end{figure}

Next we focus on cycles of type ${\bf N_1} \otimes {\bf N_2}$ with ${\bf N_i} = {\bf I} , {\bf II}$. If ${\bf N_1} = {\bf N_2}$, the cycle is, after choosing the proper sign, {\it sLag} with calibration $\Re(\Omega_2)$, and otherwise the calibration is $\Im(\Omega_2)$. Since these cycles fulfil $\Im(x_1) = \Im(x_2) = 0$, it is convenient to draw them in the real $x_1$-$x_2$ plane. Note that the compactification of this plane is a $T^2$ since one needs to add circles at infinity to glue together the boundaries at $x_i = \pm\infty$. The zero locus of $y$ describes four horizontal and four vertical lines (at $x_i = 0, \pm1, \infty$) in the real $x_1$-$x_2$ plane and divides it into 16 rectangles, where eight rectangles correspond to {\it sLag} cycles and the remaining eight are {\it Lag} but not {\it sLag}, see figure \ref{Fig:LagsInX1X2Plane}.
 In fact, each cycle consists of two identical rectangles differing only by the sign of $y$ and glued together at the boundary (where $y=0$), so the cycles look like a $S^2$ or more precisely like a $T^2 / \Z_2$. Therefore, they clearly represent fractional cycles as in equation~\eqref{Eq:Def-frac}. 
The parameters $\sigma_i \in \{0,1\}$ are adjustable since each cycle exists either undisplaced (label {\bf a} $\Leftrightarrow \sigma_i = 0$) or displaced (label {\bf b} $\Leftrightarrow \sigma_i = 1$) from the origin, cf.\ table~\ref{tab:T2LagrangianLinesUntilted}. This determines the choice of exceptional cycles in equation~\eqref{Eq:Def-PiZ2}, however their signs, encoded in $\tau^{\Z_2}$, $\tau_1$ and $\tau_2$, are not clear as long as we are at the orbifold point. This issue will be resolved on the deformation in section \ref{Sss:SquareT4Z2Integrals}.

We now turn to {\it sLag} cycles on the deformation space. These will on the one hand include deformations of the bulk and fractional cycles that we just encountered on the orbifold. On the other hand, we hope to see exceptional cycles that stem from the deformation itself and are shrunk to zero volume on the orbifold. For simplicity, we start with the deformation of just one $\Z_2$ fixed point which will be at position $(3,3)$ on $T^2 \times T^2$ since this corresponds to the origin in the $x_1$-$x_2$ plane\footnote{This deformation is equivalent to the deformation of any other fixed point as can be seen by discrete shifts on the torus.}. The hypersurface equation \eqref{Eq:DefT4Z2Hypersurface} with the expressions for $F_i$, $\delta F_i^{(\alpha)}$ given in \eqref{Eq:DeformMonomialBasis} and zeros of $y$ at the $\Z_2$ fixed points $\epsilon_2=-1$, $\epsilon_3=0$ and $\epsilon_4=+1$ now becomes
\begin{equation}
 \label{Eq:HyperSurfaceOneDeform}
 \begin{array}{rcl}
 y^2 &=& F_1(x_1,v_1) \cdot F_2(x_2,v_2) - \varepsilon \cdot \delta F_1^{(3)}(x_1,v_1) \cdot \delta F_2^{(3)}(x_2,v_2) \\
 &=& v_1 v_2 \left( x_1^2 - v_1^2 \right) \left(  x_2^2 - v_2^2 \right) \left( x_1 x_2 - \varepsilon v_1 v_2 \right) \\
 & \stackrel{v_i \equiv 1}{=}& \left( x_1^2 - 1 \right) \left(  x_2^2 - 1 \right) \left( x_1 x_2 - \varepsilon \right) \,.
 \end{array}
\end{equation}
The involution $\sigma_{\mathcal{R}}:(x_i,v_i) \mapsto (\overline{x}_i,\overline{v}_i)$ clearly requires $\varepsilon \in \R$, which confirms the fact that orientifolding removes half of the complex structure moduli. The effect of the deformation on the fixed locus of $\sigma_{\mathcal{R}}$ is best visualised in the real $x_1$-$x_2$ plane, where we find that two formerly separated zero-lines of $y$ ($x_1=0$ and $x_2=0$) now combine into one zero-curve (namely $x_1x_2 = \varepsilon$), see figure \ref{Fig:LagsInX1X2PlaneDeformed1}. If $\varepsilon > 0$,  the cycles ${\bf I_b} \otimes {\bf I_b}$ and ${\bf II_b} \otimes {\bf II_b}$ start to shrink with growing $\varepsilon$. Although this by itself does not imply anything about the volume, we can already conclude that the $\Z_2$ eigenvalue in \eqref{Eq:Def-PiZ2} is negative\footnote{This is in analogy to the result when blowing up the singularity using the K\"ahler modulus $\mathfrak{v}$ in table \ref{Tab:closed-OR-spectrum}, where one finds a relation between ordinary divisors $D \simeq \Pi^\text{frac}$, inherited divisors $R \simeq \Pi^\text{bulk}$ and exceptional divisors $E \simeq \Pi^{Z_2}$ of the form $2 D \sim R - E$, see e.g.\ \cite{Lust:2006zh,Blaszczyk:2011hs}. }, $(-1)^{\tau^{\Z_2}} = -1$, since at the point $\varepsilon=1$ the cycle vanishes completely and thus its volume must decrease with the deformation. This fact tells us furthermore that the fractional cycle $e_{33}$, which now has finite volume, has the same calibration as the bulk part $\Pi_{13}$ (or $-\Pi_{24}$) because both cycles are topologically components of the {\it sLag} cycle ${\bf I_b} \otimes {\bf I_b}$ (or ${\bf II_b} \otimes {\bf II_b}$). In section \ref{Sss:SquareT4Z2Integrals} we will confirm this quantitatively. 

The cycles ${\bf I_b} \otimes {\bf II_b}$ and ${\bf II_b} \otimes {\bf I_b}$ on the other hand seem to have merged. This is necessary because the calibration of $\Pi_{14}$ (or $\Pi_{23}$) differs from the one of $e_{33}$, and thus a fractional cycle of the form $\frac{1}{2}(\Pi_{14} \pm e_{33} \pm \ldots)$ (or $\frac{1}{2}(\Pi_{23} \pm e_{33} \pm \ldots)$) cannot be {\it sLag}, cf. also the discussion of supersymmetry and combinatorics of fractional cycles in section~\ref{S:sLags}. Instead, the newly created cycle must be the union of two fractional cycles, one with bulk part $\Pi_{14}$ and one with $\Pi_{23}$, such that the exceptional part $e_{33}$ cancels out.

For the exceptional cycles we want to choose a similar ansatz as in the local case in section \ref{Ss:sLagLocalDeform} with the involution $\sigma:x_1 \leftrightarrow \overline x_2$. Globally, however, this only works if we at the same time exchange the $\bC^*$ scalings and in particular the other homogeneous coordinates, $\sigma: v_1 \leftrightarrow \overline v_2\,, \ y \mapsto \pm \overline y$. For the fixed set of this involution we can rewrite \eqref{Eq:HyperSurfaceOneDeform}, after a proper choice of branch, as
\begin{equation}
  \left( \frac{\Im(y)}{\sqrt{ \left( x_1^2 -1 \right) \cdot \left( \overline{x}_1^2 - 1 \right) }} \right)^2 + \left| x_1 \right|^2 = \varepsilon\,,
\end{equation}
which still is the equation of a $S^2$ as long as $|x_1| < 1$. This is ensured when we consider small deformations with $\varepsilon < 1$. This cycle is {\it sLag} with calibration $\Re(\Omega_2)$ with $\Omega_2$ as in \eqref{Eq:HolomorphicTwoFormGlobal}, and its position in figure \ref{Fig:LagsInX1X2PlaneDeformed1} is indicated by the dashed line. In case of a deformation with $\varepsilon<0$, one finds a similar picture with an orientfold-odd cycle of calibration $\Im(\Omega_3)$.

This clearly explains why we observe  fractional {\it sLag} cycle with bulk part $\Pi_{13}/2$, but not with e.g.\ $\Pi_{14}/2$: it has a different calibration from the exceptional part $e_{33}$. One can engineer the fractional cycle $\Pi_{14}/2$ by cutting the cycle ${\bf I_b} \otimes {\bf II_b} + {\bf II_b} \otimes {\bf I_b}$  along $x_1 = x_2$. The resulting boundary is a circle, and it divides the exceptional cycle into two hemispheres. Thus, we can take one such hemisphere to close the cycle, which will be of the form e.g.\ $\frac{1}{2} \left( \Pi_{14} \pm e_{33} \pm \ldots \right)$ where the $\pm$ in front of the deformed exceptional cycle $e_{33}$ comes from the ambiguity of which hemisphere is used. These fractional cycles have the same behaviour under the orientifold as required in the generalised T-dual to the Gimon--Polchinski model in table~\ref{tab:GP-model}, namely 
\begin{equation}
\Pi^\text{frac}_a = \frac12 \left( \Pi^\text{bulk}_a + \Pi^{\Z_2}_a \right) \,, \qquad
\Pi^\text{frac}_{a'} = \frac12 \left( \Pi^\text{bulk}_a - \Pi^{\Z_2}_a \right) \,.
\end{equation} 


The whole discussion so far applied to the case of just one deformation in positive direction, $\varepsilon_{33} > 0$. The exceptional cycle that appears is {\it sLag} with calibration $\Re(\Omega_2)$, and the adjacent fractional cycles 
(${\bf I_b} \otimes {\bf I_b}$ and ${\bf II_b} \otimes {\bf II_b}$) of same calibration and with negative exceptional contribution show up as {\it sLag} cycles made out of one component, whereas fractional cycles of different calibration had to merge to stay {\it sLag}. If we deform in negative direction, i.e.\ $\varepsilon_{33} < 0$, we find the same picture just with switched roles. The exceptional cycle (in this case at $x_1 = - \overline x_2$) is now calibrated w.r.t.\ $\Im(\Omega_2)$, and we find that equally calibrated {\it sLag}s (here ${\bf I_b} \otimes {\bf II_b}$ and ${\bf II_b} \otimes {\bf I_b}$) appear with negative exceptional part whereas the cycles ${\bf I_b} \otimes {\bf I_b}$ and ${\bf II_b} \otimes {\bf II_b}$ have merged together. 

Since the deformation happens locally for small $\varepsilon$, this result applies to each singularity when we consider a general deformation. More precisely,  each fixed point can be deformed in two directions such that the exceptional cycle becomes calibrated w.r.t.\ the real or imaginary part of $\Omega_2$. Before the deformation, there is a transversal crossing of the $y=0$ locus with itself, indicating a singularity. Afterwards, the crossing point is (locally) replaced  by two smooth non-intersecting curves, cf. the hyperbolas in figure~\ref{Fig:LagsInX1X2PlaneDeformed2}. 
Switching on all deformations leads in general to a rather complicated cycle structure. As an example, in figure \ref{Fig:LagsInX1X2PlaneDeformed2} we show the horizontal (${\bf I_i} \otimes {\bf I_j}$) and vertical (${\bf II_i} \otimes {\bf II_j}$) two-cycles, where all singularities are deformed such that the exceptional cycles have the same calibration. 
The corresponding cycles are listed in table \ref{Tab:sLagsInFullDeform}. We see that all cycles with calibration $\Im(\Omega_2)$ have merged to one connected {\it sLag} cycle (white area in figure~\ref{Fig:LagsInX1X2PlaneDeformed2}),  whereas those calibrated with $\Re(\Omega_2)$ are fractional cycles with negative exceptional contribution.

 \begin{minipage}[ht]{\textwidth}
  \begin{minipage}[b]{0.49\textwidth}
    \centering
\includegraphics[scale=0.38]{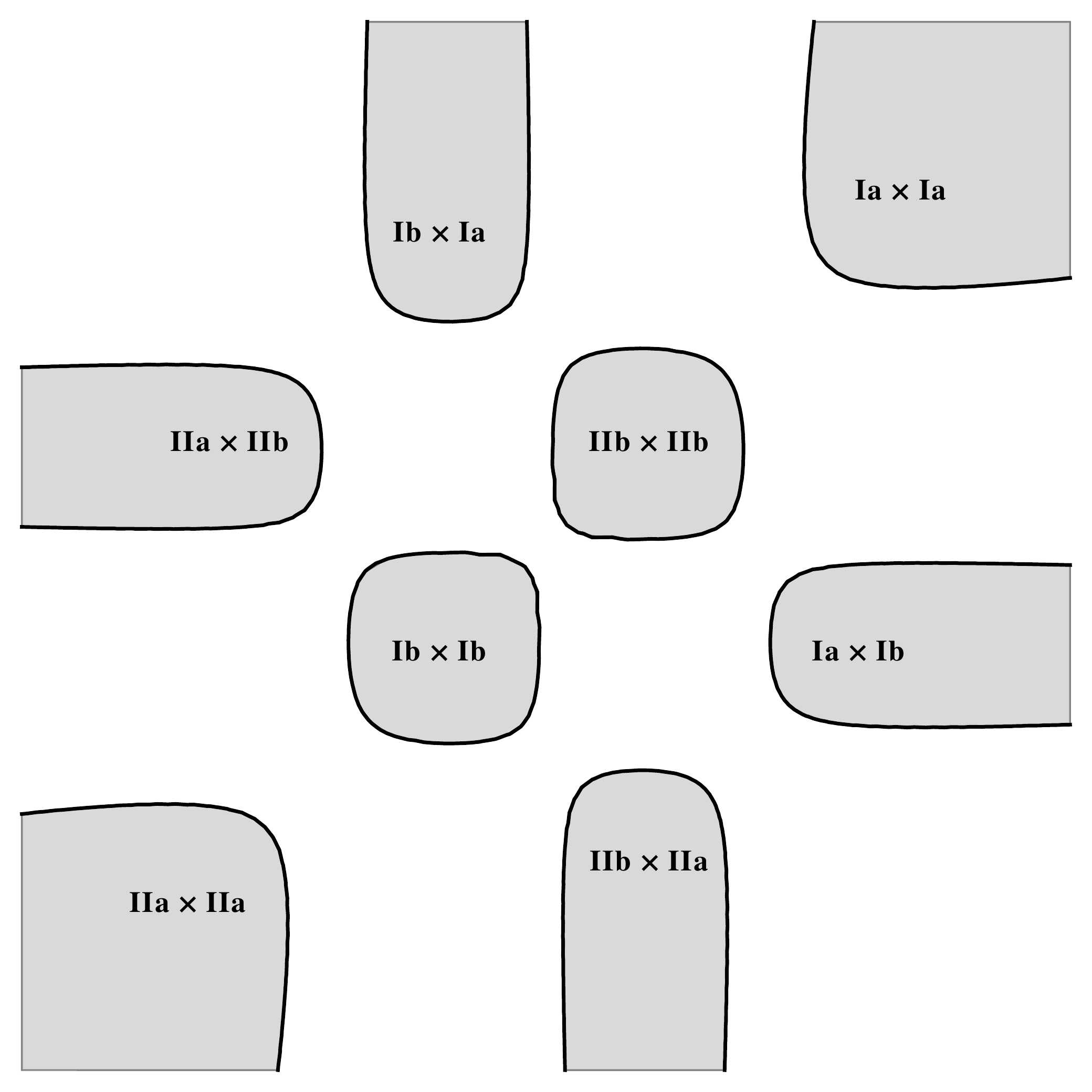} 
    \captionof{figure}{Projection of {\it sLag} cycles on the real $x_1$-$x_2$ plane with all 16 singularities deformed. The white region represents a cycle which is the union of all fractional cycles with calibration $\Im(\Omega_2)$.}
\label{Fig:LagsInX1X2PlaneDeformed2}     
  \end{minipage}
  \hfill
 \begin{minipage}[b]{0.49\textwidth}
\scriptsize
\begin{tabular}{|c|c|}
\hline
\muc{2}{|c|}{\bf {\it (s)Lag} cycles on the deformed $T^4/\Z_2$}
\\ \hline \hline
 ${\bf N_1} \otimes {\bf N_2}$ & $\Pi$ \\
 \hline \hline
 ${\bf III} \otimes {\bf III}$ & $\Pi_{13}$ \\
 \hline
 ${\bf IV} \otimes {\bf IV}$ & $-\Pi_{24}$ \\
 \hline
 ${\bf III} \otimes {\bf IV}^{\dagger}$ & $\Pi_{14}$ \\
 \hline
 ${\bf IV} \otimes {\bf III}^{\dagger}$ & $\Pi_{23}$ \\
 \hline \hline
 ${\bf I_a} \otimes {\bf I_a}$ & $\frac12\left( \Pi_{13} \!-\! e_{11} \!-\! e_{12} \!-\! e_{21} \!-\! e_{22} \right) $ \\
 \hline
 ${\bf I_a} \otimes {\bf I_b}$ & $\frac12\left( \Pi_{13} \!-\! e_{14} \!-\! e_{13} \!-\! e_{24} \!-\! e_{23} \right) $ \\
 \hline
 ${\bf I_b} \otimes {\bf I_a}$ & $\frac12\left( \Pi_{13} \!-\! e_{41} \!-\! e_{42} \!-\! e_{31} \!-\! e_{32} \right) $ \\
 \hline
 ${\bf I_b} \otimes {\bf I_b}$ & $\frac12\left( \Pi_{13} \!-\! e_{44} \!-\! e_{43} \!-\! e_{34} \!-\! e_{33} \right) $ \\
 \hline
 ${\bf II_a} \otimes {\bf II_a}$ & $\frac12\left( -\Pi_{24} \!-\! e_{11} \!-\! e_{14} \!-\! e_{41} \!-\! e_{44} \right) $ \\
 \hline
 ${\bf II_a} \otimes {\bf II_b}$ & $\frac12\left( -\Pi_{24} \!-\! e_{12} \!-\! e_{13} \!-\! e_{42} \!-\! e_{43} \right) $ \\
 \hline
 ${\bf II_b} \otimes {\bf II_a}$ & $\frac12\left( -\Pi_{24} \!-\! e_{21} \!-\! e_{24} \!-\! e_{31} \!-\! e_{34} \right) $ \\
 \hline
 ${\bf II_b} \otimes {\bf II_b}$ & $\frac12\left( -\Pi_{24} \!-\! e_{22} \!-\! e_{23} \!-\! e_{32} \!-\! e_{33} \right) $ \\
 \hline
 $\sum\limits_{i,j=a,b} \!\!\!\! \left( {\bf I_i} \otimes {\bf II_j} + {\bf II_i} \otimes {\bf I_j} \right)^{\dagger}  $ & $2 \left( \Pi_{14} + \Pi_{23} \right) $ \\
 \hline
\end{tabular}
     \captionof{table}{List of horizontal and vertical {\it sLag} cycles on the fully deformed $T^4/\Z_2$. Cycles marked with a $\dagger$ are {\it sLag} with calibration $\Im(\Omega_2)$.}
 \label{Tab:sLagsInFullDeform} 
\vspace{15pt}
   \end{minipage}
  \end{minipage}

\subsubsection[Integrals of $\Omega_2$ and numerical results]{Integrals of $\boldsymbol{\Omega_2}$ and numerical results}\label{Sss:SquareT4Z2Integrals}

In the following part, we will compute integrals of the holomorphic two-form $\Omega_2$ over the cycles that we just found on the deformed spaces. This is done as a function of the deformation parameters in order to show the dependence of these integrals on the deformation. The results will allow us to determine the cohomology classes of these cycles and in particular the exceptional parts of the fractional cycles.

Since we are integrating closed forms over closed cycles, the integrals are fully determined by the cohomology class of the two--form and the homology class of the cycle in question. More precisely, we expand the two--form class in a cohomology basis, such that the basis elements are Poincar\'e duals of integral cycles (indicated by square brackets in the equations below). For $K3$ as the deformed $T^4 / \Z_2$ orbifold, such a basis is provided by the (Poincar\'{e} duals to the) six bulk cycles and the 16 exceptional ones,
\begin{equation}
\label{Eq:Omega2Topo}
 [\Omega_2] = \sum_{i=1}^{b_2=22} \mathfrak{z}_i [\Pi_i] \quad \stackrel{\text{Def}(T^2/\Z_2)}{=} \sum_{i<j} \mathfrak{z}^\text{untw.}_{ij} \left[\Pi_{ij} \right] - \sum_{\alpha\beta} \mathfrak{z}^\text{tw.}_{\alpha\beta} \left[ e_{\alpha\beta} \right] \,.
\end{equation}
Integrals are then expressed in terms of the topological intersection numbers given in equations \eqref{Eq:BulkIntersection} and \eqref{Eq:ExceptionalCycleIntersection}. 
The complex structure is chosen such that $\mathfrak{z}^\text{untw.}_{12}=\mathfrak{z}^\text{untw.}_{34}=0$, which reflects the choice of K\"ahler form $J_{1,1}^{\text{K\"ahler}}\propto dz_1 \wedge d \overline{z}_1 + dz_2 \wedge d\overline{z}_2$. Furthermore, at the orbifold point ($\mathfrak{z}^\text{tw.}_{\alpha\beta} \equiv 0$), we choose $\mathfrak{z}^\text{untw.}_{13}=-\mathfrak{z}^\text{untw.}_{24}=-i\mathfrak{z}^\text{untw.}_{14} = -i\mathfrak{z}^\text{untw.}_{23}$ corresponding to a square torus with $\Omega_2 = dz_1 \wedge dz_2 $.

At the orbifold point, integrals of $\Omega_2$ can be traced back to the product of two integrals of the holomorphic one-form $\Omega_1$ defined in~\eqref{Eq:HolomorphicOneForm} over one-cycles on the elliptic curve. These lead to elliptic integrals\footnote{The elliptic integral is defined as $K(k) = \int_0^{\pi/2} \left( 1 - k^2 \sin^2(\phi) \right)^{-1/2} \  d\phi$.} $K(k)$ which cannot be solved analytically. For example, we find for a horizontal cycle ${\bf I_a} $  on the square torus 
\begin{equation}
\label{Eq:IntOmega1}
 \int\limits_{\bf I_a} \Omega_1 = 2 \int\limits_1^\infty \frac{dx}{\sqrt{x^3-x}} =  2 \sqrt{2} K\left( 1/\sqrt{2} \right)  \,.
\end{equation}
We see that the integrals over all horizontal or vertical cycles have the same value, up to signs and factors of $i$, namely
\begin{equation}
 \label{Eq:IntOmega2Orbifold}
 \int\limits_{{\bf N_1}\otimes{\bf N_2}}  \Omega_2 = i^n \cdot 4 K(1/\sqrt2)^2 \approx i^n \cdot 13.75\ldots \,, \qquad \text{for } {\bf\N_i} = {\bf I_{a/b}} , {\bf II_{a/b}} \,.
\end{equation}

All integrals over {\it sLag} cycles on the deformed spaces are here evaluated numerically: We start with the case of just one deformation as in figure \ref{Fig:LagsInX1X2PlaneDeformed1}. Due to the symmetries  $x_1 \leftrightarrow x_2$ and $(x_1,x_2) \mapsto (-x_1,-x_2)$, we do not have to consider each cycle on its own but just a representative sample. The most interesting cycles are those which are directly related to the deformation, i.e.\ the exceptional cycles, and those fractional cycles which intersect the singularity, since they have contributions from the deformation modulus $\mathfrak{z}^\text{tw.}_{33}$. However, we also look at other cycles away from the deformed singularity in order to see if the parameter $\varepsilon_{33}$ plays a role there. In the following we will discuss the integrals on the various two-cycles.

 \begin{minipage}[ht]{\textwidth}
  \begin{minipage}[t]{0.49\textwidth}
    \centering
\includegraphics[scale=0.6]{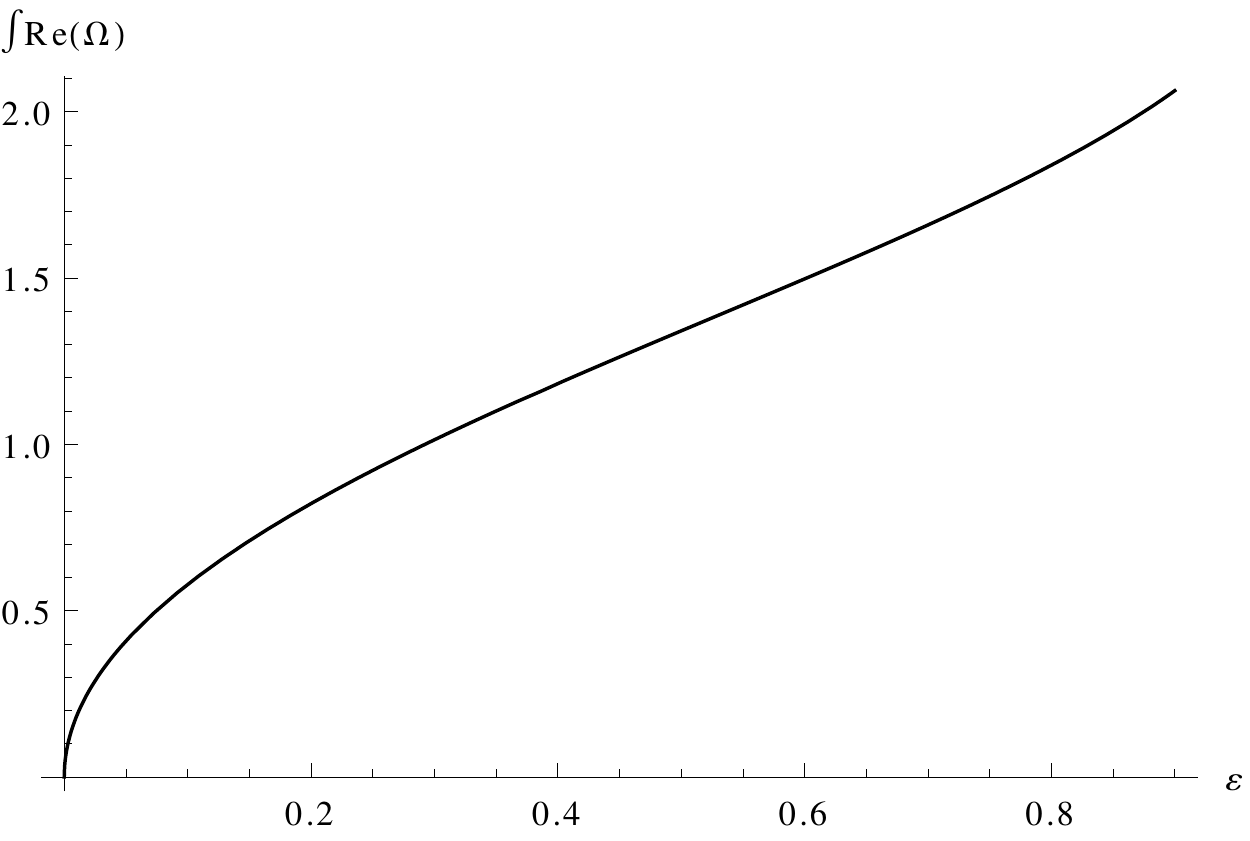} 
    \captionof{figure}{Integral of $\Re(\Omega_2)$ over the exceptional cycle $e_{33}$ for a single deformation, depending on the deformation parameter $\varepsilon_{33}$.  
    $\Omega_2$ is normalised to the value at the orbifold point.}
\label{Fig:IntegralOmega3}    
  \end{minipage}
  \hfill
  \begin{minipage}[t]{0.49\textwidth}
    \centering
\includegraphics[scale=0.6]{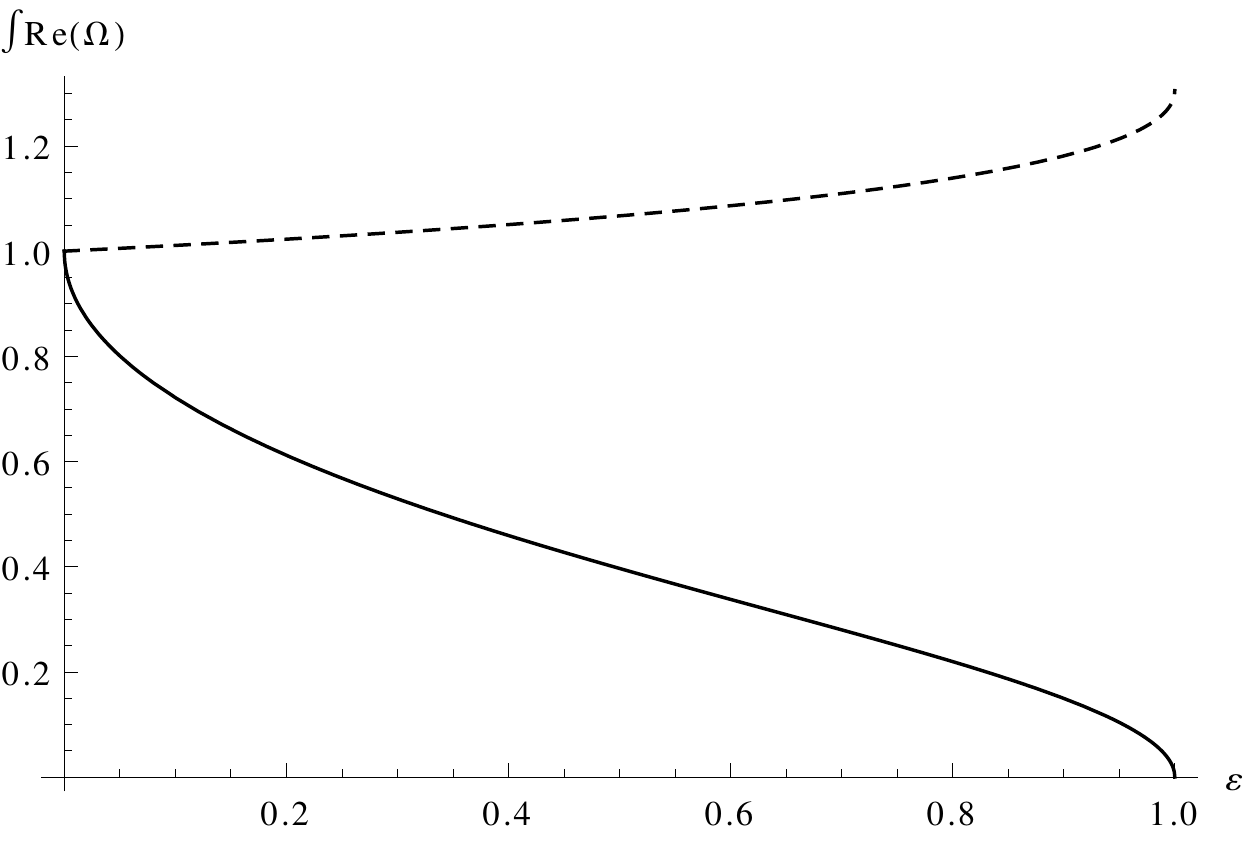} 
    \captionof{figure}{Integrals of $\Re(\Omega_2)$ over fractional {\it sLag} cycles for a single deformation, depending on the deformation parameter $\varepsilon_{33}$. The solid curve shows the result for cycles which intersect the deformed singularity (e.g.\ ${\bf I_b} \otimes {\bf I_b}$) while the dashed curve is for the other ones. 
    }
\label{Fig:IntegralOmega1}     
  \end{minipage}
  \hfill
 \end{minipage}
 %
\begin{itemize}
 \item Figure \ref{Fig:IntegralOmega3} shows the integral of $\Re(\Omega_2)$ over the exceptional cycle $e_{33}$. From the topological expansion \eqref{Eq:Omega2Topo} we expect $\int\Re(\Omega_2) = 2 \, \mathfrak{z}^\text{tw.}_{33}$. 
Computing the integral directly using spherical coordinates on $e_{33}$ leads to $\mathfrak{z}^\text{tw.}_{33} =  2\pi \sqrt{\varepsilon_{33}} + \mathcal{O}(\varepsilon_{33}^{3/2})$.
\item Next we consider the integral over the cycle ${\bf I_b} \otimes {\bf I_b}$ (or equivalently ${\bf II_b} \otimes {\bf II_b}$) which is visualised by the solid curve in figure \ref{Fig:IntegralOmega1}. For positive values of the deformation parameter, we see a non-analytic behaviour at the origin and an analytic computation shows indeed that 
\begin{equation}
 \int\limits_{{\bf I_b} \otimes {\bf I_b}} \Re(\Omega_2) = 4 K(1/\sqrt2)^2 - 2 \pi \sqrt{\varepsilon_{33}} + \mathcal{O}(\varepsilon_{33}) \,,
\end{equation}
which confirms that the cycle has a contribution $- 1/2 \cdot e_{33}$, i.e.\ it is a fractional cycle with negative $\Z_2$ eigenvalue.
\item The other cycles of calibration $\Re(\Omega_2)$, i.e.\ all other grey cycles in figure \ref{Fig:LagsInX1X2PlaneDeformed1}, turn out to give the same value for the integral, see the dashed curve in figure \ref{Fig:IntegralOmega1}, although these cycles are not always related by symmetries in any obvious way. This follows from the fact that these cycles have the same bulk part, namely $\Pi_{13}/2$ or $-\Pi_{24}/2$, on the underlying square torus. However, the integrals are not constant but grow linearly with $\varepsilon_{33}$, and we find
\begin{equation}
\mathfrak{z}^\text{untw.}_{13} = -\mathfrak{z}^\text{untw.}_{24} = 4 K(1/\sqrt{2})^2 + 2\pi \frac{\Gamma(3/4)^2}{\Gamma(1/4)^2}\varepsilon_{33} + \mathcal{O}(\varepsilon_{33}^2)\,.
\end{equation}
\item Finally, we consider the integral of $\Im(\Omega_2)$ over the cycles which are now as well calibrated by $\Im(\Omega_2)$. The results are shown in figure \ref{Fig:IntegralOmega2}. The integral on the merged cycle ${\bf II_b} \otimes {\bf I_b} + {\bf I_b} \otimes {\bf II_b}$ (solid curve) turns out to be exactly twice the integral over each of the other cycles (dashed curve), which again are all equal. This result confirms that the merged cycle has no exceptional contribution and thus is the union of two fractional cycles with opposite $\Z_2$ eigenvalue. Furthermore, we again observe a linear change, 
\begin{equation}
\mathfrak{z}^\text{untw.}_{14} = \mathfrak{z}^\text{untw.}_{23} = 4 K(1/\sqrt{2})^2 - 2\pi \frac{\Gamma(3/4)^2}{\Gamma(1/4)^2}\varepsilon_{33} + \mathcal{O}(\varepsilon_{33}^2)\,.
\end{equation}
\end{itemize}
 \begin{minipage}[ht]{\textwidth}
  \begin{minipage}[t]{0.49\textwidth}
    \centering
\includegraphics[scale=0.6]{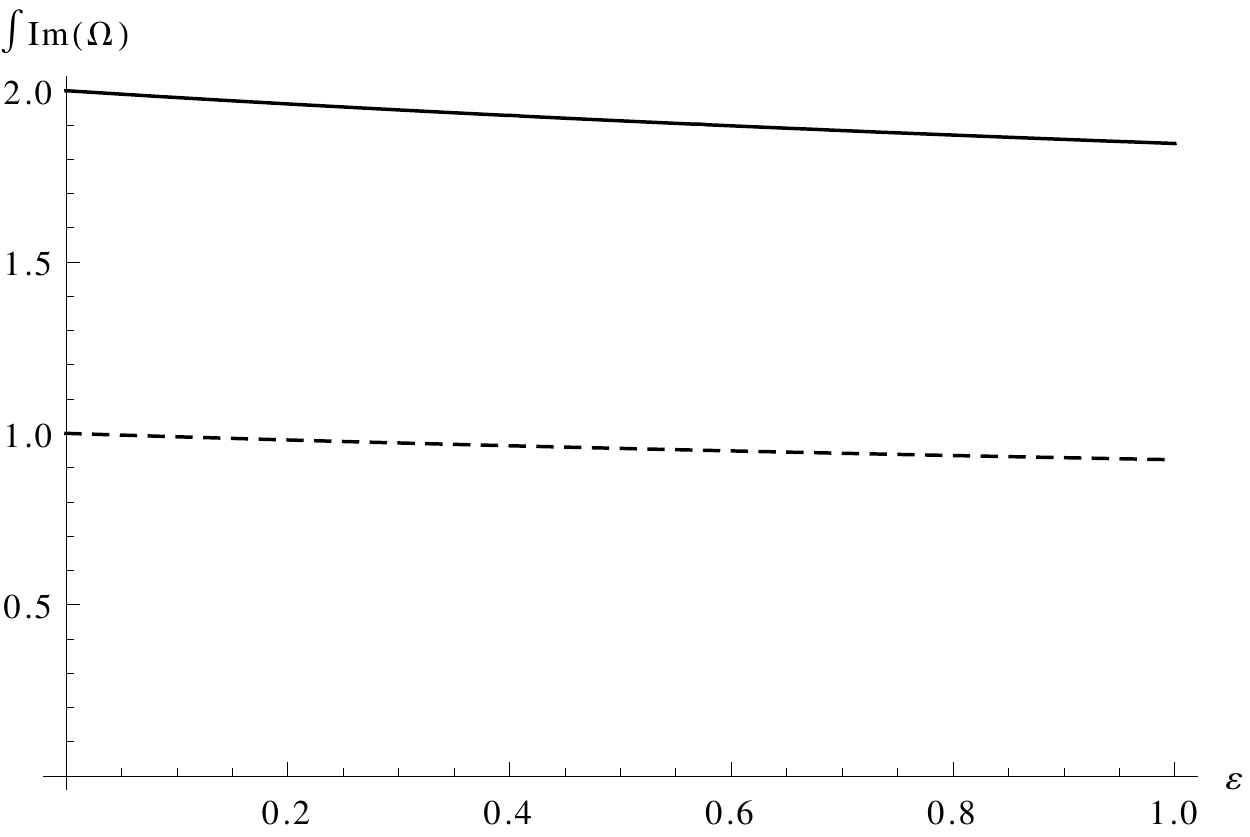} 
    \captionof{figure}{Integrals of $\Im(\Omega_2)$ over {\it sLag} cycles (now calibrated with $\Im(\Omega_2)$) for a single deformation, depending on the deformation parameter $\varepsilon_{33}$. The solid curve shows the result for the cycle ${\bf I_b} \otimes {\bf II_b} + {\bf II_b} \otimes {\bf I_b}$, and the dashed curve shows the result for any other one as discussed in the main text. 
    }
\label{Fig:IntegralOmega2}     
  \end{minipage}
  \hfill
  \begin{minipage}[t]{0.49\textwidth}
    \centering
\includegraphics[scale=0.6]{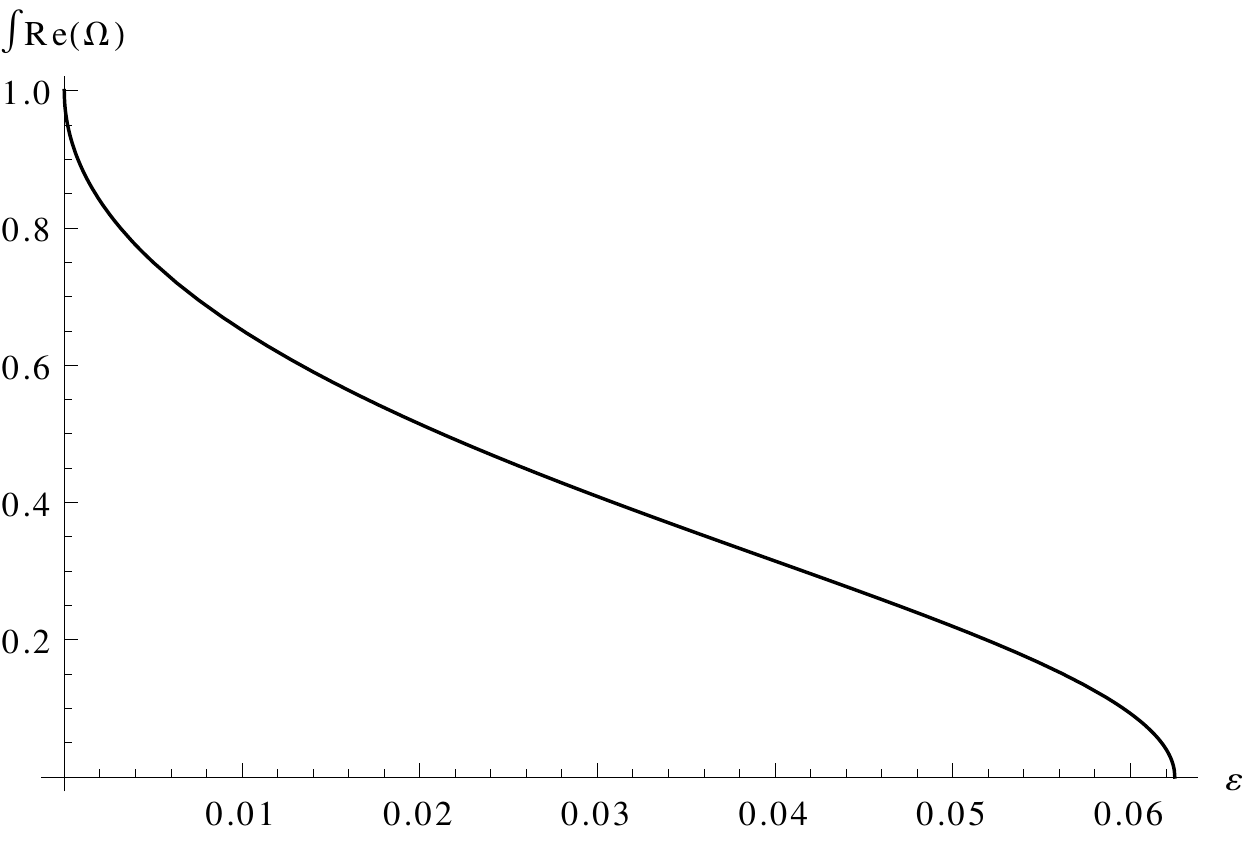} 
    \captionof{figure}{Integrals of $\Re(\Omega_2)$ over fractional {\it sLag} cycles for a full deformation of all $\Z_2$ fixed points, depending on the universal deformation parameter $\varepsilon$.     }
\label{Fig:IntegralOmega4}     
  \end{minipage}
 \end{minipage}

To sum up, we have seen that the modulus $\mathfrak{z}^\text{tw.}_{33}$ corresponds in lowest order to the square root of the deformation parameter $\varepsilon_{33}$, whereas the untwisted moduli $\mathfrak{z}^\text{untw.}_{ij}$ are also influenced by the deformation but in higher order, $\mathfrak{z}^\text{untw.}_{ij} = \text{const.} + \mathcal{O}(\varepsilon_{33})$. Another result is that at $\varepsilon_{33}=1$, we obtain a new singularity and thus leave the geometric phase of a deformed orbifold singularity. This is indicated by the non-analytic behaviour of the integrals in figure \ref{Fig:IntegralOmega1} and the vanishing of the fractional cycle ${\bf I_b} \otimes {\bf I_b}$ (or ${\bf II_b} \otimes {\bf II_b}$)
at $\varepsilon_{33}=1$.

Finally, we discuss an example of a multiple deformation. We choose the deformation as shown in figure \ref{Fig:LagsInX1X2PlaneDeformed2} with all singularities equally deformed, i.e.\ $\varepsilon_{\alpha\beta} \equiv \varepsilon$ for all $\alpha$ and $\beta$. Due to the shift symmetry on the underlying torus, all fractional cycles with calibration $\Re(\Omega_2)$, i.e.\ the fractional cycles in table \ref{Tab:sLagsInFullDeform}, lead to the same integral, which is plotted in figure \ref{Fig:IntegralOmega4}. 
Qualitatively it shows the same behaviour as the integral for just one deformation in figure \ref{Fig:IntegralOmega1}. In the series expansion around $\varepsilon=0$, the factor of the $\sqrt\varepsilon$ term is four times larger compared to the single deformation. This proves that the cycle under consideration is a fractional cycle whose four exceptional parts come with a minus sign as stated in table \ref{Tab:sLagsInFullDeform}. Furthermore, the point where the integral becomes zero is exactly at $\varepsilon=1/16$. This is not fully surprising since the true deformation modulus $\mathfrak{z}$ corresponds, at least to lowest order, to the square root of $\varepsilon$ divided by the before mentioned factor of four. The surprising point is rather that this factor four also holds for finite values of $\varepsilon$.


\subsection[sLags on the deformed $T^6 / \Z_2 \times \Z_2 $ on square tori]{sLags on the deformed $\boldsymbol{ T^6 / \Z_2 \times \Z_2}$ on square tori}\label{Ss:SquareT6Z22}

Since the main phenomenological interest lies in compactifications to four dimensions, we also discuss {\it sLag} cycles on the orbifold $T^6 / \Z_2 \times \Z_2 $ with discrete torsion and its deformations, in extension of the results of section \ref{Ss:SquareT4Z2}. The orbifold perspective for this case was presented in section \ref{Ss:T6Z2Z2N}. We briefly analyze the structure of {\it sLag} cycles in this case and show the results for various types of deformations. For simplicity we choose the complex structure of the two-tori to be $\rho = 1$ in the notation of section~\ref{S:sLags}, i.e.\ the tori are of square shape.

The topological expansion of the holomorphic three-form reads
\begin{equation}
 \label{Eq:Omega3Topo}
 \left[ \Omega_3 \right] = \sum_{i=1}^2 \sum_{j=3}^4 \sum_{k=5}^6 \mathfrak{c}_{ijk} \left[\Pi_{ijk} \right] - \sum_{i=1}^{3} \sum_{\alpha,\beta=1}^4 \left(  \mathfrak{c}^i_{\alpha\beta} \left[ \varepsilon^{(i)}_{\alpha\beta} \right] + \mathfrak{\tilde c}^i_{\alpha\beta} \left[ \tilde\varepsilon^{(i)}_{\alpha\beta} \right] \right) \,.
\end{equation}
However, only $h_{21}=51$, see equation \eqref{Eq:T6Z22HodgeNumbers}, of the $104$ parameters $\mathfrak{c}$ are independent complex structure moduli: two of them can be absorbed by complex rescaling, and one half of the remaining parameters $\mathfrak{c}$ are determined by the prepotential on the complex structure moduli space, see e.g.\ \cite{Becker:2007zj}. The orientifolding leads to $51$ real complex structure parameters, cf.\ table \ref{Tab:closed-OR-spectrum}.

\subsubsection{Cycle structure}\label{Sss:SquareT6Z22Cycles}

We start at the orbifold point, where we find bulk cycles and fractional cycles that descend from the underlying torus cycles in table \ref{tab:T2LagrangianLinesUntilted}. The most basic cycles to work with are again horizontal and vertical ones, and we focus here on cycles which pass through the fixed planes, i.e.\ cycles of the form ${\bf N_1} \otimes {\bf N_2} \otimes {\bf N_3}$ with ${\bf N_i} = {\bf I_{a/b}},  {\bf II_{a/b}}$. Such a cycle is {\it sLag} with calibration $\Re(\Omega_3)$ if the number of horizontal one-cycles in the factorisation (${\bf N_i} = {\bf I_{a/b}}$) is odd, whereas the cycle is {\it sLag} with calibration $\Im(\Omega_3)$ if the number of horizontal cycles is even. These cycles lie in the real $x_1$-$x_2$-$x_3$ three-plane, and their boundaries are given by the zero locus of $y$ in equation \eqref{Eq:T6Z22Hypersurface}. This zero locus factorises into $3 \times 4$ two-planes at $x_i=-1,0,1,\infty$, $i=1,2,3$, in analogy to figure \ref{Fig:LagsInX1X2Plane}. In order to obtain exceptional cycles with finite volume, one has to switch on the corresponding complex structure deformations, which we will discuss step by step.

The simplest class of deformations are those which deform only fixed planes from one twisted sector, e.g.\ $\varepsilon^{1}_{\alpha\beta} = \varepsilon^2_{\alpha\beta} \equiv 0$.\footnote{Observe that $\varepsilon^{i}_{\alpha\beta}$ denotes the deformation parameter of the $\Z_2^{(i)}$ singularity through which the exceptional three-cycles $\varepsilon^{(i)}_{\alpha\beta}$, $\tilde{\varepsilon}^{(i)}_{\alpha\beta}$ pass.} In this case every, three-cycle can be written as a product of a two-cycle on $\text{Def}(T^4_{(3)}/\Z_2)$, see section \ref{Sss:SquareT4Z2Cycles}, and a one-cycle on $T^2_{(3)}$, see table \ref{tab:T2LagrangianLinesUntilted}, plus its image\footnote{In the hypersurface formalism, the orbifold image is automatically included.} under the remaining $\Z_2$. Accordingly, the holomorphic three-form factorises, $\Omega_3 = \Omega_2 \wedge \Omega_1$, as well as its integral over any three-cycle. 
This simple situation is suitable for describing the example of $\text{Vol}(\Pi^{\text{rigid}}_a + \Pi^{\text{rigid}}_{a'})$ for a rigid three-cycle perpendicular to the exotic O6-plane as in equation~\eqref{Eq:Z2Z2-frac-Pis-one-Z2-only}.

The discussion of a generic deformation turns out to be difficult because the functions $\varepsilon_{\alpha\beta\gamma}\left( \varepsilon^i_{\alpha\beta} \right)$, which are required to ensure that the full space contains $64$ conifold singularities, cf.\ section \ref{Ss:GlobalDeform}, are hard to determine. However, we discuss two cases in which this is possible:
\begin{itemize}
 \item When we deform three fixed planes from three different sectors, which all intersect in one codimension three singularity, the hypersurface equation can be traced back to the equation for the local deformation \eqref{Eq:C3Z22Hypersurface}. 
 \item When all $3 \times 16$ fixed planes are deformed equally, we can use the symmetries that are preserved by that deformation, to find an expression for $\varepsilon_{\alpha\beta\gamma}$.
\end{itemize}

\subsubsection{Deformation of three fixed planes}\label{Sss:T6Z22-3Deform}

Here we choose the deformation of the three fixed planes that intersect at the fixed point $(3,3,3)$ at $z_{i} =(1+i)/2$, with the same magnitude. For this we set $\varepsilon_{33}^1= \varepsilon_{33}^2 = \varepsilon_{33}^3 =: \varepsilon$ and find $\varepsilon_{333} = 2 \varepsilon^{3/2}$, while the other parameters are zero. The hypersurface equation then simplifies to
\begin{equation}
 \label{Eq:T7Z22-3Deform}
  y^2 = \left( x_1^2-1 \right) \cdot \left( x_2^2-1 \right) \cdot \left( x_3^2-1 \right) \cdot \left( x_1 x_2 x_3 - \varepsilon \left( x_1 + x_2 + x_3 \right) + 2 \varepsilon^{3/2} \right) \,.
\end{equation}

The fractional, horizontal or vertical, cycles then fall into three categories:
\begin{itemize}
 \item There are cycles which do not intersect any of the deformed singularities. In the topological expansion of the integral of $\Omega_3$, they do not receive contributions from the twisted deformation modulus that we switch on, and thus we expect their volume to be constant to leading order with a subleading linear dependence on $\varepsilon$ in analogy to figure~\ref{Fig:IntegralOmega2}. The fractional three-cycle $\Pi^{\text{rigid}}_a \simeq {\bf I_a} \otimes {\bf II_a} \otimes {\bf II_a}$ in equation~\eqref{Eq:Z2Z2-rigid-example} is exactly of this type.
 \item For cycles, which intersect one of the singularities the discussion is similar to the $T^4/\Z_2$ case in section \ref{Ss:SquareT4Z2}. For example, the cycle ${\bf I_{a}}\otimes{\bf I_{b}}\otimes{\bf I_{b}}$ has the same calibration as $\varepsilon^{(1)}_{33}$ and turns out to be a fractional cycle of the form $\frac{1}{4}\left( \Pi_{135} - \varepsilon^{(1)}_{33} \pm \ldots \right)$.

  The cycles ${\bf I_{a}}\otimes{\bf I_{b}}\otimes{\bf II_{b}}$ and ${\bf I_{a}}\otimes{\bf II_{b}}\otimes{\bf I_{b}}$ on the other hand are differently calibrated from a possible exceptional part and thus merge to a cycle of the form $\frac{1}{4}\left( \Pi_{136} + \Pi_{145} \pm \ldots \right)$.
 \item The cycles, which intersect all three singularities, are the most interesting ones. They are shown in table \ref{Tab:CyclesThreeDeform}. The cycles which are products of only horizontal (group 1) or vertical (group 4) one-cycles have the same calibration as each of their exceptional components. Thus they constitute a fractional {\it sLag} cycle. The mixed cycles (groups 2 and 3) on the other hand contain exceptional contributions which are differently calibrated than the bulk part and thus cannot be fractional {\it sLag} cycles on their own. Instead, the three cycles in each group merge to one larger cycle such that the exceptional parts of wrong calibration cancel out. The resulting cycle is therefore {\it sLag} with calibration $\Im(\Omega_3)$ for group 2 and calibration $\Re(\Omega_3)$ for group 3. Accordingly, the zero locus of $y$ splits into three branches, see figure \ref{Fig:T6Z22DeformationVisualisation}. Note that the two branches that enframe the cycle from group 2 intersect at the conifold point $x_1=x_2=x_3=\sqrt{\varepsilon}$.

The fractional cycle in equation~\eqref{Eq:Z2Z2-rigid-example} constitutes an explicit example of group 3 if one uses the shift symmetry on the underlying $(T^2)^3$ to replace the $\Z_2$ fixed point locus (3,3,3) 
and corresponding deformation parameters by those at point (1,1,1).
\end{itemize}
 \begin{minipage}[ht]{\textwidth}
 \begin{minipage}[t]{0.59\textwidth}
 \vspace{-190pt}
\begin{tabular}{|c|c|c|}
\hline
\!\!Group\!\! & ${\bf N_1} \otimes {\bf N_2} \otimes {\bf N_3}$ & $4 \Pi^\text{frac} \supset$\\
\hline \hline
1 & ${\bf I_{b}}\otimes{\bf I_{b}}\otimes{\bf I_{b}}$ & $\Pi_{135} \pm  \varepsilon^{(1)}_{33}  \pm  \varepsilon^{(2)}_{33} \pm \varepsilon^{(3)}_{33}  $ \\
 \hline \hline
\multirow{3}{*}{2} & ${\bf II_{b}}\otimes{\bf I_{b}}\otimes{\bf I_{b}}$ & $ \Pi_{235}  \pm  \tilde\varepsilon^{(1)}_{33} \pm \varepsilon^{(2)\dagger}_{33} \pm \varepsilon^{(3)\dagger}_{33}  $ \\
 \cline{2-3}
& ${\bf I_{b}}\otimes{\bf II_{b}}\otimes{\bf I_{b}}$ & $ \Pi_{145} \pm \varepsilon^{(1)\dagger}_{33}  \pm  \tilde\varepsilon^{(2)}_{33} \pm \varepsilon^{(3)\dagger}_{33}  $ \\
 \cline{2-3}
& ${\bf I_{b}}\otimes{\bf I_{b}}\otimes{\bf II_{b}}$ & $\Pi_{136} \pm \varepsilon^{(1)\dagger}_{33} \pm \varepsilon^{(2)\dagger}_{33}  \pm  \tilde\varepsilon^{(3)}_{33}  $ \\
 \hline \hline
\multirow{3}{*}{3} & ${\bf II_{b}}\otimes{\bf II_{b}}\otimes{\bf I_{b}}$ & $-\Pi_{245} \pm \tilde\varepsilon^{(1)\dagger}_{33} \pm \tilde\varepsilon^{(2)\dagger}_{33}  \pm  \varepsilon^{(3)}_{33}  $ \\
 \cline{2-3}
& ${\bf I_{b}}\otimes{\bf II_{b}}\otimes{\bf II_{b}}$ & $ -\Pi_{146}  \pm  \varepsilon^{(1)}_{33} \pm \tilde\varepsilon^{(2)\dagger}_{33} \pm \tilde\varepsilon^{(3)\dagger}_{33}  $ \\
 \cline{2-3}
& ${\bf II_{b}}\otimes{\bf I_{b}}\otimes{\bf II_{b}}$ & $ -\Pi_{236} \pm \tilde\varepsilon^{(1)\dagger}_{33}  \pm  \varepsilon^{(2)}_{33} \pm \tilde\varepsilon^{(3)\dagger}_{33}  $ \\
 \hline \hline
4 & ${\bf II_{b}}\otimes{\bf II_{b}}\otimes{\bf II_{b}}$ & $ -\Pi_{246}  \pm  \tilde\varepsilon^{(1)}_{33}  \pm  \tilde\varepsilon^{(2)}_{33}  \pm  \tilde\varepsilon^{(3)}_{33}$ \\
  \hline
\end{tabular}
     \captionof{table}{{\it sLag} cycles for three deformed singularities. Cycles in groups 1 and 3 are calibrated w.r.t.\ $\Re(\Omega_3)$ and cycles in groups 2 and 4 are calibrated w.r.t.\ $\Im(\Omega_3)$. A $\dagger$ on an exceptional cycle indicates that it has a different calibration from the bulk part. Only exceptional cycles at deformed singularities are explicitly given.}
\label{Tab:CyclesThreeDeform}
   \end{minipage}
   \hfill
  \begin{minipage}[t]{0.39\textwidth}
 \includegraphics[scale=0.3]{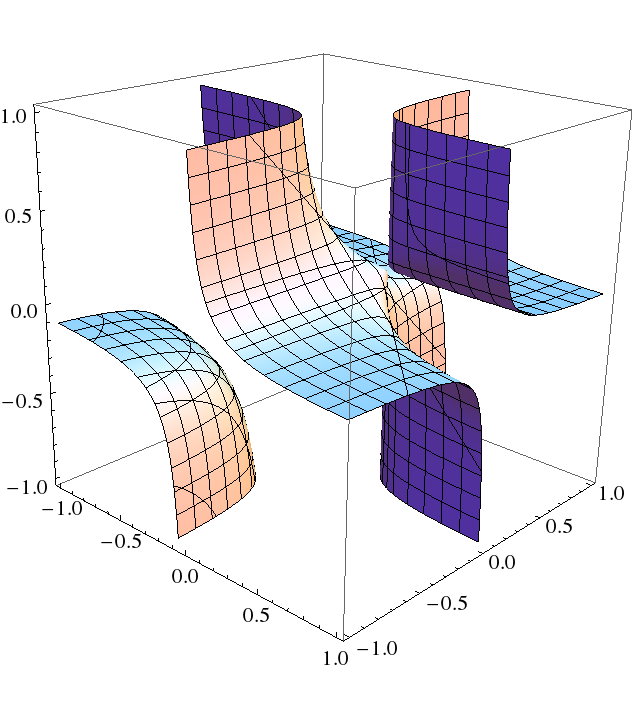}
    \captionof{figure}{Zero locus of $y$ for three deformations in the region $-1<x_i<1$. Due to the permutation symmetry of $(x_1,x_2,x_3)$ the axes are not labelled. The three branches separate the four {\it sLag} cycles which correspond to the groups in table \ref{Tab:CyclesThreeDeform}.}
 \label{Fig:T6Z22DeformationVisualisation}
  \end{minipage}
  \end{minipage}

\subsubsection{Deformation of all fixed planes}\label{Sss:T6Z22-AllDeform}

Finally we discuss the deformation of all 48 fixed planes. To do so, we set all deformation parameters equal, i.e.\ $\varepsilon^i_{\alpha\beta} \equiv \varepsilon$ for all $i=1,2,3$, $\alpha,\beta=1,\ldots,4$. The hypersurface equation~(\ref{Eq:DeformT6Z22}) then becomes
\begin{equation}
 \label{Eq:T6Z22AllDeform}
 y^2 = \prod_{i=1}^3 x_i(x_i^2-1) 
 - \ \ \ 
 \varepsilon 
 \!\!\!\!\!\!\sum_{(i,j,k) = \sigma(1,2,3) \atop \sigma \text{ cyclic}} \!\!\!\!  x_i(x_i^2-1) (x_j^2+1)^2 (x_k^2+1)^2 + 2 \, \varepsilon^{3/2}\prod_{i=1}^3 (x_i^2+1)^2  \,.
\end{equation}
We indeed find $64$ conifold singularities at $x_i=\hat x$, $i=1,2,3$ where $\hat x$ is one of the solutions of $\hat{x}^3-\hat{x}=\varepsilon (\hat{x}^2+1)^2$. For small deformations, we observe the following horizontal and vertical {\it sLag} cycles:
\begin{itemize}
 \item All purely vertical cycles (${\bf II_{a/b}}^{\otimes 3}$) start to shrink, which indicates that they are fractional cycles whose exceptional part has negative sign, i.e.\ $\frac{1}{4}\left( \Pi_{246} - \sum_{i,\alpha\beta} \tilde{\varepsilon}^{(i)}_{\alpha\beta} \right)$. Furthermore, the contributing exceptional cycles  must be {\it sLag} with same calibration as the purely vertical cycles, namely $\Im(\Omega_3)$. At $\varepsilon=1/64$ the purely vertical cycles disappear.
 \item Cycles which are purely horizontal (${\bf I_{a/b}}^{\otimes 3}$) also shrink, but not as fast as the vertical ones. This shows that they are fractional cycles also with negative exceptional contribution, schematically $\frac{1}{3}\left( \Pi_{135} - \sum_{i,\alpha\beta} {\varepsilon}^{(i)}_{\alpha\beta}  \right)$. In fact, they disappear at $\varepsilon=1/16$. The discrepancy to the point where the purely vertical cycles disappear could be a signal that the exceptional cycles $\tilde\varepsilon^{(i)}_{\alpha\beta} $ have bigger volume than the exceptional cycles $\varepsilon^{(i)}_{\alpha\beta} $.
 \item The cycles with one vertical and two horizontal factors (${\bf II_{a/b}} \otimes {\bf I_{a/b}}^{\otimes 2}$ and permutations) have merged to one connected {\it sLag} cycle with calibration ${\Im(\Omega_3)}$. One finds that this large cycle also shrinks and disappears at $\varepsilon=1/16$ which can only be explained by negative exceptional contributions from $\tilde\varepsilon^{(i)}_{\alpha\beta} $. The contributions from the cycles $\varepsilon^{(i)}_{\alpha\beta} $ cancel out.
\item All cycles with one horizontal and two vertical factors (${\bf I_{a/b}} \otimes {\bf II_{a/b}}^{\otimes 2}$ and permutations) such as the one in equation~(\ref{Eq:Z2Z2-rigid-example})
have also merged in order to form one connected {\it sLag} cycle with calibration ${\Re(\Omega_3)}$. However, it is not clear what sign their exceptional part has since at $\varepsilon \ge 1/16$ they fill the whole real $x_i$ three-plane.
\end{itemize}
In addition, one observes that all but the purely vertical cycles intersect the conifold singularities.

\subsection{Towards more general D-brane configurations}\label{Ss:MoreGeneral}

\paragraph{Tilted tori}
The whole discussion of sections \ref{Ss:SquareT4Z2} and \ref{Ss:SquareT6Z22} was performed for orbifolds with underlying square tori. One advantage was that one could describe many horizontal and vertical {\it sLag} cycles in the real $x_i$ plane. However, many interesting orbifolds are based on tilted tori. The most simple example of $T^4/\Z_2$ with tilted tori in section~\ref{Ss:T4Z2N} can be viewed as built on square tori rotated by $e^{\pm i \pi/4}$, with the consequence that the {\it sLags} can be constructed from products of the cycles {\bf VII} and {\bf VIII} (plus some exceptional cycles)
in table~\ref{tab:T2LagrangianLinesUntilted}. These consideration directly carry over to the $T^6/\Z_4$ orientifolds of~\cite{Blumenhagen:2002gw}.

Tilted tori are also of particular importance, if the orbifold group contains a factor of $\Z_3$, which also includes e.g.\ the $\Z_6^{(\prime)}$~\cite{Honecker:2004kb,Honecker:2004np,Bailin:2006zf,Gmeiner:2007we,Gmeiner:2007zz,Bailin:2007va,Gmeiner:2008xq,Bailin:2008xx,Bailin:2011am}  and $\Z_6^{(\prime)} \times \Z_2$~\cite{Forste:2010gw,Honecker:2012qr,Honecker:2013kda,EckerHoneckerStaessens:2014} orientifolds of phenomenological interest. In case of a factorisable torus, each lattice must be the $A_2$ Lie lattice, which is a special case of a tilted torus. A computational challenge in discussing deformations lies the fact that two of the $\Z_2$ fixed points are in the complex $x$ plane, and the {\it sLag} lines, which intersect these fixed points, are described by circle equations, see table~\ref{tab:T2LagrangianLinesTilted}. Accordingly, the equations for these {\it sLags} become even more involved when switching on deformations. 
 
\paragraph{Enhancing the orbifold group}

The description of toroidal orbifolds as a complete intersection of hypersurfaces, which allows for a resolution of the singularities, is known for all factorisable orbifolds~\cite{Blaszczyk:2011hs}. A description that allows for deformations is only known for $\Z_2$ and $\Z_3$ singularities in four compact dimensions and for $\Z_2 \times \Z_2$ and $\Z_3 \times \Z_3$ singularities in six compact dimensions. However, many interesting D-brane models are based on e.g.\ the $\Z_6\times \Z_2$~\cite{Forste:2010gw,EckerHoneckerStaessens:2014} or the $\Z_6^{\prime} \times \Z_2$~\cite{Forste:2010gw,Honecker:2012qr,Honecker:2013kda} orientifold. 
These orbifolds with D6-branes  of phenomenological interest are chosen with discrete torsion, and some of their singularities can be deformed and give rise to three-cycles as discussed in appendices~\ref{Sss:Z2Z6} and~\ref{Sss:Z2Z6p}. 
One possibility to treat deformations on these orbifolds is to take the $\Z_2 \times \Z_2$ subgroup discussed in the present article and mod out another $\Z_3$ by hand. On the elliptic curve in equation (\ref{Eq:DefEllipticCurve}), such an $\Z_3$ acts by $x \mapsto e^{2\pi i /3} x$.  This will put some restrictions on the coefficients in the hypersurface equation, namely the elliptic curves must be shaped like the $A_2$ lattice ($g_2 = 0$), and $\Z_2$ fixed points which are mapped onto each other under $\Z_3$ must be equally deformed. Then the singularities in all $\Z_2$ sectors can be deformed, but the fixed points of the higher order orbifold group elements stay singular. For the $T^6/\Z_6' \times \Z_2$ orbifold with discrete torsion, this ansatz could potentially be expanded by adding blow-ups of codimension two singularities in the $\Z_3$ and $\Z_6$ twisted sectors, cf. the discussion in appendix~\ref{Sss:Z2Z6p}.
For $T^6/\Z_6 \times \Z_2$ with discrete torsion, however, no such clear separation of singularities exists, cf. appendix~\ref{Sss:Z2Z6}.

\paragraph{Cycles with higher winding numbers $\boldsymbol{(n,m)}$}

In this work we restricted to {\it sLag} cycles that were realised as fixed planes on antilinear involutions in the homogeneous coordinates. On the square torus, this gave us cycles with winding numbers $(n,m)$ not exceeding $
\pm 1$. Topologically, more general cycles can be constructed as sums of the horizontal and vertical ones, but this does not give us a {\it sLag} representative cycle. On the elliptic curve, one can draw these cycles in the complex $x$-plane and see their winding numbers in the winding around the points where $y=0$. To compute integrals of the holomorphic volume form, one can choose a cycle with the same winding properties, which is easier to parametrise. Due to the closedness of the holomorphic volume form, the result is independent  of the concrete cycle, but in general such a cycle will not have the {\it sLag} property. In particular, it is not clear if such a {\it sLag} representative also exists when deformations are switched on. One such cycle is shown in figure \ref{Fig:41cycle1}.

 \begin{minipage}[ht]{\textwidth}
  \begin{minipage}[t]{0.49\textwidth}
    \centering
\includegraphics[scale=0.7]{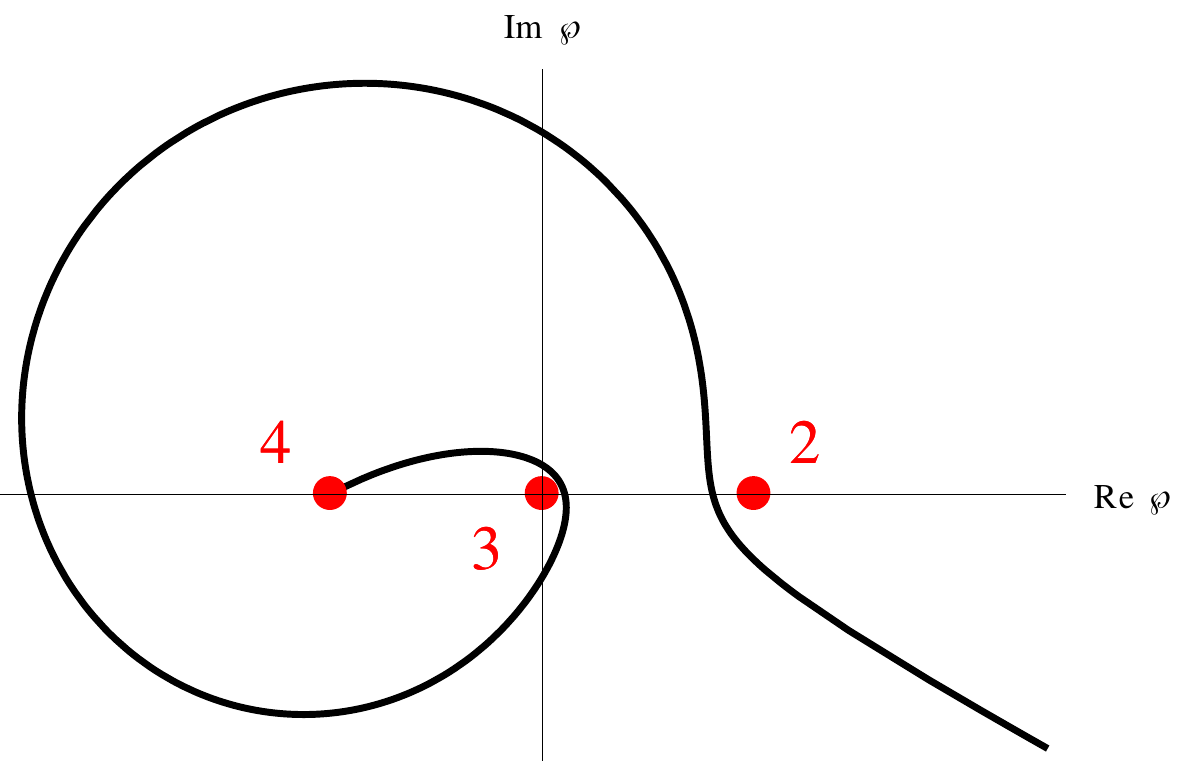} 
  \end{minipage}
  \hfill
  \begin{minipage}[t]{0.49\textwidth}
    \centering
\includegraphics[scale=0.7]{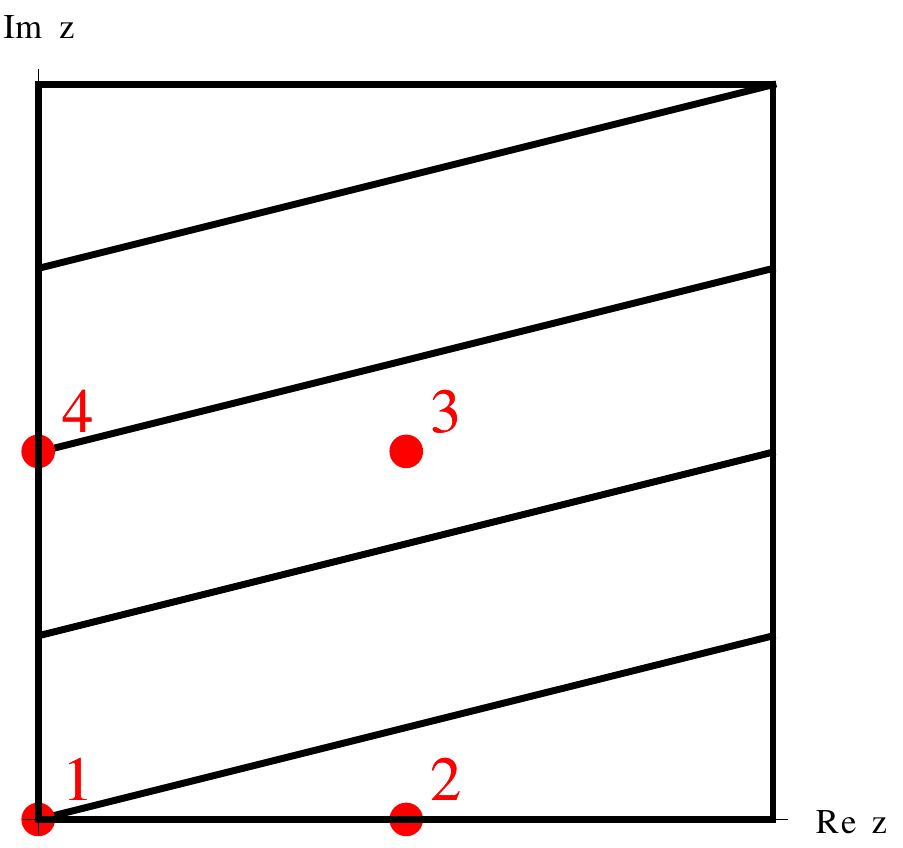} 
  \end{minipage}  
    \captionof{figure}{Visualisation of a $(n,m)=(4,1)$ cycle in the complex $\wp(z)$ plane (left figure) and in the fundamental $z$-domain (right figure). The $\Z_2$ fixed point at $z=0$ (no. 1), lies at $\wp(z)=\infty$. Note that due to the double cover structure, the cycle in the $\wp$ plane corresponds to a curve from $-1$ to $\infty$ which has the same winding behaviour as half the cycle in the $z$-plane.}
\label{Fig:41cycle1}  
 \end{minipage}

\paragraph{Choice of orientifold involution}

In traditional orientifold models on toroidal orbifolds, the orientifold involution was chosen such that the O-planes topologically only contained bulk parts. However, here we have seen that, for the simplest antiholomorphic action on the homogeneous coordinates ($\sigma: (x_i,v_i) \mapsto (\overline{x}_i,\overline{v}_i)$), one can obtain different sets of orientifold fixed planes. If $\sigma: y \mapsto \overline{y}$ and the deformation was such that $\varepsilon >0$, the $\sigma$ fixed set was the union of all purely horizontal and purely vertical fractional cycles of calibration $\Re(\Omega)$, each with negative exceptional part , see e.g.\ figure \ref{Fig:LagsInX1X2PlaneDeformed2}. If, on the other hand, $\sigma: y \mapsto -\overline{y}$ (equivalently one could deform in $\varepsilon<0$ direction), the fixed set is one large cycle which consists of all mixed horizontal-vertical cycles such that the exceptional parts cancel out. Thus, it seems that depending on the direction of the deformation we get either O-planes with or without exceptional part. In the orbifold CFT construction T-dual to the generalised Gimon--Polchinski model of section \ref{Sss:T4Z2}, one was required to use fractional cycles whose exceptional part gets a sign w.r.t\ the bulk part under $\Omega\mathcal{R}$ which stay only {\it Lag}, but loose the {\it sLag} property under deformations, or in other words, such deformations would break ${\cal N}=1$ SUSY. 

One can build models on the deformed smooth spaces in which the O-plane has exceptional contributions and where fractional cycles are {\it sLag}, and ask the question to what these would correspond in the orbifold limit. 

Furthermore, we have found involutions $(x_1,v_1)\leftrightarrow(\overline{x}_2,\overline{v}_2)$ whose fixed set contains (integer multiples of) exceptional cycles. It would be interesting to see also in this case if such involutions have a correspondence of a fully fledged string theory compactification at the orbifold point.

\section{Discussion and Conclusions}\label{S:Conclusions}

In this article, we focussed on {\it sLag} cycles in type II orientifold models on $T^4/\Z_2$ and $T^6/\Z_2\times\Z_2$ orbifolds. In particular, we were interested in the cycles when deforming the $\Z_2$ fixed points or fixed planes. We used the formalism of describing these orbifolds as  hypersurfaces in some toric variety, which allowed us to switch on deformations. Within this framework, a certain set of {\it sLag} cycles was described as fixed set under antiholomorphic involutions, which are potential candidates for orientifold symmetries. We saw how these symmetries set restrictons on the complex structure deformation parameters. Whereas at the orbifold point, one could only see the bulk part of a fractional cycle, on the deformation one could, depending on the deformation itself, identify the exceptional part. To this end, it was useful to explicitly compute integrals of the holomorphic volume form over these cycles. Furthermore, due to the {\it sLag} property, these integrals are equal to the volume of these cycles which in turn gives information about e.g.\ the tree level gauge coupling in the low-energy effective field theory. 
As expected, the value of the gauge coupling hardly changed for a D-brane away from the deformed singularity, whereas D-branes wrapped on the deformation locus showcased a power law dependence on the relevant modulus.

Another result is that certain sets of cycles had to merge with each other, because their fractional part had a different calibration from the bulk part. This would imply that such a deformation is only allowed if all cycles within such sets have the same number of D-branes wrapped around them. This is also visible when writing down the D-terms for the U(1) factors which descend from such D-branes. 

One observation was that the more deformations are switched on, the more complicated are the precise descriptions of the cycles. E.g.\ the boundaries of the projection to the real plane become non-trivial to describe. 
Moreover, the observation in section~\ref{Sss:T6Z22-AllDeform} that three-cycles parallel to the $\OR$-invariant plane behave differently under deformation from those parallel to some $\OR\Z_2^{(k)}$-invariant plane seems to suggest that T-duality arguments used to permute all four kinds of O6-planes cannot be applied to deformed orbifolds.

As a next step, we plan to generalize this formalism to more general D-brane models, ideally to phenomenologically appealing ones. This will include orbifolds of tilted tori and orbifold groups of higher order as well as D-branes with more general winding numbers. In addition, the computation of integrals of the holomorphic volume form become highly involved on more general deformations, which is a problem that will be investigated in the future~\cite{BlaszczykHoneckerKoltermann:2014}. It will also be interesting to see if the discrete symmetries of~\cite{Honecker:2013hda,Honecker:2013kda} survive after deformation. Moreover, the deformations are also expected to affect the perturbative Yukawa couplings in models with fractional D6-branes~\cite{Honecker:2012jd,Honecker:2012fn,Honecker:2013kda} as well as the non-perturbative couplings generated by D2-brane instantons~\cite{Blumenhagen:2006xt,Ibanez:2006da,Cvetic:2007ku,Blumenhagen:2009qh}. 

Finally, the formalism of describing {\it sLag} cycles on deformed toroidal orbifolds is {\it a priori} a purely geometric construction which might also be applicable to other string / D-brane models.

\noindent
{\bf Acknowledgements:} 
It is a pleasure to thank Wieland Staessens for helpful discussions.
This work is partially supported by the {\it Cluster of Excellence `Precision Physics, Fundamental Interactions and Structure of Matter' (PRISMA)} DGF no. EXC 1098,
the DFG research grant HO 4166/2-1, the DFG Research Training Group {\it `Symmetry Breaking in Fundamental Interactions'} GRK 1581,
and the Research Center {\it `Elementary Forces and Mathematical Foundations' (EMG)} at JGU Mainz.

\vspace{10mm}

\appendix
\section[$\Z_2$ Twisted Sectors of $T^4/\Z_6$ and $T^6/\Z_6^{(\prime)} \times \Z_2$]{$\boldsymbol{\Z_2}$ Twisted Sectors of $\boldsymbol{T^4/\Z_6}$ and $\boldsymbol{T^6/\Z_6^{(\prime)} \times \Z_2}$}\label{A:Z6}

In this appendix, we briefly summarise the changes compared to the $T^4/\Z_2$ and $T^6/\Z_2 \times \Z_2$ orbifolds of sections~\ref{Ss:T4Z2N} and~\ref{Ss:T6Z2Z2N} if an additional $\Z_3$ symmetry is imposed. 
For each of the three orbifolds discussed here, the identification of $\Z_3$ images leads to a reduced number of blow-up or deformation moduli.

\subsection[$\Z_2$ twisted sectors of $T^4/\Z_6$ with shift vector $\vec{v}=\frac{1}{6}(1,-1)$]{$\boldsymbol{\Z_2}$ twisted sectors of $\boldsymbol{T^4/\Z_6}$ with shift vector $\boldsymbol{\vec{v}=\frac{1}{6}(1,-1)}$}\label{Sss:T4Z6}

In the context of tilted tori, we have seen in section~\ref{Ss:T4Z2N} that the number of independent blow-ups or deformations is reduced by ${\cal R}$-identifications of different $\Z_2$ fixed points. 
If the point group $\Z_{2N}$ (with generator $\theta$) is larger than just $\Z_2$, additional identifications occur as exemplified in figure~\ref{Fig:Z6lattice} for $\Z_6$.
The $\Z_3$ subsymmetry enforces the shape of tilted tori to be that of hexagonal $SU(3)$ lattices with two different crystallographically allowed ${\cal R}$-invariant axes,
and the number of independent factorisable bulk two-cycles is reduced from four to two,
\begin{equation}\label{Eq:Def-Z6-bulk}
\Pi^{\text{bulk}}_a = X^a \tilde{\varrho}_1 + Y^a \tilde{\varrho}_2
\qquad
\text{with}
\quad
\left\{\begin{array}{c}
\tilde{\varrho}_1= \sum_{i=0}^5 \theta^i (\pi_1 \otimes \pi_3) , \quad \tilde{\varrho}_2= \sum_{i=0}^5 \theta^i (\pi_1 \otimes \pi_4)
\\ (X^a,Y^a) = (n_1^a n_2^a - m_1^a m_2^a \, , \, n_1^a m_2^a + m_1^a n_2^a + m_1^a m_2^a)
\end{array}\right.
.
\end{equation}
The bulk supersymmetry condition in the language of equation~(\ref{Eq:Def-Z}) with \linebreak \mbox{${\cal Z}_a = \frac{2}{\sqrt{3}}  \prod_{k=1}^2 e^{-\pi i \tilde{\phi}_k} \left(n_k^a + e^{\frac{\pi i }{3}} m_k^a \right)$}
and $\tilde{\phi}_k=0, \frac{1}{6}$ per {\bf A}- and {\bf B}-lattice orientation, respectively, projects onto a one-dimensional subspace of one-cycles.
While naively there exist three different choices for $T^2_{(1)} \times T^2_{(2)}$, the number of tensor multiplets $n_T=1$ on {\bf AA} and {\bf BB} and $n_T=3$ on {\bf AB}~\cite{Blumenhagen:2002wn} points to only two physically
distinct background choices.

There exist three different kinds of twist sectors for the $T^4/\Z_6$ orbifold: the origin (point 1 in figure~\ref{Fig:Z6lattice}) is fixed under the full $\Z_6$ symmetry
and thus supports five exceptional divisors $f_{11}^{(i)}$ ($i=1 \ldots 5$) with the intersection form given by  $-C(A_5)$. Points 2 and 3 are fixed under the $\Z_3$ subgroup and support eight exceptional divisors 
($d_{12}^{(i)} \stackrel{\Z_2}{\longleftrightarrow} d_{13}^{(i)}$, $d_{21}^{(i)} \stackrel{\Z_2}{\longleftrightarrow} d_{31}^{(i)}$,  $d_{22}^{(i)} \stackrel{\Z_2}{\longleftrightarrow} d_{33}^{(i)}$ and $d_{23}^{(i)} \stackrel{\Z_2}{\longleftrightarrow} d_{32}^{(i)}$ for $i=1,2$) with intersection form $-C(A_2)$ per $d_{\alpha\beta}^{(i)}$ for fixed $\alpha\beta$.
Finally, points $4\stackrel{e^{i\pi/3}}{\longrightarrow} 5\stackrel{e^{i\pi/3}}{\longrightarrow} 6$ are fixed under the $\Z_2$ subgroup and support one exceptional divisor $\epsilon_{k}$ per triplet of $\Z_2$ fixed points $e_{\alpha\beta}$ with self-intersection number $-2$ per fixed $\alpha\beta$,
\begin{equation}\label{Eq:Def_epsilon_T4Z6}
\begin{array}{lll}
 \epsilon_0= 3\, e_{11},  & \epsilon_1 = e_{41} + e_{51} + e_{61}, \qquad &  \epsilon_2 = e_{14} + e_{15} + e_{16},  \\
  \epsilon_3 = e_{44} + e_{56} + e_{65}, \qquad  & \epsilon_4 = e_{45} + e_{54} + e_{66}, \qquad & \epsilon_5 = e_{46} + e_{55} + e_{64} ,
\end{array}
\end{equation}
where we have formally singled out $e_{11}$ from $f_{11}^{(i)}$ as the exceptional divisor at the origin that is associated to the $\Z_2$ twisted sector.

As in the $T^4/\Z_2$ case, any fractional two-cycle on $T^4/\Z_6$ consists of a given bulk cycle with the combinatorics of assigning exceptional cycles in dependence of the even- or oddness of $(n_k,m_k)$ 
given in table~\ref{Tab:Reference_Point+Signs} for convenience.
\begin{SCtable}
$\begin{array}{|c|ccc|}\hline
\muc{4}{|c|}{\text{\bf Assignment of prefactors $1$ or $(-1)^{\tau_i^a}$ }}
\\\hline\hline
(n_i^a,m_i^a) & \;\;\text{(o,o)} \;\;\stackrel{\pi/3}{\longrightarrow} & \;\text{(o,e)} \;\;\stackrel{\pi/3}{\longrightarrow} & \text{(e,o)}
\\\hline\hline
\sigma_i^a=0 & \quad \tarh{1}{6} \quad \longrightarrow  & \:\:\:\tarh{1}{4} \quad \longrightarrow & \tarh{1}{5}
\\\hline
\sigma_i^a=1 & \quad \tarh{4}{5} \quad \longrightarrow & \:\:\:\tarh{5}{6} \quad \longrightarrow  & \tarh{6}{4}   
\\\hline
\end{array}$
\caption{Consistent assignment of the reference point (upper entry)  and the second $\Z_2^{(i)}$ fixed point (lower entry) contributing with  sign factor  $+1$ or $(-1)^{ \tau_i^a}$ to 
$\Pi^{\Z_2^{(j),j\neq i}}_a$ in dependence of the even- or oddness of the toroidal wrapping numbers, cf.~\cite{Honecker:2012qr} for details.
}
\label{Tab:Reference_Point+Signs}
\end{SCtable}
The subtlety concerning the correct identification of the $\Z_2$ twisted cycle in the origin can be avoided in D7-brane models by restricting to some non-vanishing displacement $(\vec{\sigma}) \neq (\vec{0})$,
but we refrain from discussing deformations of orbifold singularities at this point. The same reasoning holds for deformations  of the $T^6/\Z_6 \times \Z_2$ orbifold in section~\ref{Sss:Z2Z6}, 
whereas for $T^6/\Z_6' \times \Z_2$ there exists a clear geometrical separation of $\Z_2$ singularities from $\Z_6$ and $\Z_3$ singularities as discussed in detail in appendix~\ref{Sss:Z2Z6p}. 

To close the discussion of the $T^4/\Z_6$ orbifold, the counting of independent deformations or blow-ups proceeds as follows: the $\Z_3$ identifications in equation~(\ref{Eq:Def_epsilon_T4Z6}) reduce the number of $\Z_2$ 
fixed points from 16 to six independent ones. The orientifold action
\begin{equation}
\begin{aligned}
\text{all lattices:} \quad& \epsilon_0 \stackrel{{\cal R}}{\longrightarrow} -\epsilon_0 ,\qquad \epsilon_1 \stackrel{{\cal R}}{\longrightarrow} -\epsilon_1 ,\qquad \epsilon_2 \stackrel{{\cal R}}{\longrightarrow} -\epsilon_2 ,\qquad\\
{\bf AA:} \quad &\epsilon_3 \stackrel{{\cal R}}{\longrightarrow} -\epsilon_3 ,\qquad \epsilon_4 \stackrel{{\cal R}}{\longleftrightarrow} -\epsilon_5,\\
{\bf AB:} \quad & \epsilon_5 \stackrel{{\cal R}}{\longrightarrow} -\epsilon_5 ,\qquad  \epsilon_3 \stackrel{{\cal R}}{\longleftrightarrow} -\epsilon_4,\\
{\bf BB:} \quad & \epsilon_4 \stackrel{{\cal R}}{\longrightarrow} -\epsilon_4 ,\qquad  \epsilon_3 \stackrel{{\cal R}}{\longleftrightarrow} -\epsilon_5,
\end{aligned}
\end{equation}
leads to one more identification such that, in the end, five deformations remain independent.

\subsection[$\Z_2 \times \Z_2$ twisted sectors of $T^6/\Z_6 \times \Z_2$ with $\vec{v}=\frac{1}{6}(1,-1,0)$ and $\vec{w}=\frac{1}{2}(0,1,-1)$]{$\boldsymbol{\Z_2 \times \Z_2}$ twisted sectors of $\boldsymbol{T^6/\Z_6 \times \Z_2}$ with $\boldsymbol{\vec{v}=\frac{1}{6}(1,-1,0)}$ and $\boldsymbol{\vec{w}=\frac{1}{2}(0,1,-1)}$}\label{Sss:Z2Z6}

The $T^6/\Z_6 \times \Z_2$ orbifold with discrete torsion contains - besides the $\Z_2 \times \Z_2$ sectors in section~\ref{Sss:Z2Z2} -  also $T^4/\Z_6 \times T^2$ as a subsector, as can be seen from the Hodge numbers,
\begin{equation}
\begin{aligned}
h_{11} =19 =& 3_{\text{bulk}} + (2 \times 4)_{\Z_6'} + 8_{\Z_3},\\
h_{21} =19 =& 1_{\text{bulk}} + 2_{\Z_6} + 2_{\Z_3} + \left(2 \times 4 + 6 \right)_{\Z_2},
\end{aligned}
\end{equation}
where $\Z_6'$ corresponds to the sector twisted by $\vec{v}+\vec{w}$. 

Any {\it factorisable} bulk three-cycle takes the form~\cite{Forste:2010gw},
\begin{equation}
\begin{aligned}
&\Pi^{\text{bulk}}_a = P^a \, \varrho_1 + Q^a \, \varrho_2 + U^a \, \varrho_3 + V^a \, \varrho_4 \\
&\qquad \qquad \text{with} \quad  \left\{\begin{array}{c} \varrho_{i} = 2 \, \tilde{\varrho}_i \otimes \pi_5 , \quad \varrho_{i+2} = 2 \, \tilde{\varrho}_i \otimes \pi_6 \\(P^a, Q^a, U^a,V^a) = ( X^a \, n_3^a \, , \, Y^a n_3^a \, , \,  X^a \, m_3^a \, , \, Y^a m_3^a )\end{array} \right. ,
\end{aligned}
\end{equation}
and $X^a,Y^a$ and $\tilde{\varrho}_i$ as defined in equation~(\ref{Eq:Def-Z6-bulk}). The bulk supersymmetry conditions are derived from 
${\cal Z}_a = \left[ \frac{2}{\sqrt{3}}  \prod_{k=1}^2 e^{-\pi i \tilde{\phi}_k} \left(n_k^a + e^{\frac{\pi i }{3}} m_k^a \right) \right]\times  \frac{n_3^a R_1^{(3)} + i \, \tilde{m}_3^a R_2^{(3)}}{\sqrt{R_1^{(3)}R_2^{(3)}}}$ 
with again $\tilde{\phi}_k=0, \frac{1}{6}$ per {\bf A}- and {\bf B}-lattice orientation, respectively.

The three-cycles in the $\Z_2^{(3)}$ twisted sector are obtained from $T^4/\Z_6$ by tensoring with a toroidal one-cycle along $T^2_{(3)}$,
\begin{equation}
\varepsilon^{(3)}_{\alpha} = 2 \, \epsilon_{\alpha} \otimes \pi_5 , \qquad  \tilde{\varepsilon}^{(3)}_{\alpha} = 2 \, \epsilon_{\alpha}  \otimes \pi_6  ,
\end{equation}
with the $\Z_3$ orbits $\epsilon_{\alpha}$ of $\Z_2$ fixed points defined in equation~(\ref{Eq:Def_epsilon_T4Z6}).
A basis of exceptional three-cycles in the $\Z_2^{(1)}$ twisted sector is given by~\cite{Forste:2010gw},
\begin{equation}
\mbox{\resizebox{0.9\textwidth}{!}{%
$\varepsilon^{(1)}_{\alpha}=2 \, \left(\pi_1 \otimes e^{(1)}_{4\alpha} + \pi_{-2} \otimes e^{(1)}_{6\alpha} + \pi_{2-1} \otimes e^{(1)}_{5\alpha}  \right) , \quad
\tilde{\varepsilon}^{(1)}_{\alpha} = 2 \, \left(\pi_2 \otimes e^{(1)}_{4\alpha} + \pi_{1-2} \otimes e^{(1)}_{6\alpha} + \pi_{-1} \otimes e^{(1)}_{5\alpha} \right)
$}},
\end{equation}
and the $\Z_2^{(2)}$ twisted sector has an identical basis upon permutation of two-torus indices.
The number of independent complex structure deformations is thus reduced from $3_{\text{bulk}} + (3 \times 16)_{\Z_2}$ for $T^6/\Z_2 \times \Z_2$ with discrete torsion
to $1_{\text{bulk}}  + (6 + 2 \times 4)_{\Z_2}$ on $T^6/\Z_6 \times \Z_2$ with discrete torsion due to the $\Z_3$ subsymmetry.

The multiplicities $(h_{11}^-,h_{11}^+)$ of K\"ahler moduli and closed string vector multiplets computed in~\cite{Forste:2010gw} (cf.~also table~\ref{Tab:closed-OR-spectrum}) provide a first hint that the choice of \textbf{\textit{a}}- or \textbf{\textit{b}}-type lattice on $T^2_{(3)}$ leads to physically inequivalent string models. The assumption of only two equivalent lattice backgrounds is further confirmed by a closer inspection of RR tadpole cancellation and supersymmetry conditions as well as massless open string spectra~\cite{EckerHoneckerStaessens:2014}. Focussing on the {\textbf{AA\textit{a}/\textit{b}}} formulation, the orientifold projection on the exceptional three-cycles at $\Z_2$ fixed points is given in table~\ref{Tab:Z2Z6-exOrientifoldImages},
\begin{table}[ht]
\renewcommand{\arraystretch}{1.3}
\begin{center}
\begin{equation*}
\begin{array}{|c||c|c||c|c|}\hline
\multicolumn{5}{|c|}{\OR \; \text{\bf  on exceptional three-cycles for $T^6/\Z_6 \times \Z_2$ on AA\textit{a}/\textit{b}}}
\\\hline\hline
k & \OR (\varepsilon^{(k)}_{\alpha} )& \OR(\tilde{\varepsilon}^{(k)}_{\alpha}) & \alpha=\alpha' & \alpha \leftrightarrow \alpha'
\\\hline\hline
1,2 &  - \eta_{(k)} \, \varepsilon^{(k)}_{\alpha'} &  \eta_{(k)} \left( \tilde{\varepsilon}^{(k)}_{\alpha'} -\varepsilon^{(k)}_{\alpha'} \right) & 1,4 & \begin{array}{c} 2,2+2b \\ 3,3-2b\end{array}
\\\hline
3 & \eta_{(3)} \left( -\varepsilon^{(3)}_{\alpha'} + (2b) \tilde{\varepsilon}^{(3)}_{\alpha'} \right) & \eta_{(3)} \, \tilde{\varepsilon}^{(3)}_{\alpha'} & 0,1,2,3  & 4,5 
\\\hline
\end{array}
\end{equation*}
 \end{center}
\caption{Orientifold projection on exceptional three-cycles in the $\Z_2^{(k)}$ twisted sector on the {\textbf{AA\textit{a}/\textit{b}}} lattice orientation of $T^6/(\Z_6 \times \Z_2 \times \OR)$ with discrete torsion. The
sign factor \mbox{$\eta_{(k)} \equiv \eta_{\OR} \eta_{\OR\Z_2^{(k)}}$} depends on the choice of the exotic O6-plane. }
\label{Tab:Z2Z6-exOrientifoldImages}
\end{table}
from which we can read off the $h_{21}^{\text{twisted}}$ ${\cal R}$-even cycles in dependence of the choice of exotic O6-plane. 
These are exactly the cycles to which the independent complex structure deformations in table~\ref{Tab:closed-OR-spectrum} are associated.

\subsection[$\Z_2 \times \Z_2$ twisted sectors of $T^6/\Z_6' \times \Z_2 $ with $\vec{v}=\frac{1}{6}(1,1,-2)$ and $\vec{w}=\frac{1}{2}(0,1,-1)$]{$\boldsymbol{\Z_2 \times \Z_2}$ twisted sectors of $\boldsymbol{T^6/\Z_6' \times \Z_2}$ with $\boldsymbol{\vec{v}=\frac{1}{6}(1,1,-2)}$ and $\boldsymbol{\vec{w}=\frac{1}{2}(0,1,-1)}$}\label{Sss:Z2Z6p}

The Hodge numbers on the $T^6/\Z_6' \times \Z_2$ orbifold with discrete torsion,
\begin{equation}
\begin{aligned}
h_{11} =15=& 3_{\text{bulk}} + (3 \times 1)_{\Z_6} + 9_{\Z_3} ,\\
h_{21} =15 =& (3 \times 5)_{\Z_2},
\end{aligned}
\end{equation}
display a clear separation of blow-up (K\"ahler) modes in the $\Z_6$ and $\Z_3$ twisted sectors and complex structure deformations in the $\Z_2$ twisted sectors.
Any bulk three-cycle can be expanded as $\Pi^{\text{bulk}}_a = X_a \, \varrho_1 + Y_a \, \varrho_2$ with 
\begin{equation}
\begin{aligned}
\varrho_1 =& \sum_{k=0}^5 \sum_{l=0}^1 \theta^k \omega^l (\pi_1 \otimes \pi_3 \otimes \pi_5), \qquad  X^a = n_1^a n_2^a n_3^a - m_1^a m_2^a m_3^a - \sum_{i \neq j \neq k \neq i} n_i^a m_j^a m_k^a,\\
\varrho_2 =& \sum_{k=0}^5 \sum_{l=0}^1 \theta^k \omega^l (\pi_1 \otimes \pi_3 \otimes \pi_6),\qquad  Y^a = \sum_{i \neq j \neq k \neq i} \left(n_i^a n_j^a m_k^a + n_i^a m_j^a m_k^a \right),
\end{aligned}
\end{equation}
where $\theta$ and $\omega$ are the generators of $\Z_6'$ and $\Z_2$, respectively, and the bulk supersymmetry condition is encoded in ${\cal Z}_a = \left(\frac{2}{\sqrt{3}}\right)^{3/2}  \prod_{k=1}^3 e^{-\pi i \tilde{\phi}_k} \left(n_k^a + e^{\frac{\pi i }{3}} m_k^a \right)$, cf.~\cite{Forste:2010gw} for details.

The $T^6/\Z_6' \times \Z_2$ orbifold with discrete torsion has three equivalent $\Z_2^{(k)}$ twisted sectors with a basis of exceptional three-cycles given by
\begin{equation}
\mbox{\resizebox{0.92\textwidth}{!}{%
$\begin{aligned}
\varepsilon^{(k)}_1 =2 \, \left( e^{(k)}_{41} -  e^{(k)}_{61} \right) \otimes \pi_{2k-1} +2 \, \left( e^{(k)}_{61} -  e^{(k)}_{51} \right) \otimes \pi_{2k},
&\quad \tilde{\varepsilon}^{(k)}_1 =2 \, \left( e^{(k)}_{51} -  e^{(k)}_{61} \right) \otimes \pi_{2k-1} +2 \, \left( e^{(k)}_{41} -  e^{(k)}_{51} \right) \otimes \pi_{2k}, \\
\varepsilon^{(k)}_2 =2 \, \left( e^{(k)}_{14} -  e^{(k)}_{16} \right) \otimes \pi_{2k-1} +2 \, \left( e^{(k)}_{16} -  e^{(k)}_{15} \right) \otimes \pi_{2k}, 
&\quad  \tilde{\varepsilon}^{(k)}_2 =2 \, \left( e^{(k)}_{15} -  e^{(k)}_{16} \right) \otimes \pi_{2k-1} +2 \, \left( e^{(k)}_{14} -  e^{(k)}_{15} \right) \otimes \pi_{2k}, \\
\varepsilon^{(k)}_3 =2 \, \left( e^{(k)}_{44} -  e^{(k)}_{66} \right) \otimes \pi_{2k-1} +2 \, \left( e^{(k)}_{66} -  e^{(k)}_{55} \right) \otimes \pi_{2k},
&\quad \tilde{\varepsilon}^{(k)}_3 =2 \, \left( e^{(k)}_{55} -  e^{(k)}_{66} \right) \otimes \pi_{2k-1} +2 \, \left( e^{(k)}_{44} -  e^{(k)}_{55} \right) \otimes \pi_{2k}, \\
\varepsilon^{(k)}_4 =2 \, \left( e^{(k)}_{45} -  e^{(k)}_{64} \right) \otimes \pi_{2k-1} +2 \, \left( e^{(k)}_{64} -  e^{(k)}_{56} \right) \otimes \pi_{2k},
&\quad \tilde{\varepsilon}^{(k)}_4 =2 \, \left( e^{(k)}_{56} -  e^{(k)}_{64} \right) \otimes \pi_{2k-1} +2 \, \left( e^{(k)}_{45} -  e^{(k)}_{56} \right) \otimes \pi_{2k}, \\
\varepsilon^{(k)}_5 =2 \, \left( e^{(k)}_{46} -  e^{(k)}_{65} \right) \otimes \pi_{2k-1} +2 \, \left( e^{(k)}_{65} -  e^{(k)}_{54} \right) \otimes \pi_{2k},
&\quad \tilde{\varepsilon}^{(k)}_5 =2 \, \left( e^{(k)}_{54} -  e^{(k)}_{65} \right) \otimes \pi_{2k-1} +2 \, \left( e^{(k)}_{46} -  e^{(k)}_{54} \right) \otimes \pi_{2k}.
\end{aligned}$}}
\end{equation}

As demonstrated in~ \cite{Honecker:2012qr} in terms of the RR tadpole cancellation and supersymmetry conditions as well as massless spectra, 
only two orientations {\bf AAA} and {\bf BBB} provide physically inequivalent models. Moreover, only {\bf AAA} is suitable for D6-brane model building
due to the largest possible rank of 16 in combination with completely rigid D6-branes without matter in the (anti)symmetric representation.
The orientifold projection on exceptional three-cycles is again associated to complex structure deformations as discussed in section~\ref{S:sLags}.
\begin{table}[h!]
\renewcommand{\arraystretch}{1.3}
  \begin{center}
\begin{equation*}
\begin{array}{|c|c||c|c|}\hline
\multicolumn{4}{|c|}{\OR \; \text{\bf  on exceptional three-cycles on $T^6/\Z_6' \times \Z_2$ on {\bf AAA}}}
\\\hline\hline
 \OR ( \varepsilon^{(k)}_{\alpha}) & \OR ( \tilde{\varepsilon}^{(k)}_{\alpha}) & \alpha=\alpha' & \alpha \leftrightarrow \alpha'
\\\hline\hline
  - \eta_{(k)} \, \varepsilon^{(k)}_{\alpha'} & \eta_{(k)} \left( \tilde{\varepsilon}^{(k)}_{\alpha'} -\varepsilon^{(k)}_{\alpha'} \right) &  1,2,3 & 4,5
\\\hline 
\end{array}
\end{equation*}
\end{center}
\caption{Orientifold projection of the exceptional three-cycles from the $\Z_2^{(k), k \in \{1,2,3\}}$ twisted sector on the {\bf AAA} background of 
$T^6/(\Z_6' \times \Z_2 \times \OR)$ with discrete torsion.  In the example below $\eta_{\OR}=-1 = \eta_{(k), k \in\{1,2,3\}}$ is chosen.}
\label{Tab:Z2Z6p-OR-ex}
\end{table}
According to table~\ref{Tab:Z2Z6p-OR-ex}, for the choice $\eta_{\OR}=-1$ of the exotic O6-plane, the  relevant ${\cal R}$-even cycles are $\varepsilon^{(k)}_{\alpha}$
for $\alpha \in \{1,2,3\}$ plus $(\varepsilon^{(k)}_4 + \varepsilon^{(k)}_5)$ and $(-\varepsilon^{(k)}_4 + \varepsilon^{(k)}_5 +2 \tilde{\varepsilon}^{(k)}_4 -2 \tilde{\varepsilon}^{(k)}_5)$.

The global five-stack Pati-Salam model~\cite{Honecker:2012qr} in table~\ref{tab:Z6pZ2-PSmodel} displays some key features discussed in section~\ref{S:sLags}: the branes $a,b,c$ have identical bulk three-cycles and thus 
identical tree level gauge couplings at the orbifold point, while the dependence on deformation moduli differs due to the diverse choice of sign factors. D6-brane $d$ is chosen such 
that its gauge coupling only experiences the deformation of the $\Z_2$ fixed point orbits involving $(\varepsilon^{(k)}_{\alpha}, \tilde{\varepsilon}^{(k)}_{\alpha})$ for $k=1,2$ and $\alpha=4,5$,
and only $\varepsilon^{(3)}_3$.
Finally, D6-brane $e$ is parallel to the $\OR$-invariant orbit, but due to the choice $\tau_1^e=\tau_2^e \neq 0$ of discrete Wilson lines, the rigid three-cycle is not orientifold invariant. The 
associated gauge coupling only depends on the deformation moduli in the $\Z_2^{(3)}$ twisted sector.
 \mathtabfix{
\begin{array}{|c|c|c|c|c|c|c|c|}\hline
\muc{6}{|c|}{\text{\bf  D6-brane configuration of a global Pati-Salam model on $T^6/\Z_6' \times \Z_2$}}
\\\hline\hline
x & \!\!\!\left(\!\!\!\begin{array}{c} n_1,m_1 \\ n_2,m_2 \\ n_3,m_3 \end{array}\!\!\!\right)\!\!\! & (\vec{\sigma}) & \Z_2 & (\vec{\tau})
& \Pi_x + \Pi_{x'}
\\\hline\hline
\begin{array}{c} a\\ b\\c \end{array} & \left(\!\!\!\begin{array}{c} 0,1\\1,0\\1,-1\end{array}\!\!\!\right) & (\vec{1}) & \!\!\!\begin{array}{c} (+++)\\(--+)\\(-+-) \end{array}\!\!\!  & \!\!\!\begin{array}{c} (0,0,1)^T\\(0,1,1)^T\\(1,0,1)^T \end{array}\!\!\! &
\frac{\rho_1}{2} + \frac12  
{ \scriptsize \left\{\!\!\!\begin{array}{cc}
\varepsilon^{(1)}_3   -  \varepsilon^{(1)}_5 - \tilde{\varepsilon}^{(1)}_4  + \tilde{\varepsilon}^{(1)}_5  +     \varepsilon^{(2)}_3    -  \varepsilon^{(2)}_5 -  \tilde{\varepsilon}^{(2)}_4 + \tilde{\varepsilon}^{(2)}_5  -  \varepsilon^{(3)}_3    & a \\
- \varepsilon^{(1)}_3 + \varepsilon^{(1)}_4  -   \tilde{\varepsilon}^{(1)}_4  + \tilde{\varepsilon}^{(1)}_5  - \varepsilon^{(2)}_3  + \varepsilon^{(2)}_5 +  \tilde{\varepsilon}^{(2)}_4 - \tilde{\varepsilon}^{(2)}_5  +  \varepsilon^{(3)}_3  - \varepsilon^{(3)}_5  -  \tilde{\varepsilon}^{(3)}_4 + \tilde{\varepsilon}^{(3)}_5 & b \\
 - \varepsilon^{(1)}_3   +  \varepsilon^{(1)}_5 + \tilde{\varepsilon}^{(1)}_4   - \tilde{\varepsilon}^{(1)}_5  -  \varepsilon^{(2)}_3  +  \varepsilon^{(2)}_4   -  \tilde{\varepsilon}^{(2)}_4 + \tilde{\varepsilon}^{(2)}_5  + \varepsilon^{(3)}_3  +  \varepsilon^{(3)}_4 + \varepsilon^{(3)}_5  & c \end{array}\right. }
\\\hline
d & \left(\!\!\!\begin{array}{c} -1,2\\2,-1\\1,-1\end{array}\!\!\!\right) & (\vec{1}) & (--+) &  \left(\!\!\!\begin{array}{c}0\\0\\1  \end{array}\!\!\!\right)  &
\frac{3 \, \rho_1}{2}  + \frac{- \varepsilon^{(1)}_4  +  \varepsilon^{(1)}_5  + 2 \, \left\{ \tilde{\varepsilon}^{(1)}_4 -  \tilde{\varepsilon}^{(1)}_5 \right\} - \varepsilon^{(2)}_4   +  \varepsilon^{(2)}_5  + 2 \,  \left\{ \tilde{\varepsilon}^{(2)}_4 - \tilde{\varepsilon}^{(2)}_5 \right\}  -   \varepsilon^{(3)}_3}{2}
\\\hline
e &\left(\!\!\!\begin{array}{c} 1,0\\1,0\\1,0 \end{array}\!\!\!\right) &  \left(\!\!\!\begin{array}{c}1\\1\\0\end{array}\!\!\!\right)  & (+--) &  (\vec{1}) &
\frac{\rho_1}{2}+ \frac{  \varepsilon^{(3)}_3 -  \varepsilon^{(3)}_4  + \tilde{\varepsilon}^{(3)}_4 -  \tilde{\varepsilon}^{(3)}_5  }{2}
\\\hline
\end{array}
}{Z6pZ2-PSmodel}{D6-brane data for a global Pati-Salam model on five stacks for $\eta_{\OR}=-1$ of the $T^6/\Z_6' \times \Z_2$ orientifold with discrete torsion.
The corresponding gauge couplings $g_{SU(N)_x}^{-2} \propto \text{Vol}(\Pi_a + \Pi_{a'})$ only depend on a subset of deformation moduli as displayed in the last column.}

\vspace{10mm}


\addcontentsline{toc}{section}{References}
\bibliographystyle{ieeetr}

\end{document}